\documentclass[prd,twocolumn,nofootinbib,aip,preprintnumbers,amsmath,amssymb,reprint]{revtex4-1}
\usepackage{natbib,hyperref}
\usepackage{amsmath}
\usepackage{stmaryrd}
\usepackage{braket}
\usepackage{graphicx}
\usepackage{dcolumn}
\usepackage{tabularx}
\usepackage{bm}
\usepackage{epsfig,color,xspace,multirow,xr,bbold}
\usepackage[all]{xy}
\usepackage{setspace}
\usepackage{url}
\usepackage[normalem]{ulem}
\usepackage{threeparttable, tablefootnote}
 
\usepackage{natbib,hyperref}
\usepackage{float}
\usepackage{placeins}
\usepackage[utf8]{inputenc}
\usepackage[dvipsnames]{xcolor}

\newcommand{\rxnone}[1]{ROO$\cdot$ $\xrightarrow{\rm TS1}$ $\cdot$QOOH}
\newcommand{\rxntwo}[1]{ROO$\cdot$ $\xrightarrow{\rm TS2}$ alkene+ $\cdot$OOH}
\newcommand{\rxnthree}[1]{$\cdot$QOOH$\xrightarrow{\rm TS3}$ alkene+ $\cdot$OOH}
\newcommand{\rxnfour}[1]{$\cdot$QOOH$\xrightarrow{\rm TS4}$ $cy$-ether+ $\cdot$OH}
\usepackage[none]{hyphenat}
\usepackage{calligra}
\usepackage{xtab,afterpage,longtable}
\usepackage[version=4]{mhchem}

\ExplSyntaxOn
\keys_define:nn {mhchem}
 {
  arrow-min-length .code:n =
   \cs_set:Npn \__mhchem_arrow_options_minLength:n { {#1} } 
 }
\ExplSyntaxOff

\begin{document}

\title[]{
Stereo-Electronic Factors Influencing the Stability of Hydroperoxyalkyl Radicals:
Transferability of Chemical Trends across Hydrocarbons and {\it ab initio} Methods
}

\author{Saurabh Chandra Kandpal}
\affiliation{Tata Institute of Fundamental Research, Hyderabad 500046, India}

\author{Kgalaletso P. Otukile}
\affiliation{Department of Chemistry, University of the Free State, PO Box 339, Bloemfontein 9300, South Africa}

\author{Shweta Jindal}
\affiliation{Tata Institute of Fundamental Research, Hyderabad 500046, India}

\author{Salini Senthil}
\affiliation{Tata Institute of Fundamental Research, Hyderabad 500046, India}

\author{Cameron Matthews}
\affiliation{Department of Chemistry, University of the Free State, PO Box 339, Bloemfontein 9300, South Africa}

\author{Sabyasachi Chakraborty}
\affiliation{Tata Institute of Fundamental Research, Hyderabad 500046, India}

\author{Lyudmila V. Moskaleva}
\email{moskaleval@ufs.ac.za}
\affiliation{Department of Chemistry, University of the Free State, PO Box 339, Bloemfontein 9300, South Africa}

\author{Raghunathan Ramakrishnan}
\email{ramakrishnan@tifrh.res.in}
\affiliation{Tata Institute of Fundamental Research, Hyderabad 500046, India}

\date{\today}

\begin{abstract}
The hydroperoxyalkyl radicals ($\cdot$QOOH) are known to play a significant role in combustion and tropospheric processes, yet their direct spectroscopic detection remains challenging. In this study, we investigate molecular stereo-electronic effects influencing the kinetic and thermodynamic stability of a $\cdot$QOOH along its formation path from the precursor, alkylperoxyl radical (ROO$\cdot$), and the depletion path resulting in the formation of cyclic ether + $\cdot$OH. We focus on reactive intermediates encountered in the oxidation of acyclic hydrocarbon radicals: ethyl, isopropyl, isobutyl, tert-butyl, neopentyl, and their alicyclic counterparts: cyclohexyl, cyclohexenyl, and cyclohexadienyl. We report reaction energies and barriers calculated with the highly accurate method Weizmann-1 (W1) for the channels:
\ce{ROO$\cdot$ <=> $\cdot$QOOH},
\ce{ROO$\cdot$ <=> alkene {+} $\cdot$OOH}, 
\ce{$\cdot$QOOH <=> alkene {+} $\cdot$OOH}, and
\ce{$\cdot$QOOH <=> cyclic ether {+} $\cdot$OH}. Using W1 results as a reference, we have systematically benchmarked the accuracy of popular density functional theory (DFT), composite thermochemistry methods, and an explicitly correlated coupled-cluster method.
We ascertain inductive, resonance, and steric effects on the overall stability of $\cdot$QOOH and computationally investigate the possibility of forming more stable species. With new reactions as test cases, we probe the capacity of various {\it ab initio} methods to yield quantitative insights on the elementary steps of combustion. 
\end{abstract}

\maketitle

\section{Introduction}\label{sec_intro}
Hydrocarbon combustion is a complex process that involves multiple interconnected chemical reactions between hydrocarbon (RH) and atmospheric molecular oxygen (O$_2$) over a wide range of temperatures\cite{franklin1967mechanisms,hucknall2012chemistry,bugler2015revisiting}. These highly exothermic reactions occur through a free radical chain mechanism involving the typical steps: initiation, propagation, and termination.\cite{mayo1968free, orlando2012laboratory}. Hydrocarbon radicals (R$\cdot$) are key intermediates in these mechanisms. They react with O$_2$ to form alkyl peroxy radicals (ROO$\cdot$), which can either propagate the combustion chain by reacting with another hydrocarbon molecule 
or isomerize to form short-lived $\cdot$QOOH radicals. These carbon-centred radicals are generally unstable and undergo unimolecular decomposition, releasing $\cdot$OH and $\cdot$OOH \cite{zador2011kinetics}.

Due to competing decomposition and isomerization pathways, their direct spectroscopic detection has been a longstanding challenge. 
Therefore, in most cases, they are detected indirectly through an analysis of their dissociation products. Thus far, only a few of these 
elusive reactive intermediates relevant to combustion or atmospheric 
chemistry have been observed experimentally\cite{hansen2021watching, savee2015direct, welz2012direct, taatjes2013direct, su2013infrared}. 
Notably, the existence of short-lived $\cdot$QOOH radical was only confirmed by Savee \textit{et al.}\cite{savee2015direct} in 2015, thanks to the unique chemistry of the substrate employed in the investigation. 

\begin{figure*}
        \centering
        \includegraphics[width=1.0\linewidth]{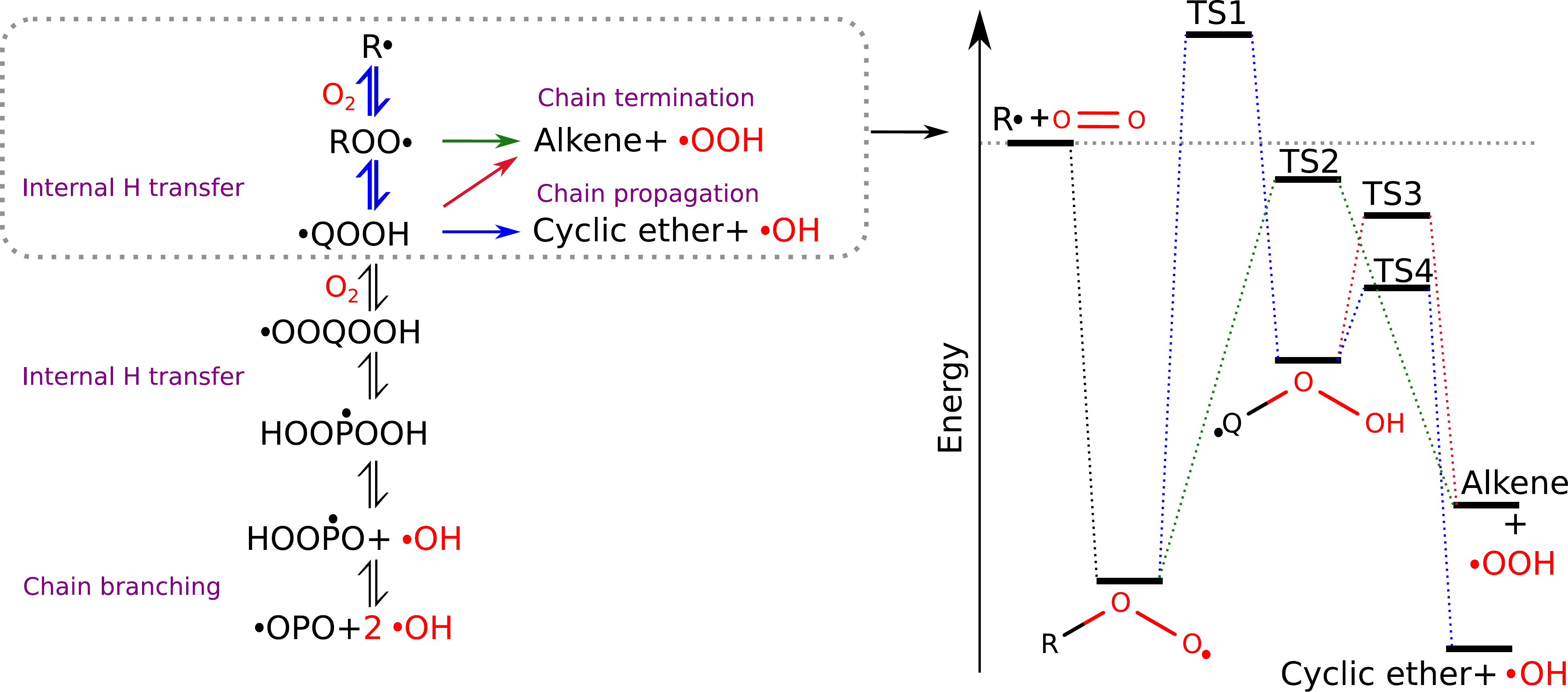}
        \caption{
        Schematic representation of the low-temperature hydrocarbon combustion mechanism and the associated reactive intermediates adapted
        from the work of Z\'{a}dor \textit{et al.}\cite{zador2011kinetics}.
        The reactions investigated in this study are shown in a box on 
        the left, while their corresponding energy profiles are depicted on the right.
        Parent hydrocarbons denoted R, Q, or P contain one less C–H bond following the
        sequence: R$\rightarrow$Q$\rightarrow$P\cite{osborn2017reaction}.
       }
        \label{fig:ReactionScheme}
\end{figure*}

Among the diverse pathways involved in low-temperature combustion, as depicted in FIG.~\ref{fig:ReactionScheme}, a few energetically, kinetically and statistically (wherein different starting species result in the same final product), favourable routes are important and lead to different ultimate products. These pathways depend on several factors, such as experimental conditions, stability of intermediates, transition state (TS) energies, as well as the functional groups on the substrate\cite{rotavera2021influence}

In the case of short-lived intermediates that evade experimental detection, high-accuracy {\it ab initio} quantum chemistry methods offer a means to identify the TSs, the reaction paths\cite{moore2017fate,weidman2018high} and describe their behavior\cite{rienstra2000c2h5+,xing2020kinetics}. Additionally, {\it ab initio} modelling allows for the examination of the influence of substitutions on overall reaction outcomes, making it a valuable tool for understanding the reactivity of diverse species and designing selective and efficient reactions\cite{rotavera2021influence}. Further, when it is challenging to associate experimental spectra of transient species with multiple potential isomers, computational findings provide valuable insights\cite{alessandrini2023computational,watrous2022theoretical}. For modelling low-temperature combustion reactions of small hydrocarbons, coupled cluster singles doubles with perturbative triples approximation, CCSD(T), has yielded highly accurate predictions\cite{xu2019theoretical,xi2021theoretical}. For larger molecules, composite wavefunction theories have demonstrated the best computational efficiency for reproducing experimental formation enthalpies of stable species\cite{moskaleva2000unimolecular,deyonker2007computational,davis2011ab,dorofeeva2013accurate,karton2016computational,das2021critical} and for accurate modelling of barrier heights\cite{curtiss2010assessment,karton2012assessment,wheeler2008thinking}. Among different composite wavefunction theories, the Weizmann-1 method (W1) has shown ``kJ/mol accuracy'' when evaluated against extensive test sets of experimental thermochemical data, with mean errors lower than the experimental uncertainty\cite{parthiban2001assessment}. This approach has been successful in benchmarking barrier heights for non-hydrogen-transfer reactions\cite{zhao2005benchmark}, investigating H-atom abstraction between carbon-centered radicals\cite{coote2004reliable}, and for
studying various reactions\cite{parthiban2001benchmark,karton2015accurate,yu2020benchmark,pu2005benchmark}. 
However, the W1 method is extremely computationally expensive and is not practical to use for high-throughput data generation and large-scale 
screening of reaction networks.

The development of large-scale, data-enhanced, \textit{ab initio}, automated frameworks such as the reaction mechanism generator\cite{gao2016reaction} has significantly advanced our understanding of combustion chemistry\cite{pio2022automatically,dewey2023reaction,harper2011comprehensive,farina2023automating}. Future progress in this direction requires the high-throughput generation of large datasets, which often necessitates identifying 
efficient theoretical protocols\cite{himthani2020big,kaisler2013big}. Despite advances in artificial intelligence-driven molecular 
design, the ability to design entire reactions and predict quantitatively accurate rate constants remains elusive due 
to the broad context of the datasets used to train the statistical models\cite {ramakrishnan2015big,ramakrishnan2015many,zaspel2018boosting}. Therefore, carefully planned rational design studies are necessary to bridge this gap and to probe the transferability of chemical effects across molecules\cite{harvey2016molecular,lockwood2022advances}.

In this study, we investigate a series of hydrocarbons to explore the impact of stereochemical effects, such as inductive, strain, and resonance, on the thermodynamic and kinetic stability of $\cdot$QOOH.
In practically relevant systems, $\cdot$QOOH is unstable, but it is essential through which pathway it decomposes: $\cdot$OH or $\cdot$OOH. The former leads to chain propagation, whereas the latter leads to chain termination. The addition of a second O$_2$ would lead to chain branching.
It is now generally thought that the reactions happening via $\cdot$QOOH dominate low-temperature
chain branching. Hence, the formation and reactions
of these $\cdot$QOOH molecules are central to autoignition chemistry.
To this end, we calculate the energy profiles highlighted in FIG.~\ref{fig:ReactionScheme} using high-accuracy {\it ab initio} methods. We apply accurate wavefunction methods and also study the predictive performance of popular density functional theory models and calibrate them to minimize the impact of systematic errors, as well as the effects of spin contamination. We identify the best-performing functionals that could be used in future studies of similar systems.
We specifically focus on the pathways for intramolecular H-transfer 
in ROO$\cdot$ to form $\cdot$QOOH via 5- and 6-membered TS, providing insights into the influence of steric strain on the activation energy barrier, which further affects the temperature at which autoignition can take place. Furthermore, we explain the impact of stereo-electronic effects on the stability of $\cdot$QOOH with respect to competing decomposition pathways. Finally, we leverage the insights acquired from model systems as guiding principles to propose novel reactions proceeding via stable $\cdot$QOOH intermediates. These stable radicals could exhibit an enhanced lifetime, making them suitable for experimental characterization.

The rest of this article is organized as follows: (i) We describe the W1 method's implementation and other computational details in `II. METHODS'. (ii) We apply the W1 method to establish accurate reaction energies and barriers for a benchmark dataset discussed in `III. SELECTION OF HYDROCARBONS'. (iii) In `III A. Benchmarks for reaction energies and barriers', several methods offering various levels of accuracy are identified by benchmarking against the W1 results. 
(iv) In `III B and C', we apply the best DFT and coupled-cluster variant
to probe the transferability of chemical effects influencing the
reaction energetics. (v) For the species showing a 
stronger propensity to form thermodynamically and kinetically stable $\cdot$QOOH intermediates (starting from ROO$\cdot$), we study 
their corresponding dissociation channels. Finally, we conclude by commenting on the scope of the main results of this article.

\section{Methods}\label{sec_comp}

\subsection{W1 calculations}

\begin{figure*}
        \centering
        \includegraphics[width=1.0\linewidth]{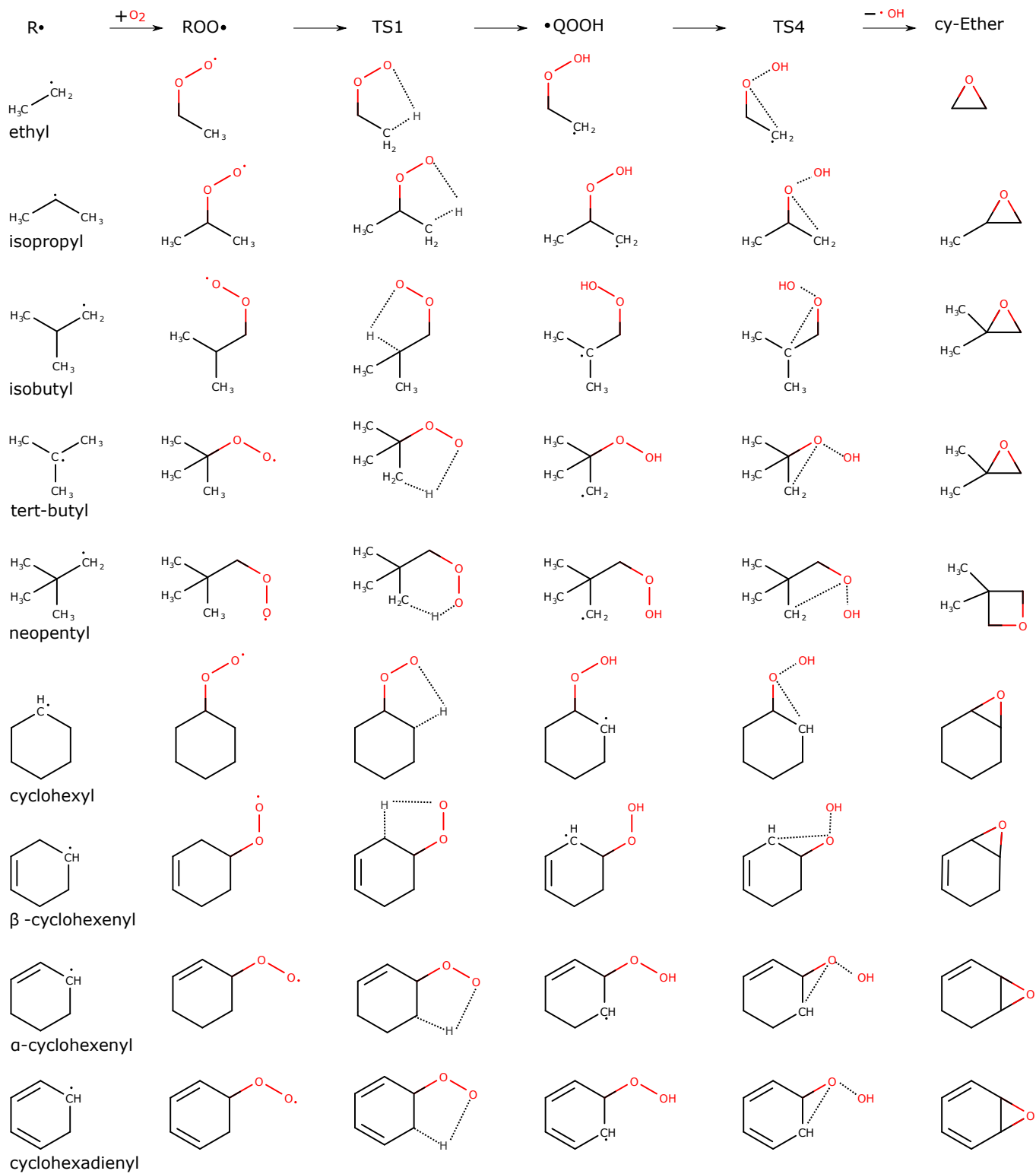}
        \caption{Steps and species involved in RH combustion
        of various R$\cdot$ systems considered in this study. The species are classified at the top in terms of their role in the general reaction mechanism. For all systems, 
        complete reactions that also include TS2 and TS3 are shown in the SI.}
        \label{fig:dataset}
\end{figure*}

Reaction energies and barriers of the elementary steps
in the combustion of the alkanes shown in FIG.~\ref{fig:dataset} were calculated using the W1 method\cite{martin1999towards}. All energies are quantified using
the zero-Kelvin internal energy, $U_0$, which is the sum of the total electronic energy, $E_{\rm ele}$, and the zero-point 
vibrational energy (ZPVE):
\begin{eqnarray}
    U_0 = E_{\rm ele} + {\rm ZPVE}.
\end{eqnarray}
For calculating the ZPVE of TS, the imaginary mode is ignored\cite{ochterski2000thermochemistry,frisch1981stability}. 
In the original implementation of the W1 method, the main ingredient of $E_{\rm ele}$ 
is the Hartree--Fock (HF) limit, which
for open-shell systems is determined in a spin-unrestricted fashion ({\it i.e.} UHF). 
For an open-shell system, the electronic energy at the W1 level, $E_{\rm ele}^{\rm W1}$, is defined as 
 \begin{eqnarray}
     E_{\rm elec}^{\rm W1}&=& E_{\rm UHF}+\Delta E_{\rm CCSD}+ \Delta E_{\rm T}+ \Delta E_{\rm CV,Rel}
 \end{eqnarray}
The individual terms are:\\
(i) Complete-basis set (CBS) limit of spin-unrestricted Hartree--Fock (UHF) energy estimated as  
    \begin{eqnarray}\label{HF}    
     E_{\rm UHF} & =& E_{\rm UHF/AVQZ}+\nonumber \\
     & & \frac{(E_{\rm UHF/AVQZ}-E_{\rm UHF/AVTZ})^{2}}{(4/3)^{5}-1}
    \end{eqnarray}\\
(ii) Basis set extrapolated valence-electron correlation energy calculated at the  coupled-cluster with singles and doubles (CCSD) 
level with the  frozen-core (FC) approximation
     \begin{eqnarray}\label{CCSD}
      \Delta E_{\rm CCSD} & = & \Delta E_{\rm CCSD/AVQZ}+ \nonumber \\
      & & \frac{\Delta E_{\rm  CCSD/AVQZ}-{\Delta E_{\rm CCSD/AVTZ}}}{(4/3)^{3.22}-1},
  \end{eqnarray}
  where $\Delta$ signifies that the correlation energy is obtained after subtracting the HF energy from the CCSD total energy for
  a given basis set.\\
(iii) Basis set extrapolated  post-CCSD correlation energy determined using FC-CCSD with perturbative triples correction, CCSD(T),  
 \begin{eqnarray}\label{CCSD(T)}
     \Delta E_{\rm T} & = & \Delta E_{\rm T/AVTZ}+ \nonumber \\
     & & \frac{\Delta E_{\rm T/AVTZ}-\Delta E_{\rm T/AVDZ}}{(3/2)^{3.22} -1},
 \end{eqnarray}
 where $\Delta E_{\rm T}= \Delta E_{\rm CCSD(T)}- \Delta E_{\rm CCSD}$ with the corresponding basis set. \\
 (iv) Higher-order effects such as core-valence (CV) correlation effects 
 and relativistic corrections (Rel) (collectively referred to as $E_{\rm CV,Rel}$).
  \begin{eqnarray}\label{CV}
    E_{\rm CV, Rel} & = & E_{\rm CCSD(T,Full)/MTsmall/DKH} - \nonumber \\
    & & E_{\rm CCSD(T,FC)/MTsmall} 
 \end{eqnarray}
In the above equations, the basis set AVXZ (X=D, T, or Q) is cc-pVXZ for H and 
aug-cc-pVXZ for C and O, DKH is Douglas-Kroll-Hess scalar relativistic approximation \cite{jansen1989revision}, 
MTsmall (Martin-Taylor) \cite{martin1999towards} basis set is the uncontracted 
cc-pVTZ basis set with additional two tight $d$ and an $f$ functions.

Since UHF calculations are susceptible to spin contamination, a parametric correction is applied to obtain the final spin-corrected energy. 
We denote internal energy determined at this level as $U_0^{\rm W1sc}$, where
`sc' denotes spin correction.  
The expression for $U_0^{\rm W1sc}$ is given by:
\begin{eqnarray}
U_0^{\rm W1sc} = U_0^{\rm W1,UHF}+\Delta E^{\rm spin}, 
\end{eqnarray}
where the spin correction term is defined as\cite{barnes2009unrestricted}:
\begin{eqnarray}
\Delta E^{\rm spin}= -6.28mE_{h} \times \Delta \left\langle S^{2}\right\rangle_{\rm UHF}. 
\end{eqnarray}
We also determined W1 energetics by calculating the basis set correction using the restricted open-shell HF (ROHF) limit, which does not require a spin correction. The internal energy determined at this level is denoted as $U_0^{\rm W1}$. 
For the free-radical species involved in the combustion of ethane and isopropane,
we calculated $U_0^{\rm W1sc}$ and $U_0^{\rm W1}$ using Gaussian-16 suite of programs\cite{g16} and Molpro respectively. 
$U_0^{\rm W1}$ in Molpro\cite{werner2015molpro} was calculated by the procedure described in the preceding paragraph using a script.
For all species involved in the combustion of ethane and isopropane,
we found $U_0^{\rm W1}$ to show excellent agreement with $U_0^{\rm W1sc}$ as shown in TABLE~\ref{tab:Molpro v/s G16}. All the individual deviations lie in the small range: 0.0--0.31 kcal/mol. 
For hydrocarbons with $>3$ C atoms, since adding O$_2$ increases the number of heavy atoms further, we found W1sc calculations to be computationally intractable. Since the W1 calculations performed with Molpro offer excellent computational memory management, we use $U_0^{\rm W1}$ for systems shown in FIG.~\ref{fig:dataset}.
 \begin{table}[H]
\centering
\caption{
Comparison of relative internal energies, $\Delta U_0$ obtained using
W1 and W1sc methods for the species shown in FIG.~\ref{fig:ReactionScheme} 
based on ethane and isopropane parent hydrocarbons.
 All values are reported after subtracting the energy of R$\cdot$+O$_2$, and are in kcal/mol. 
}
\begin{tabular}[t]{ l rrr r rrr}
\hline
 \multicolumn{1}{l}{Reaction}  & \multicolumn{4}{l}{RH = ethane} & \multicolumn{3}{l}{RH = isopropane} \\
\cline{2-4} \cline{6-8}
&  \multicolumn{1}{l}{W1} & \multicolumn{1}{l}{W1sc} & \multicolumn{1}{l}{$\Delta$} & ~~ & \multicolumn{1}{l}{W1} & \multicolumn{1}{l}{W1sc} & \multicolumn{1}{l}{$\Delta$}
\\
\hline 
ROO$\cdot$&-32.82&-32.66&-0.16 &&-34.58&-34.5&-0.08\\
$\cdot$QOOH&-16.20&-15.89&-0.31 &&-17.42&-17.33&-0.09\\
$cy$-ether+$\cdot$OH&-32.58&-32.47&-0.11& &-34.79&-34.69&-0.10\\
alkene+$\cdot$OOH&-13.29&-13.19&-0.10&&-13.28&-13.19&-0.09\\
TS1&4.05&4.14&-0.09  & &1.63&1.72&-0.09\\
TS2&-1.44&-1.44&0.00 &  &-3.42&-3.43&0.01\\
TS3&0.44&0.40&0.04 &   &-1.09&-1.14&0.05\\
TS4&-1.52&-1.82&0.30 &  &-4.74&-4.99&0.25\\
\hline
\end{tabular}
\label{tab:Molpro v/s G16}
\end{table}%
With W1 as the reference method, we evaluate the accuracy of the reaction energetics obtained with the composite wave function theories (cWFTs): G4\cite{curtiss2007gaussian}, G4(MP2)\cite{curtiss2007gaussian_a}, and CBS-QB3\cite{montgomery1994complete}. Furthermore, we benchmark the performance of 16 density functional approximations (DFAs) of varying rigour, namely,
\begin{itemize}
\item hybrid generalized gradient approximation (hGGA): M05-2X\cite{zhao2006design}, M06-2X\cite{zhao2008m06}, MN12-L\cite{peverati2012improved}, MN15-L\cite{yu2016mn15}, B3LYP\cite{becke1993becke, lee1988development}
\item range-separated hybrid functionals (RSH): $\omega$B97M-V\cite{mardirossian2016omega}, $\omega$B97X-D\cite{chai2008long}
\item double hybrid DFAs: $\omega$B97X-2\cite{chai2009long}, B2PLYP\cite{grimme2006semiempirical}, B2PLYPD\cite{schwabe2007double}, B2PLYPD3\cite{grimme2011effect}, mPW2PLYP\cite{schwabe2006towards}, mPW2PLYPD\cite{schwabe2007double}, PBE0DH\cite{bremond2011seeking}, PBEQIDH\cite{bremond2014communication}, DSDPBEP86\cite{kozuch2011dsd}.
\end{itemize}
Two DFAs, $\omega$B97X-2 and $\omega$B97M-V, are benchmarked with the Resolution of Identity (RI) approximation. 
These are referred to as RI-$\omega$B97X-2 and RI-$\omega$B97M-V in the text. For DFAs, single-point energies were calculated using the aug-cc-pVQZ basis set. In addition, we also consider CCSD(T)-F12a\cite{adler2007simple} in combination with the VDZ-F12 basis set, which has been extensively used to study species involved in low-temperature combustion \cite{xu2019theoretical, li2023comprehensive, doner2022stereoisomer, doner2023unimolecular}. The F12 method is an explicitly correlated version of the canonical CCSD(T) method involving a factor that depends on the explicit two-electron distance, $r_{12}$.

\subsection{Computational details}
For all molecules considered in this work, we generated initial structures by starting with the simplified molecular input line entry system (SMILES)\cite{weininger1988smiles} representations, 
which were converted to three-dimensional structures and relaxed with the universal force field (UFF)\cite{rappe1992uff} using the program OpenBabel\cite{o2011open}. In the present study, we aim to evaluate the performance of various ab initio methods and explore the transferability of chemical trends across different molecules; hence, we do not examine the role of stereoisomers in combustion chemistry. However, diastereomers of similar thermodynamic stability can result in reaction paths of different kinetic natures influencing the modelling of reaction rates\cite{doner2021isomer,danilack2021diastereomers}.
The geometries of all molecular species considered in this study were optimized using the hybrid-DFT method, B3LYP\cite{becke1993becke} along with the basis set, cc-pVTZ+d\cite{dunning2001gaussian} because this combination is employed in the W1 thermochemistry method.
We performed initial searches for the TS structures using the Berny algorithm\cite{schlegel1982optimization} (using the keyword {\tt opt(ts,calcall)}) with the B3LYP method and a small basis set 6-31G. This approach converged towards stationary structures with one imaginary frequency for systems studied in this work. However, when subsequently optimized with the Berny algorithm at the B3LYP/cc-pVTZ+d level, some of these structures exhibited convergence failure. In such cases, we used the B3LYP/6-31G structures as the initial guess and located the TS using the synchronous transit-guided quasi-Newton (STQN) approach\cite{peng1993combining} through keyword {\tt opt=QST3}.
Following geometry optimizations, harmonic vibrational frequency analyses were performed 
to ascertain the dynamic character of the
 minima and TS on the potential energy landscape. 
 All TS structures were subsequently verified through intrinsic reaction coordinate (IRC) calculations\cite{fukui1981path}. 
 
 In all geometry relaxations, force constant matrices  were calculated for eigenvector-following through the keywords
 {\tt calcfc} or {\tt calcall}. In cases where  
 calculations failed to converge, integration grids were improved from {\tt ultrafine} to {\tt superfine} settings to achieve convergence. 
 The ZPVE determined using the B3LYP/cc-pVTZ+d level used harmonic frequencies scaled by 0.985\cite{martin1999towards}, while G4, G4(MP2), and CBSQB3 methods employed scaling factors optimized for the respective models.
All geometry optimizations, harmonic frequency calculations, DFA calculations, barring $\omega$B97M-V and $\omega$B97X-2, and G4, G4(MP2), CBS-QB3, and W1sc calculations were performed using Gaussian-16. $\omega$B97M-V and $\omega$B97X-2 calculations were done using ORCA program\cite{neese2012orca}. 
For RI-$\omega$B97M-V and RI-$\omega$B97X-2, {\tt RIJK} keyword was used,
where RI is used for both the Coulomb integrals and HF Exchange integrals. CCSD(T)-F12a/cc-pVDZ-F12 and W1 calculations based on the ROHF limit were performed using Molpro.

\subsection{Dressed-atom corrections for systematic errors in DFAs}\label{dfa}
DFAs provide an inexpensive alternative to model reaction enthalpies. However, they suffer from systematic errors that grow with system size. A popular empirical method to address this error is the `dressed-atom correction scheme'\cite{grimme2005accurate,winget2004enthalpies,das2021critical}. This scheme has been used to correct formation enthalpies with experimental results as the reference. In our implementation, we target the relative energy of the reaction (R$\cdot$+O$_2$$\rightarrow{\rm P}$), where P may be a radical or TS discussed in FIG.~\ref{fig:dataset}. We model this relative energy as $\Delta U$, and using W1 reaction values as our reference, we correct DFA reaction energies as follows
\begin{equation}\label{eqn:correct}
 \Delta U^{\rm W1}_0 = \Delta U^{\rm DFA}_0 + \sum_i c_in_i   
\end{equation}
Here, $n_i$ is the number of atoms of an element in P, and $c_i$ is the correction for atom type-$i$. 
Eq. \ref{eqn:correct} was solved separately for the four energy minimum species (ROO$\cdot$, $\cdot$QOOH), $cy$-ether+$\cdot$OH and Alkene+$\cdot$OOH and the four transition states (TS1, TS2, TS3 and TS4). 
We obtained eight $c_i$ of the $i$-th atom type for each DFA that we averaged over all the species. The H, C, and O corrections for all the DFA methods benchmarked in this study are presented in TABLE \ref{tab:C_H_O corr}. 

It is important to note that across a potential energy surface, the dressed atom corrections determined for energies with respect to $\rm R\cdot+O_2$, is a constant shift. Hence, these empirical corrections improve the relative energies of all species uniformly. In contrast, the dressed atom corrections cancel out for reaction energies determined with respect to another configuration on the potential energy surface.

\begin{table}[ht]
\centering
\caption{Dressed-atom corrections for DFAs methods considered
in this study}
\small\addtolength{\tabcolsep}{1.5pt}
\begin{tabular}[t]{l rrr}
\hline
Functional    &   \multicolumn{1}{l}{H}    &\multicolumn{1}{l}{C} &\multicolumn{1}{l}{O}  \\
\hline
M052X&  0.319&-0.570&-1.435 \\
M062X& -0.049&0.063&-0.871\\
MN12L&  0.151&-0.138&2.584\\
MN15L& -0.146&0.363&-0.390\\
$\omega$B97M-V&  0.304&-0.431&-0.758\\
RI-$\omega$B97M-V&  0.304&-0.431&-0.758\\
$\omega$B97XD&  0.196&-0.241&-3.049\\
$\omega$B97X-2& -0.258&0.488&-2.081\\
RI-$\omega$B97X-2& -0.258&0.488&-2.081\\
B3LYP&  0.016&-0.060&-2.257\\
B2PLYP& -0.161&0.290&-2.175\\
B2PLYPD& -0.129&0.284&-1.891\\
B2PLYPD3& -0.149&0.318&-1.888\\
mPW2PLYP& -0.116&0.198&-2.087\\
mPW2PLYPD& -0.093&0.193&-1.881\\
PBE0DH&  0.152&-0.146&-3.269\\
PBEQIDH& -0.073&0.223&-2.922\\
DSDPBEP86& -0.349&0.641&-0.844\\
\hline 
\end{tabular}
\label{tab:C_H_O corr}
\end{table}%
\section{Selection of Hydrocarbons}
The R$\cdot$ species considered include 
ethyl ($\cdot$CH$_2$CH$_3$)\cite{rienstra2000c2h5+, desain2003measurements, miller2000theoretical, miller2001reaction, sheng2002detailed}, 
isopropyl ($\cdot$CH(CH$_3$)$_2$)\cite{desain2003measurements}, 
isobutyl ($\cdot$CH$_2$C(CH$_3$)$_2$)\cite{xing2020kinetics}, 
tert-butyl ($\cdot$C(CH$_3$)$_3$)\cite{zador2013directly, moore2017fate}, 
neopentyl ($\cdot$CH$_2$C(CH$_3$)$_3$)\cite{desain2003time}, 
cyclohexyl ($\cdot$C$_6$H$_{11}$)\cite{knepp2007theory,silke2007detailed}, 
cyclohexenyl ($\cdot$C$_6$H$_9$)\cite{koritzke2019qooh}, and 
cyclohexadienyl ($\cdot$C$_6$H$_7$)\cite{taylor2004direct} (see FIG.~\ref{fig:dataset} for more information). 
We selected these systems to explore the impact of resonance and inductive effects on the stability of the various $\cdot$QOOH species.
Additionally, these systems encompass a wide range of structural features, including 3-/4-membered cyclic ethers and 5-/6-membered TSs that exhibit varying degrees of steric strain. 

The ROO$^\cdot$ oxygen radical centre can abstract hydrogen from various carbon atoms within the molecule, leading to the formation of $\alpha$-, $\beta$-, $\gamma$-, and $\delta$-$\cdot$QOOH\cite{miyoshi2011systematic}. Consequently, a wide range of reaction paths are available for investigation. 
We do not aim to cover all possible reaction paths. So, we consistently selected one of the several possibilities to compare the same reaction step in the networks of different parent hydrocarbons.
We focused on the pathway involving $\beta$-$\cdot$QOOH for two main reasons: 
(1) Hydrocarbon radicals, such as ethyl, isopropyl, and tert-butyl, possess only $\beta$-carbon for hydrogen abstraction. 
(2) Due to the high computational cost of the W1 method, we selected a single pathway to reduce the number of species involved.  
In the case of $R$=neopentyl, the $\beta$-carbon of ROO$\cdot$ lacks an H atom for abstraction. Therefore, we selected the $\gamma$-$\cdot$QOOH pathway proceeding via a 6-membered TS for neopentyl. Furthermore, in the case of neopentyl and cyclohexadienyl, the co-product of $\cdot$OOH forming via TS2 and TS3 are dimethylcylopropane and benzene, respectively.

For the aforestated systems, the accuracy of various
DFAs, composite methods, and CCSD(T)-F12a/cc-pVDZ-F12 are compared against reference values obtained using the W1 method.
To validate the transferability of stereochemical effects on new reactions, we have also studied  isopentyl ($\cdot$CH$_2$CH$_2$CH(CH$_3$)$_2$), 
cyclohexadienylmethyl ($\cdot$CH$_2$C$_6$H$_7$) ,  
cyclohexadienylethyl ($\cdot$CH$_2$CH$_2$C$_6$H$_7$) , 
cycloheptadienyl ($\cdot$C$_7$H$_9$)), and  
cycloheptadienylmethyl ($\cdot$CH$_2$C$_7$H$_9$, $\cdot$C$_7$H$_8$(CH$_3$)) radicals. These `validation systems' are slightly larger compared to the model systems shown in FIG.~\ref{fig:ReactionScheme}, and are modeled at the CCSD(T)-F12a/cc-pVDZ-F12 level.

\begin{table*}[ht]
\centering
\caption{Relative internal energies, $\Delta U_0^{\rm W1}$ of selected species involved in the low-temperature hydrocarbon shown in FIG.~\ref{fig:dataset}. R1, R2, R3, and R4 stand for ROO$\cdot$, $\cdot$QOOH, $cy$-ether+$\cdot$OH, and Alkene+$\cdot$OOH, respectively. $\Delta U_0$ is reported as $U_0$~(species)-$U_0$~(R$\cdot$+O$_2$). All values are in kcal/mol.} 
\small\addtolength{\tabcolsep}{1.2pt}
\begin{tabular}[t]{l cccccc cccccc cccccc }
\hline
\multicolumn{1}{l}{Species} & \multicolumn{2}{l}{ethyl} & \multicolumn{2}{l}{$i$-propyl} & \multicolumn{2}{l}{$i$-butyl} & \multicolumn{2}{l}{$t$-butyl} & \multicolumn{2}{l}{$neo$-pentyl} & \multicolumn{2}{l}{{\it cy}-hexyl} & \multicolumn{2}{l}{$\alpha$-{\it cy}-hexenyl} & \multicolumn{2}{l}{{\it cy}-hexadienyl}\\
\hline 
R1       & -32.82 && -34.58 && -33.66 && -35.65 && -33.73 && -35.45 && -19.63 && -11.13\\
                           
R2      & -16.20 && -17.42 && -22.52 && -17.92 && -17.18 && -20.66 &&  -5.37 && -18.88\\

R3   & -32.58 && -34.79 && -41.78 && -36.57 && -36.73 && -38.88 && -21.12 && -18.88\\

R4 & -13.29 && -13.28 && -18.76 && -13.55 &&  ~-9.63 && -15.92 &&  ~-2.58 && -27.01\\

TS1            & ~4.05 &&   ~~1.63 &&  ~-3.29 &&   ~~0.68 &&  ~-9.01 &&   ~~1.72 &&  ~13.53 && ~17.05\\   
                      
TS2       & ~-1.44 &&  ~-3.42 &&  ~-2.02 &&  ~-5.36 &&  ~33.16 &&  ~-3.12 && ~9.56 &&  ~~~0.93\\ 
                          
TS3      & ~0.44 &&  ~-1.09 &&  ~-7.59 && ~-1.40 &&  ~32.62 &&  ~-4.52 &&   ~8.75 &&  ~~-9.38\\
                               
TS4               & ~-1.52 &&  ~-4.74 && -11.99 &&  ~-6.74 &&   ~0.71 &&  ~-8.55 &&  7.29 &&   ~~3.67\\    
                                                        
\hline
\end{tabular}
\label{tab:W1}
\end{table*}%

\section{Results and discussions}\label{sec_results}
In the low-temperature combustion paradigm, the $\cdot$QOOHs are experimentally elusive because of two reasons: 
(i) They are usually less stable than their precursor ROO$\cdot$, and 
(ii) They have fast decomposition rates. 
Nevertheless, Savee \textit{et al.} detected a $\cdot$QOOH intermediate in 2015 through prudent molecular design\cite{savee2015direct}. In this section, we discuss some molecular structural factors that affect the stability of $\cdot$QOOH and apply this knowledge as design principles to identify reactions featuring potentially long-lived $\cdot$QOOH species. Furthermore, we evaluate a range of popular density functionals and composite wavefunction-based methods with respect to their performance on the test reaction set.

\subsection{Benchmarks for reaction energies and barriers}\label{Benchmarks}
\begin{table*}[ht]
\centering
\caption{
Mean absolute deviation (MAD) for relative internal energy ($\Delta U_0$) of various species with respect to R$\cdot$+O$_2$ determined at different
levels of {\it ab initio} methods. For DFAs, MADs are reported after $\Delta U_0$ were corrected for systematic errors (denoted $c$) and for uncorrected values (denoted $u$).
MADs are reported for four minima: R1--R4, transition states TS1--TS4, and molecules defined in TABLE~\ref{tab:W1}. 
R (all) represents mean of MADs for all minima and TS (all) represents mean of MADs for all reaction barriers. 
Values in the parentheses denote standard deviations. 
`All' represents the overall mean of MADs for reaction energies and reaction barriers combined. 
All values are in kcal/mol and in the last three columns,
MADs lower than 1~kcal/mol are in bold.}
\small\addtolength{\tabcolsep}{1.2pt}
\begin{tabular}[t]{lccccccccccccc}
\hline
Functional   & $u/c$       &    R1    &R2     &R3             &R4            &TS1    &TS2    &TS3    &TS4     &      R (all)       & TS (all)    & All     \\
\hline
M052X   & $c$    &1.58 & 0.54 & 1.71 & 0.73 & 0.82 & 1.38 & 0.43 & 3.53 & 1.14 (0.94) & 1.54 (2.13) & 1.34 (1.92) \\
 M052X   & $u$  &  1.15    &2.32   &       1.01    &       2.45   & 1.90   & 4.07   & 2.63   & 6.25    & 1.74 (0.95)   & 3.71 (2.30)  & 2.72 (2.02)\\ 
M062X   & $c$    &1.66 & 1.32 & 2.34 & 0.85 & 0.50 & 0.93 & 0.85 & 4.17 & 1.54 (1.07) & 1.61 (1.74) & 1.58 (2.02) \\
M062X  & $u$   &  0.32    &0.68   &      0.74   &        1.83   &1.67   &2.72   &2.72   &6.04    &      {\bf 0.89 (1.07)}      &3.29 (1.74)     &2.09 (2.02)\\
MN12L   & $c$    &2.53 & 1.44 & 1.42 & 1.44 & 1.53 & 1.21 & 1.01 & 1.43 & 1.71 (2.11) & 1.30 (1.59) & 1.50 (1.87) \\
MN12L   & $u$  &  3.28    &6.07   &       6.96    &       7.14    &5.88   &4.89   &6.29   &6.10    &      5.86 (2.15)      &5.79 (1.57)     &5.83 (1.88)\\ 
MN15L  & $c$     &1.47 & 0.89 & 1.36 & 2.27 & 1.05 & 1.20 & 0.46 & 0.98 & 1.50 (1.66) & {\bf 0.92 (1.24)} & 1.21 (1.59) \\
MN15L  & $u$  &  1.86    &0.92   &       1.12    &       2.51    &1.10   &1.45   &0.64   &1.05    &      1.60 (1.61)      &1.06 (1.41)     &1.33 (1.64)\\   
$\omega$B97M-V   & $c$   &1.76 & 0.56 & 0.26 & 0.97 & 0.42 & 1.38 & 0.37 & 2.72 & {\bf 0.88 (1.01)} & 1.22 (1.65) & 1.05 (1.48) \\
$\omega$B97M-V  & $u$  &  0.90    &1.18   &       1.03    &       0.52    & 0.65   &1.24   &0.83   &3.56    &      {\bf 0.91 (1.04)}      &1.57 (1.78)     & 1.24 (1.56)\\ 
RI-$\omega$B97M-V  & $c$        &1.79 & 0.51 & 0.27 & 0.90 & 0.41 & 1.37 & 0.35 & 2.71 & {\bf 0.87 (1.00)} & 1.21 (1.63) & 1.04 (1.46) \\
RI-$\omega$B97M-V  & $u$      &  0.91    &1.15   &       1.12    &       0.54    &0.65   &1.23   &0.85   & 3.59   &      {\bf 0.93 (1.04)}      & 1.58 (1.75)     & 1.25 (1.54)\\   
$\omega$B97XD  & $c$    &1.77 & 1.49 & 0.93 & 1.24 & 0.95 & 2.55 & 0.74 & 3.15 & 1.36 (1.64) & 1.85 (2.26) & 1.60 (1.99) \\
$\omega$B97XD  & $u$  &  3.77    &6.85   &       6.35    &       4.29    &5.40   &2.97   &5.90   &8.67    &      5.31 (1.68)      &5.73 (2.27)     &5.52 (2.01)\\   
$\omega$B97X-2  & $c$    &1.83 & 1.17 & 1.03 & 1.01 & 1.33 & 3.22 & 1.01 & 7.09 & 1.26 (1.11) & 3.16 (3.85) & 2.21 (3.04)  \\
$\omega$B97X-2  & $u$   &  2.39    &3.02   &       3.14    &       3.99    &3.22   &1.33   &5.12   &11.21   &      3.14 (1.46)      &5.22 (3.81)     &4.18 (3.09)\\
RI-$\omega$B97X-2    & $c$       &1.85 & 1.17 & 1.02 & 1.01 & 1.34 & 3.22 & 1.02 & 7.08 & 1.26 (1.11) & 3.17 (3.85) & 2.22 (3.04)  \\
RI-$\omega$B97X-2  & $u$  &  2.40    &3.07   &       3.17    &       4.04    &3.25   &1.35   &5.18   & 11.24  &      3.17 (1.46)      &5.26 (3.81)     &4.21 (3.09)\\
B3LYP  & $c$     &1.58 & 3.14 & 3.42 & 2.52 & 1.26 & 4.44 & 2.17 & 1.22 & 2.67 (2.60) & 2.27 (2.53) & 2.47 (2.92)  \\
B3LYP   & $u$    &  6.20    &7.78   &       8.07    &       2.13    &7.16   &1.18   &2.53   &4.69    &      6.05 (2.60)      &3.50 (2.53)     &4.77 (2.92)\\ 
B2PLYP   & $c$   &0.41 & 1.40 & 0.89 & 1.75 & 0.69 & 3.99 & 0.64 & 3.31 & 1.11 (1.36) & 2.16 (2.70) & 1.63 (2.15)  \\
B2PLYP  & $u$   &  4.52    &5.83   &       5.30    &       2.70    &4.98   & 0.72   &3.89   &7.72    &      4.59 (1.43)      &4.33 (2.71)     & 4.46 (2.17)\\   
B2PLYPD  & $c$   &0.59 & 1.15 & 1.60 & 0.45 & 0.42 & 4.44 & 0.65 & 3.21 & {\bf 0.95 (1.09)} & 2.18 (2.77) & 1.56 (2.15)  \\
B2PLYPD    & $u$ &  3.03    &4.78   &       5.21    &       3.24    &3.68   &0.83   &2.96   &6.81    &      4.07 (1.16)      &3.57 (2.77)     &3.82 (2.18)\\ 
B2PLYPD3    & $c$        &0.37 & 1.18 & 1.62 & 0.75 & 0.53 & 4.42 & 0.78 & 3.16 & {\bf 0.98 (1.14)} & 2.22 (2.77) & 1.60 (2.17)  \\
B2PLYPD3  & $u$ &  3.42    &4.81   &       5.23    &       2.87    &3.83   &0.81   &2.83   & 6.77   &      4.08 (1.22)      &3.56 (2.78)     &3.82 (2.20)\\  
mPW2PLYP    & $c$        &0.62 & 0.87 & 0.36 & 1.38 & 0.71 & 2.70 & 0.42 & 3.17 & {\bf 0.81 (1.04)} & 1.75 (2.21) & 1.28 (1.74)  \\
mPW2PLYP  & $u$ &  3.76    &5.07   &       4.56    &       2.89    &4.91   &1.57   &3.99   & 7.44   &      4.07 (1.09)      &4.48 (2.21)     &4.27 (1.75)\\   
mPW2PLYPD  & $c$        &1.02 & 0.65 & 0.81 & 0.49 & 0.49 & 3.03 & 0.37 & 3.09 & {\bf 0.74 (0.92)} & 1.75 (2.25) & 1.24 (1.72)  \\
mPW2PLYPD  & $u$   &  2.68    &4.31   &       4.49    &       3.31    &3.96   &0.82   &3.32   & 6.78   &      3.70 (0.97)      &3.72 (2.25)     &3.71 (1.73)\\   
PBE0DH  & $c$    &1.60 & 0.71 & 2.10 & 1.50 & 1.05 & 1.45 & 1.80 & 3.37 & 1.48 (1.51) & 1.92 (2.08) & 1.70 (2.04)  \\
PBE0DH   & $u$    &   4.32   &6.14   &       3.80    &       5.74    &5.08   &5.22   &7.71   &9.28    &      5.00 (1.52)      &6.82 (2.10)     &5.91 (2.05)\\   
PBEQIDH  & $c$   &2.17 & 1.14 & 3.94 & 1.61 & 0.95 & 1.35 & 2.81 & 6.31 & 2.22 (1.77) & 2.85 (3.06) & 2.53 (3.12)  \\
PBEQIDH  & $u$     &   3.31   &4.35   &       1.59    &       5.23    &4.52   &4.79   &8.27   &11.77   &      3.62 (1.83)      &7.34 (3.04)     &5.48 (3.13)\\  
DSDPBEP86     & $c$      &1.60 & 1.58 & 2.04 & 0.91 & 1.24 & 2.52 & 1.55 & 7.48 & 1.53 (1.08) & 3.20 (3.85) & 2.36 (3.16)  \\
DSDPBEP86 & $u$    &   1.17   &  {0.65}  &       1.01    &       2.09    &1.08   &0.75   &3.32   & 9.25   &      1.23 (1.51)      &3.60 (3.83)     &2.41 (3.23)\\
\hline
G4           &&   1.33   & 0.98   &        0.42    &       0.86    &  0.79   &  1.34   & 0.31   &1.04    &       {\bf 0.90 (1.01)}      &  {\bf 0.87 (1.27)}     & {\bf 0.88 (1.17)}\\   
G4(MP2)        &&   2.33   &1.77   &       0.67    &       0.72    &1.42   &1.84   &0.53   & 0.69    &      1.37 (1.26)      &1.12 (1.35)     &1.25 (1.34)\\   
CBSQB3       &&   1.48   &1.46   &       0.63    &       0.28    &2.83   &1.95   &2.00   &2.09    &      {\bf 0.96 (0.81)}      &2.22 (1.13)     &1.59 (1.14)\\ 
\hline
CCSD(T)-F12a && 0.18 & 0.15 & 0.12 & 0.49 & 0.21 & 0.25 & 0.13 & 0.30 & {\bf 0.23(0.27)} & {\bf 0.22(0.25)} & {\bf 0.23(0.31)} \\
\hline
\end{tabular}
\label{tab:MAD DFT individual}
\end{table*}%

We determined the zero-point corrected total energy, also known as the
zero-Kelvin internal energy, $U_0$ for all the species encountered in the reaction scheme (FIG.~\ref{fig:ReactionScheme}) for acyclic and alicyclic hydrocarbons at the W1 level. Since energy differences are chemically more relevant than absolute total energies, we compare the relative energies, $\Delta U_0$, by subtracting the energy of the reactant R$\cdot$ + O$_2$ as presented in TABLE~\ref{tab:W1}. W1 provides exceptional accuracy when predicting energetics but is prohibitively expensive with increasing system size. The heavy computational requirement of W1 warrants the search for a computationally less-intensive method with minimal compromise in accuracy. To this extent, we provide consolidated tabulations of mean absolute deviations (MADs) with respect to the W1 predictions of reaction energies and barrier heights for all the species studied in this work (see FIG.~\ref{fig:dataset}) for 16 DFAs, three cWFTs, and  CCSD(T)-F12a/cc-pVDZ-F12 in TABLE~\ref{tab:MAD DFT individual}. CCSD(T)-F12a/cc-pVDZ-F12 performs best among all the methods studied here. MAD for minimum energy species and transition states are 0.23 kcal/mol and 0.22 kcal/mol, respectively. The overall MAD is 0.23 kcal/mol. This observation corroborates the extensive use of CCSD(T)-F12a/cc-pVDZ-F12 in the studies of combustion reactions by others\cite{xu2019theoretical, li2023comprehensive, doner2022stereoisomer, doner2023unimolecular}. The performance of other methods is described below in detail.

\textbf{Minimum energy species:} B2PLYPD3 gives the lowest MAD (0.37 kcal/mol) for ROO$\cdot$, closely followed by B2PLYP (0.41 kcal/mol). RI-$\omega$B97M-V yields the lowest MAD (0.51 kcal/mol) for $\cdot$QOOH followed by $\omega$B97M-V (0.56 kcal/mol). For $cy$-ether+$\cdot$OH, $\omega$B97M-V and RI-$\omega$B97M-V give comparably lower MADs of 0.26 kcal/mol and 0.27 kcal/mol, respectively. For alkene+$\cdot$OOH, CBSQB3 offers the lowest MAD of 0.28 kcal/mol, followed by mPW2PLYPD (0.49 kcal/mol).

\textbf{Transition states:} For the barrier along \rxnone{}, $\omega$B97M-V, RI-$\omega$B97M-V, and B2PLYPD give comparable errors of 0.42, 0.41, and 0.42 kcal/mol, respectively. For the TS along \rxntwo{}, M062X gives the lowest error of 0.93 kcal/mol, followed by MN15L (1.20 kcal/mol) and MN12L (1.21 kcal/mol). For studying \rxnthree{}, G4 delivers the best performance with a MAD of 0.31 kcal/mol, closely followed by RI-$\omega$B97M-V (0.35 kcal/mol), $\omega$B97M-V (0.37 kcal/mol) and mPW2PLYPD (0.37 kcal/mol). For the barrier in \rxnfour{}, G4(MP2) gives the lowest error of 0.69 kcal/mol, followed by MN15L (0.98 kcal/mol).

\textbf{Overall accuracy:} For relative energies of 
minimum energy species, the double-hybrid DFT method, mPW2PLYPD, yields the least MAD (0.74 kcal/mol), closely followed by the same method without dispersion correction (0.81 kcal/mol). However, without dressed atom corrections, mPW2PLYPD has an MAD of 3.70 kcal/mol, indicating that the MP2-level correlation energy contribution in this double-hybrid method largely addresses the non-systematic error 
in stable molecules; empirical corrections can correct the remaining systematic error. For the final TS, G4 has the least MAD (0.87 kcal/mol), followed by MN15L (0.92 kcal/mol). A recent study on four intramolecular H-shift reactions of peroxy radicals also found MN15L and MN12L to perform better compared to other DFT methods\cite{li2023comprehensive}. However, in our study, MN12L shows a large MAD of about 6 kcal/mol without dressed-atom corrections. For an overall performance considering both the reaction energies and reaction barriers, G4 has the lowest error of 0.88 kcal/mol among cWFTs. Among DFAs, $\omega$B97M-V and RI-$\omega$B97M-V deliver the least errors: 1.05 kcal/mol and 1.04 kcal/mol, respectively. Notably, $\omega$B97M-V has also been found to perform remarkably well for other datasets\cite{prasad2021bh9, mardirossian2018survival, mardirossian2017thirty}.

However, while discussing the energy difference between two species, i.e. reaction energies or barrier height, one should note that sometimes these methods overestimate one species and underestimate another, making the absolute energy difference between both species higher than expected. For example, as shown in TABLE~S1 and TABLE~S2 of the SI, G4 tends to overestimate the energy of ROO$\cdot$ (with respect to the starting species, R$\cdot$+O$_2$) and underestimate the energy of TS4. Similarly, RI-$\omega$B97M-V slightly overestimates ROO$\cdot$ but severely underestimates TS4. So the energy difference between ROO$\cdot$ and TS4 (i.e. $\Delta U_{\rm ROO}- \Delta U_{\rm TS4}$) for these two methods is much higher than the $\Delta U_{\rm ROO}- \Delta U_{\rm TS4}$ calculated by W1 method. In the same 
TABLE~S1, we can see that the error of CCSD(T)-F12a/cc-pVDZ-F12 is consistently small for all systems and does not seem to grow with the system size. Hence, CCSD(T)-F12a/cc-pVDZ-F12 emerges as the reliable method to study these reactions, even for larger systems.
The deviations with respect to the W1 values of individual reaction species and barriers for all the benchmark systems are shown in FIG.~S1 of the SI.

\begin{figure}
        \centering
        \includegraphics[width=1.0\linewidth]{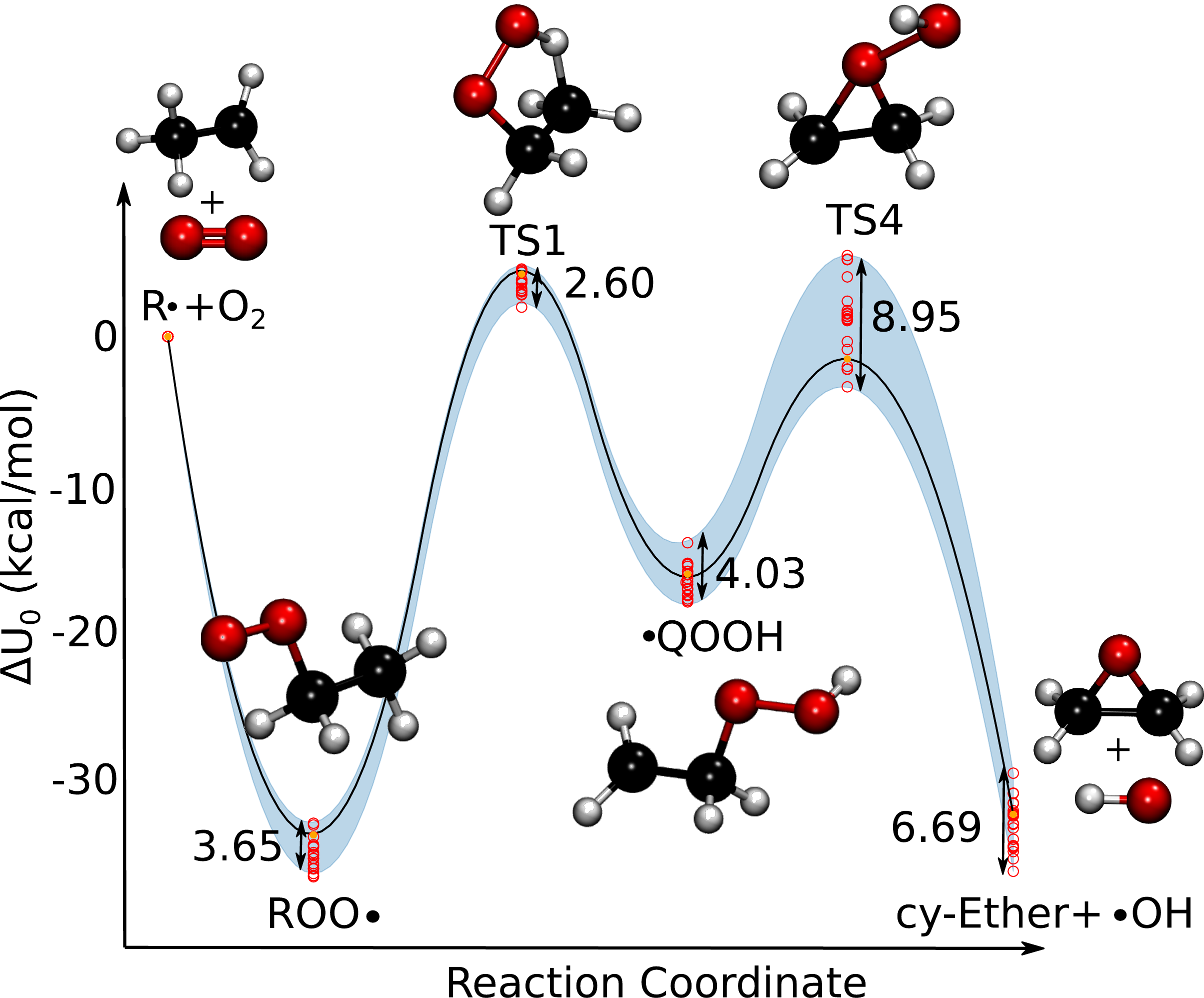}
         \caption{The variation in energy from DFAs and cWFTs for different reaction species 
         and barriers with respect to W1 for R$\cdot$ = ethyl. The black line indicates an interpolated potential energy profile
         at the W1 level while the red points correspond to the energies calculated with the DFAs including
         dressed-atom corrections. The range of energies spanned by DFAs at each minimum and maximum are also given in kcal/mol.
         }
        \label{fig:Ethyl_dft} 
    \end{figure}
    
Since we found hybrid and double-hybrid DFAs to show 
inconsistent performances across species, we illustrate the spread in their predictions along the entire reaction path for R$\cdot$ = ethyl in FIG.~\ref{fig:Ethyl_dft}. The energies of minima and TSs are connected by a smooth line for simplification. The black line represents the energy from W1, and the red points are the energy calculated by DFAs with dressed-atom corrections. We note that without dressed-atom corrections, we obtained a more considerable uncertainty across DFAs in determining the barrier heights as shown by Plata and Singleton for the Morita-Bayliss-Hillman reaction\cite{plata2015case}. The range of spread for different species varies from 2.60--8.95 kcal/mol, with predictions from some of the DFAs lying close to the reference, W1. Without dressed-atom corrections, the range of spread in DFT methods increases to 9.27--17.72 kcal/mol, indicating the effect systematic errors can have in DFT modeling, hence shedding light on the importance of empirical calibration of DFT predictions.

From the results presented above, we draw the following conclusions regarding the scope of applicability of various methods considered in this study.
\begin{enumerate}
    \item CCSD(T)-F12a/cc-pVDZ-F12a offer an excellent compromise between the speed and accuracy deviating from the reference theory W1 by less than $0.23$ kcal/mol. W1, in turn, is known to reach kJ/mol accuracy for datasets with experimental thermochemical energetics. For hydrocarbons with $>6$ C atoms, we found CCSD(T)-F12a/cc-pVDZ-F12 to require more random access memory (RAM) with longer computer run time. 
    \item For larger systems, for which  CCSD(T)-F12a/cc-pVDZ-F12 is not
    affordable, G4 is the option as it delivers MADs $<$ 1 kcal/mol for
    reaction energies and barriers considered in this benchmark analysis. 
    The individual
    steps involving CCSD(T) and HF/CBS can become the primary
    computational bottlenecks for hydrocarbons with about a dozen main group atoms. 
    \item DFT methods are ideal for studying larger systems with $>12$ or so main group
    atoms. We observed that it is necessary to calibrate DFT models by introducing
dressed-atom corrections per atom type. We found MN15L to yield consistent
results with and without such an empirical correction. Interestingly,
we also found MN15L less prone to spin-contamination errors, as 
analyzed separately in FIG.~S2 and FIG.~S3  of the SI.  
Among DFAs, $\omega$B97M-V is the best performer, especially
upon calibrating for systematic errors, delivering an overall MAD of $\thickapprox$ 1 kcal/mol. In addition, the RI approximation improves the speed 
without compromising the accuracy of the calibrated 
$\omega$B97M-V model. Hence, for modelling the elementary steps in the combustion of large hydrocarbons, RI-$\omega$B97M-V  may be applied.  
\end{enumerate}

\begin{figure}
        \centering
        \includegraphics[width=1.0\linewidth]{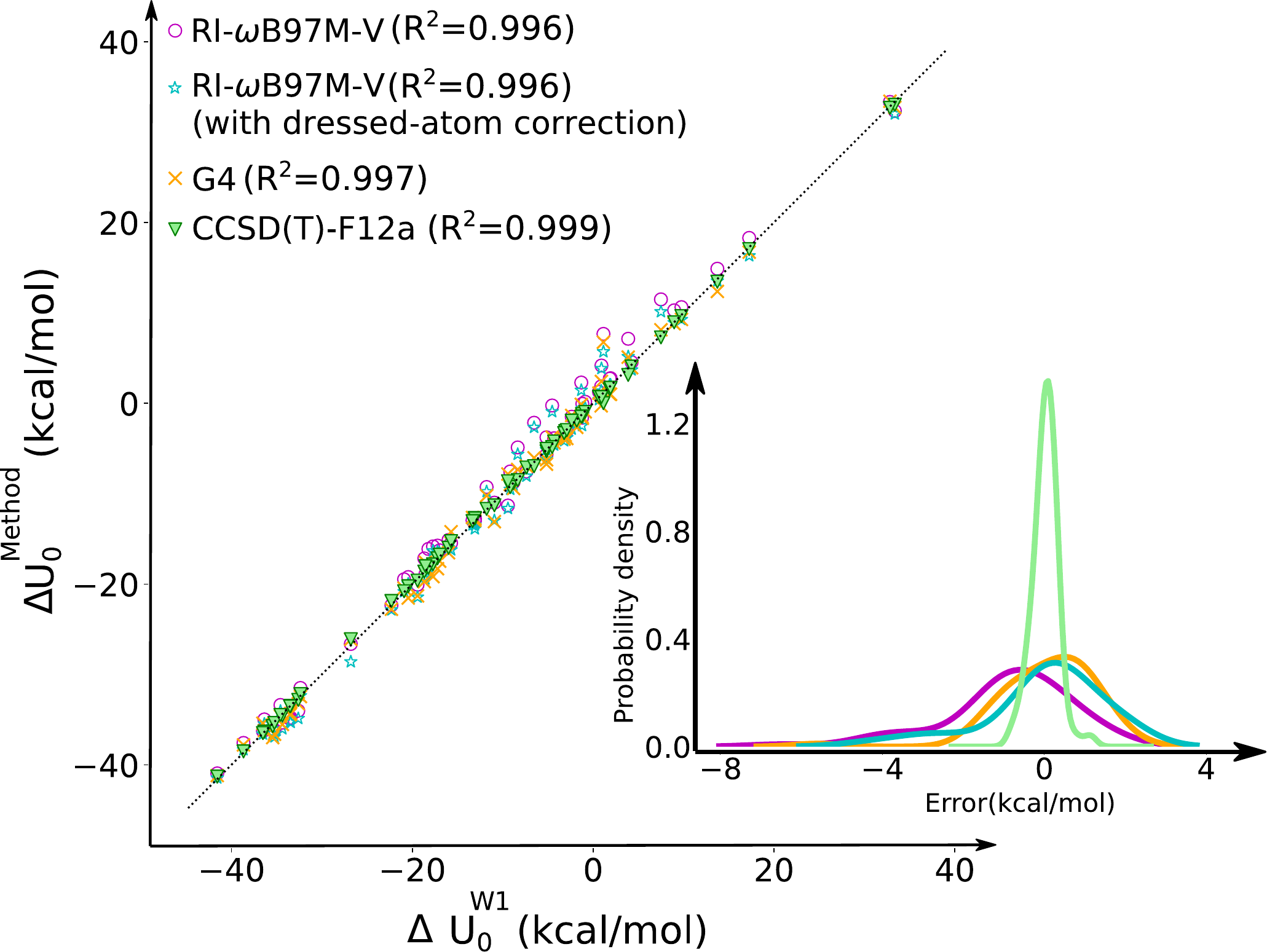}
       \caption{Scatterplots of relative internal energies ($\Delta U_{0}$), calculated with respect to the reactant R$\cdot$+O$_2$, from 
       G4 (shown in orange crosses), 
       RI-$\omega$B97M-V (pink-circle) 
       RI-$\omega$B97M-V with dressed-atom corrections (turquoise-star) and
       CCSD(T)-F12a/cc-pVDZ-F12 (green-triangle)
       are plotted alongside W1-level ($\Delta U_{0}^{\rm W1}$) for all the species represented in FIG.~\ref{fig:dataset}. Also given is the correlation 
       of determination, $R^2$.}
       \label{W1_G4}
\end{figure}

The correlations between $\Delta U_0^{\rm CCSD(T)-F12a}$, $\Delta U_0^{\rm G4}$, $\Delta U_0^{\rm RI-\omega B97M-V}$, and $\Delta U_0^{\rm RI-\omega B97M-V}$ (with dressed-atom corrections) with more-accurate reference values $\Delta U_0^{\rm W1}$ are shown in FIG.~ \ref{W1_G4}. While the error distribution is left-skewed for RI-$\omega$B97M-V, the profile shifts towards the G4 profile upon dressed-atom corrections. In the rest of this study, the details of the \rxnone{} and \rxnfour{} are explored at the CCSD(T)-F12a/cc-pVDZ-F12 level. 

\begin{figure*}
        \centering
        \includegraphics[width=1.0\linewidth]{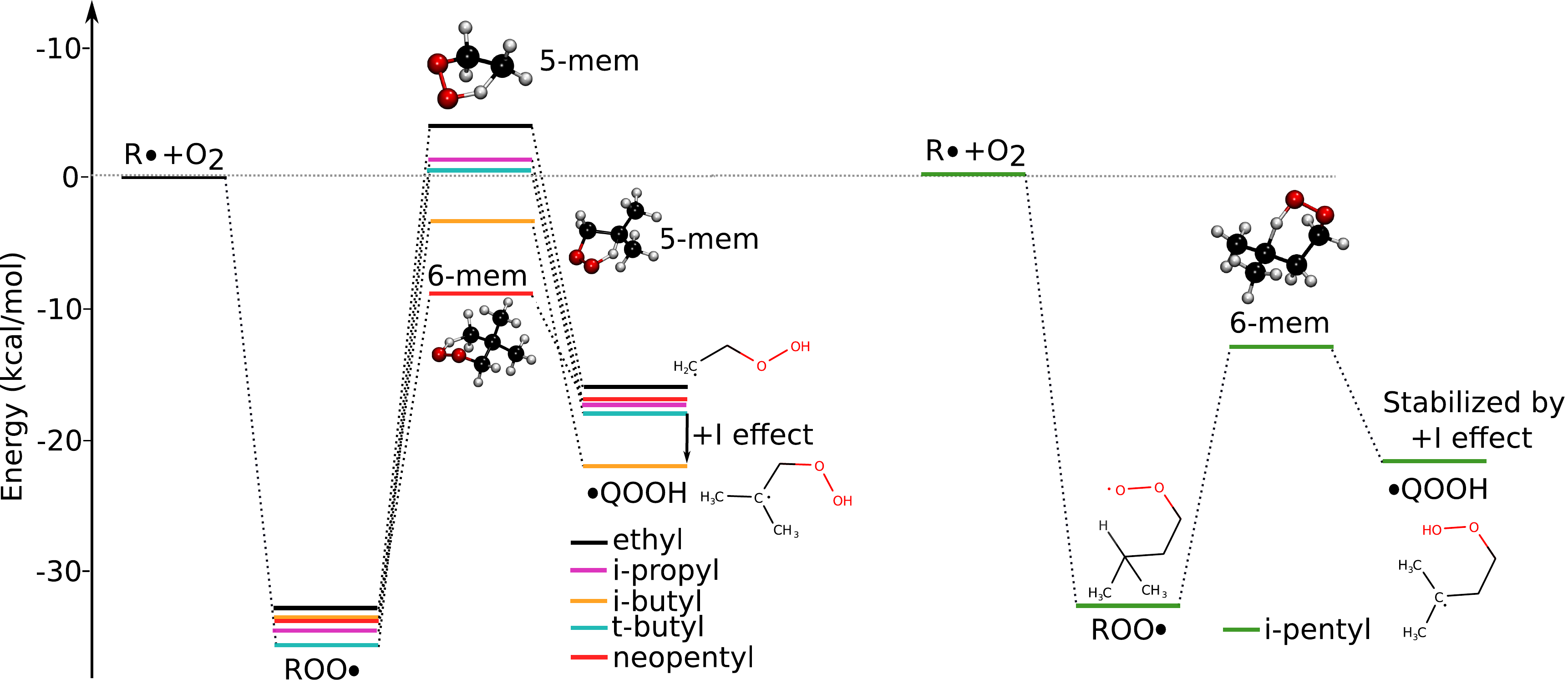}
        \caption{Energy profiles for the \rxnone{} path
        for acyclic systems. For R$\cdot$ derived from 
        small alkanes, stabilization due to the +I effect on the
        the radical site and steric strain on the TS structure are
        illustrated on the left side. The transferability of these stabilization effects is shown for the $i$-pentyl radical system on the right side.}
        \label{fig: Design alkyl}
\end{figure*}

\subsection{Acyclic hydrocarbons}
Before discussing the specific reaction steps, we will first comment on the reaction network studied in this study and shown in FIG.~ \ref{fig:ReactionScheme}. The reaction energy profile comprises two intermediates, ROO$\cdot$ and $\cdot$QOOH, connected by TS1, and three exit channels via TS2-TS4. In the following, we will focus on how the relative energies of the two intermediates and TS1 can be tuned by varying the R group. For most R groups in the test set, TS1 is the highest energy transition state. The only two exceptions are tert-butyl and neopentyl, which we will explain below.

In FIG.~\ref{fig: Design alkyl}, we investigated the relative energies of ROO$\cdot$ and $\cdot$QOOH for a series of acyclic hydrocarbons 
(R = ethyl, isopropyl, isobutyl, tert-butyl,  neopentyl) computed at the CCSD(T)-F12a/cc-pVDZ-F12 level. The energies for different species involved in the reactions are available in the SI. We find the internal energy of $\cdot$QOOH to be higher than the internal energy of ROO$\cdot$ for all the acyclic hydrocarbons, making the reactions endothermic. The reaction energies ($\Delta U$  = $U_{\rm QOOH}$ - $U_{\rm ROO}$$\cdot$ ) are similar for most hydrocarbons ($\sim$16-18 kcal/mol) except for $i$-butyl (10.61 kcal/mol). In the case of isobutyl, the increased stability of $\cdot$QOOH is due to the tertiary C radical site. Here, the inductive effect of the methyl groups stabilizes the $\cdot$QOOH radical while the other systems contain primary C radicals.

We have also examined the energy barriers for the \rxnone{} step ($\Delta U_{\rm barrier}$ = $U_{\rm TS1}$ - $U_{\rm ROO}$$\cdot$) and found that the neopentyl system has the lowest barrier height ($\Delta U_{\rm barrier}$ = 24.71 kcal/mol ) and ethyl system has the highest barrier height ($\Delta U_{\rm barrier}$ = 36.89 kcal/mol). The low barrier in the former case is due to a 6-membered TS where the deviations from the $sp^3$ structure for C atoms are negligible, ensuring minimal steric hindrance. The TS1 for other systems comprises a 5-membered ring yielding a higher barrier height. The trend in the barrier height among ethyl, isopropyl, isobutyl, and tert-butyl can be explained by the inductive effect (+I) of the methyl groups. As ethyl has the lowest +I effects among all, its barrier height is the largest, and isobutyl's barrier height is the lowest due to the stability of the TS1 by +I effects. 

By rationalizing these observations, we have identified two design principles to hold across the small acyclic hydrocarbons considered in this study. 
\begin{itemize}
    \item \textbf{Stability of $\cdot$QOOH:} The +I effect stabilizes the $\cdot$QOOH radical, but this gain in stability is still insufficient to make the \rxnone{} step exothermic. Therefore, ROO$\cdot$ is always thermodynamically more stable than $\cdot$QOOH across these systems.
    \item \textbf{Lowering of barrier TS1:} Less ring strain in the TS (6-membered compared to 5-membered) lowers the barrier height for the \rxnone{} step facilitating the product formation. Further,
    +I effect lowers the barrier even in the case of a 5-membered ring. 
\end{itemize}

To validate the transferability of these two trends to a new reaction, we studied the \rxnone{} step for the isopentyl system as shown in FIG.~\ref{fig: Design alkyl}. The energy profile diagram shows that isopentyl has a six-membered TS and a tertiary C radical, stabilized by the +I effect. As the TS has both features that provide stability, the energy barrier for $i$-pentyl is similar to that of neopentyl (featuring a 6-membered TS). In contrast, its net reaction energy is similar to isobutyl (featuring the +I effect). This observation validates the transferability of the proposed design principles based on the stability of $\cdot$QOOH and TS1 to similar acyclic hydrocarbons. The absolute energies, IRC plots, equilibrium geometries, and harmonic frequencies of all species involved are available in the SI.

\begin{figure*}
        \centering
        \includegraphics[width=1.0\linewidth]{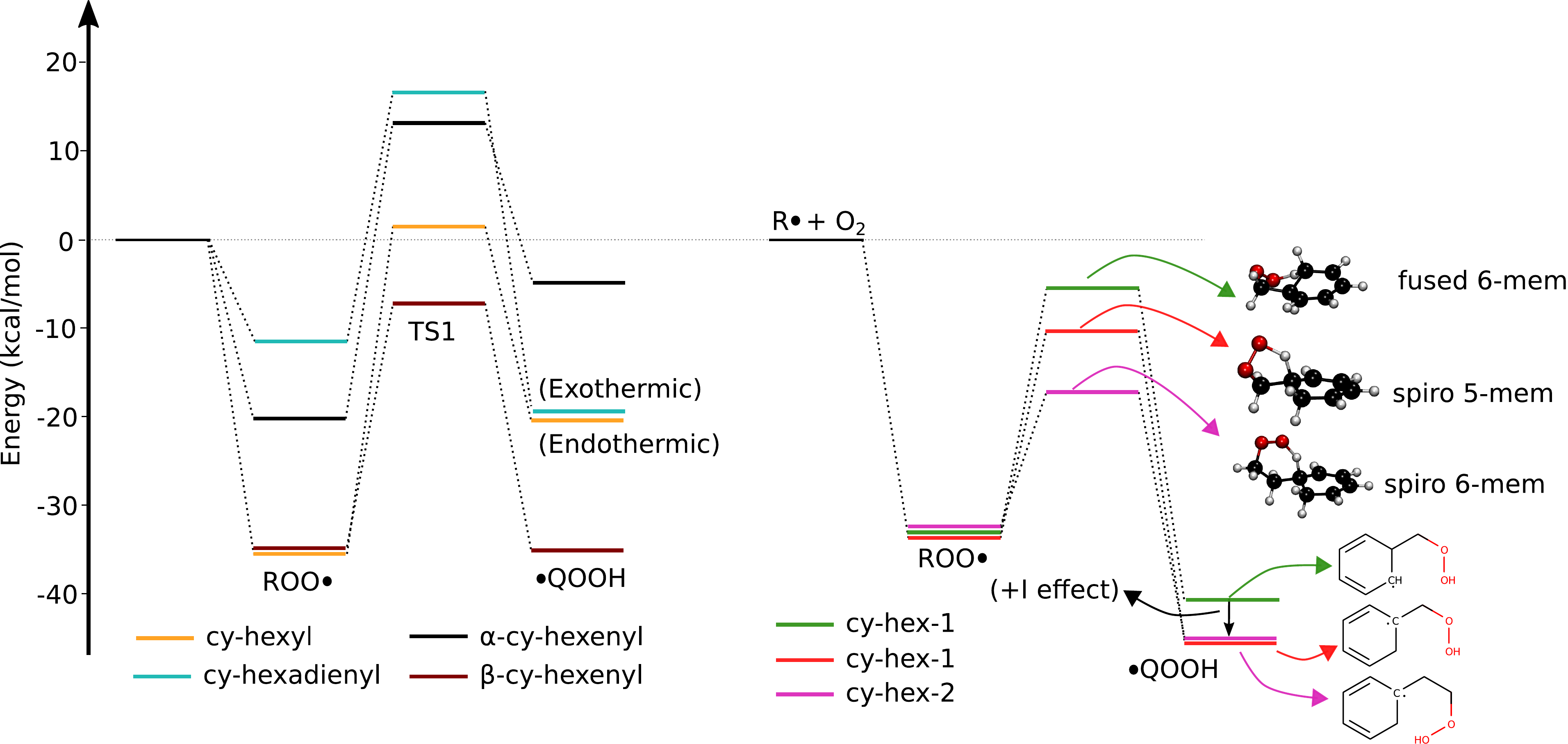}
       \caption{Energy profiles for the \rxnone{} path
        for alicyclic systems. 
        Resonance stabilization of 
        $\cdot$QOOH of 
        cyclohex-3-en-1-yl ($\beta$-$cy$-hexenyl) and 
        cyclohexa-2,4-dien-a-yl ($cy$-hexadienyl) systems
        with the introduction of double bonds conjugating with the
        radical site is shown on the left
        compared to the profile of  
        cyclohex-2-en-1-yl ($\alpha$-$cy$-hexenyl) and the
        saturated 
        cyclohexyl ($cy$-hexyl) system.  
        The transferability of resonance stabilization
        along with the +I stabilization inferred from the acyclic systems  is shown on the right for $cy$-hex-1: ($cy$-hexa-2,4-dien-1-yl)methyl and $cy$-hex-2: 2-($cy$-hexa-2,4-dien-1-yl)ethyl radical systems.}
        \label{fig: Design cyalkyl}
\end{figure*}

\subsection{Alicyclic hydrocarbons}
\begin{figure}[!hbp]
        \centering
        \includegraphics[width=1.0\linewidth]{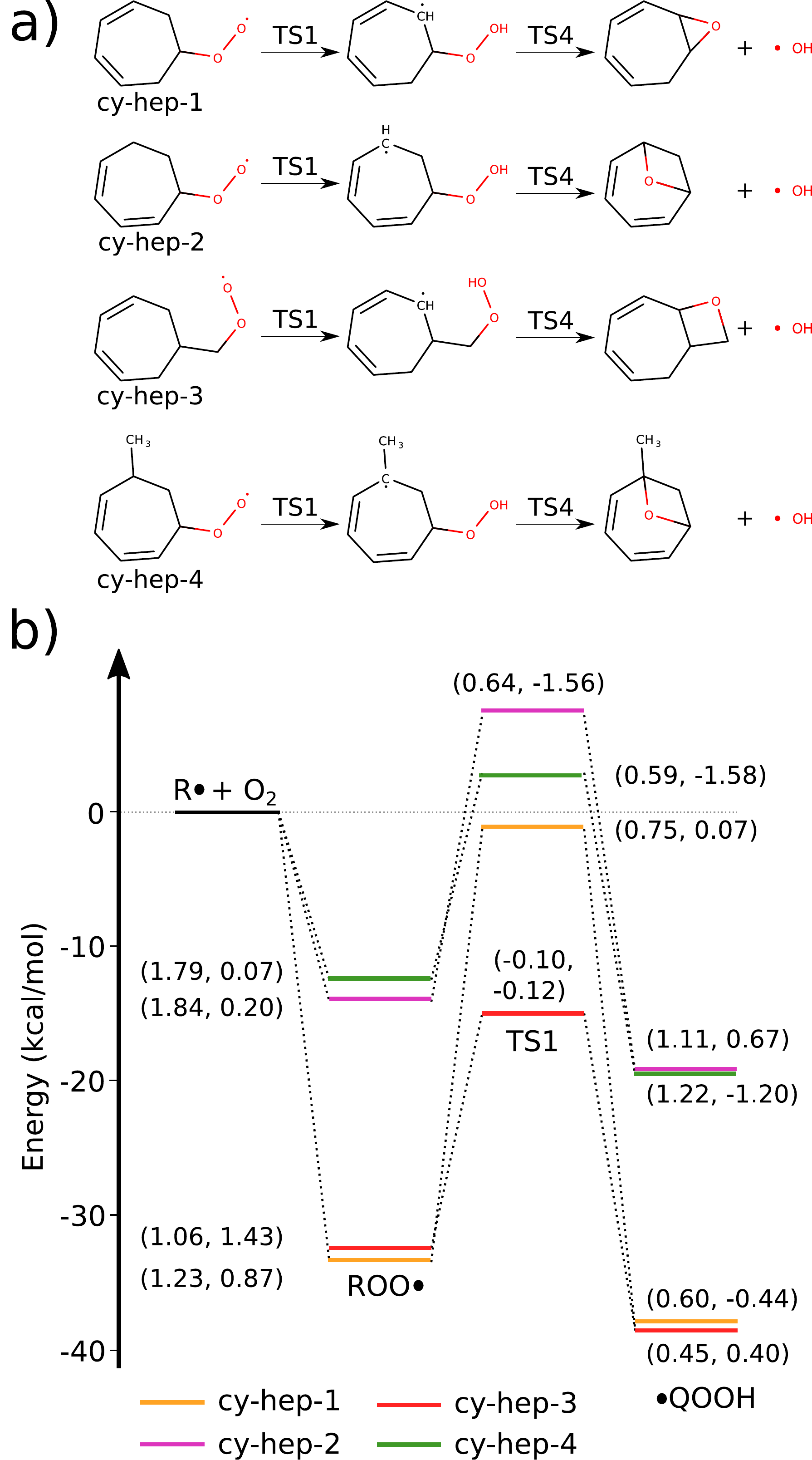}
       \caption{
       Reactions involving $\cdot$QOOH of cycloheptadienyl systems:
       a) Species involved in 
       ROO$\cdot$ $\xrightarrow{\rm TS1}$ 
       $\cdot$QOOH $\xrightarrow{\rm TS4}$ 
       $cy$-ether + $\cdot$OH steps are shown for 
       $cy$-hep-1: $cy$-hepta-3,5-dien-1-yl, 
       $cy$-hep-2: $cy$-hepta-2,4-dien-1-yl, 
       $cy$-hep-3: ($cy$-hepta-3,5-dien-1-yl)methyl, and 
       $cy$-hep-4: 6-methyl $cy$-hepta-2,4-dien-1-yl).
       b) Energy profiles for the \rxnone{} path along
       with the optimized structures of the transition 
       states are shown. 
       Deviations in CCSD(T)-F12a predictions with G4 and RI-$\omega$B97M-V (calculated as CCSD(T)-F12a - G4 and CCSD(T)-F12a - RI-$\omega$B97M-V) are given 
       in parentheses.
       }
        \label{fig: Design cyclohepta}
\end{figure}

FIG.~\ref{fig: Design cyalkyl} presents the energy profiles of \rxnone{} path for alicyclic systems (cyclohexyl, $\alpha$-cyclohexenyl, $\beta$-cyclohexenyl, cyclohexadienyl). The carbon radical site in R$\cdot$ of cyclohexadienyl and $\alpha$-cyclohexenyl is stabilized by resonance, but the oxygen radical site of ROO$\cdot$ is not. Hence, the energy of ROO$\cdot$ for cyclohexadienyl and $\alpha$-cyclohexenyl increases compared to that of ROO$\cdot$ of cyclohexyl and $\beta$-cyclohexenyl; in the latter case both R$\cdot$ and ROO$\cdot$ are not stabilized by resonance. When R$\cdot$ is stabilized by resonance, the energy difference between R$\cdot$ and ROO$\cdot$ decreases, and this behaviour can be seen in both FIG.~\ref{fig: Design cyalkyl} and FIG.~\ref{fig: Design cyclohepta}. For cyclohexyl, the \rxnone{} step is endothermic with $U_{\rm QOOH}$ - $U_{\rm ROO}$$\cdot$ = 15.12 kcal/mol. Upon introduction of a double bond in $\beta$-cyclohexenyl and two conjugate double bonds in cyclohexadienyl, the internal energy of $\cdot$QOOH drops below that of ROO$\cdot$ making the reaction exothermic with net reaction energies -0.16 kcal/mol and -6.64 kcal/mol -0.42 kcal/mol and -7.34 kcal/mol, respectively.

The $\cdot$QOOH radicals of $\beta$-cyclohexenyl and cyclohexadienyl are stabilized by resonance, making them more stable than ROO$\cdot$. It is worth noting that in the case of the $\cdot$QOOH radical of $\alpha$-cyclohexenyl lacking a resonance-stabilized radical site, the net energy is almost as endothermic as in the case of the saturated system cyclohexyl with $U_{\rm QOOH}$ - $U_{\rm ROO}$$\cdot$= 14.59 kcal/mol. With an increase in the degree of resonance, $\cdot$QOOH gains further stability, explaining the higher exothermicity in the cyclohexadienyl system compared to $\beta$-cyclohexenyl. Thus, resonance in a cyclic ring involving the radical site facilitates an exothermic \rxnone{} step. The barrier height for the three systems follows the order: cyclohexyl $>$ $\beta$-cyclohexenyl $\sim$ cyclohexadienyl. 
The difference between the energies of the saturated and unsaturated systems is negligible for the barrier compared to 
the scenario in $\cdot$QOOH indicating incomplete resonance in the TS.
Since all three TSs are five-membered, this trend does not arise from the ring strain but can be attributed to resonance. 

Based on these observations, we have identified two design principles for the alicyclic hydrocarbons. 
\begin{itemize}
    \item \textbf{Stability of $\cdot$QOOH:} Resonance stabilizes the $\cdot$QOOH radical drastically. However, conjugation
    spanning four bonds connecting the radical site makes \rxnone{} largely exothermic. 
   \item \textbf{Lowering of barrier:} The presence of partial resonance in the ring structure of the TS lowers the barrier height for the path \rxnone{}.
\end{itemize}

To validate the transferability of these effects along with the effects identified for acyclic systems, we examined the \rxnone{} step for $cy$-hex-1 system (See FIG.~\ref{fig: Design cyalkyl}). We find the
same  ROO$\cdot$ to undergo two different H-shift pathways to form 
different $\cdot$QOOHs ($cy$-hex-1-$\gamma$ (green) and $cy$-hex-1-$\beta$ (red)). The formation of tertiary (red) and secondary (green) $\cdot$QOOH radicals via 6-membered and 5-membered TS are observed. Both pathways are exothermic due to resonance stabilization, with the tertiary $\cdot$QOOH radical formation being more exothermic than the secondary radical. 
However, the barrier heights for these TSs show an unexpected trend, with the 6-membered TS being energetically higher than the 5-membered TS. This trend contradicts our previous finding that a TS with a 6-membered ring structure is expected to have a lower energy than a 5-membered ring. Further analysis revealed that the 6-membered TS in the case of $cy$-hex-1-$\gamma$ 
has a `C-C' fragment that is fused to the ring of the alicyclic hydrocarbon ($cy$-hex-1), introducing strain in both rings, while the 5-membered TS features spiro-type connectivity with bond angles close to the usual $sp^3$ arrangement resulting in less strain.  

To support this hypothesis, we studied another system, $cy$-hex-2 (see FIG.~\ref{fig: Design cyalkyl}). It forms $\cdot$QOOH by undergoing radical abstraction at the $\gamma$- position. The TS is 6-membered and is attached to the C in a spiro form, as shown in FIG.~\ref{fig: Design cyalkyl} (pink). This system exhibits a much lower barrier height than the fused system, demonstrating that the mode of attachment of the TS ring to the RH ring also plays a significant role in determining the barrier height. Overall, we find an additional trend for  
 alicyclic RHs with a 6-membered ring:
\begin{itemize}
    \item \textbf{Geometry of the TS:} The formation of a ring in spiro form is more stable than a fused form
\end{itemize}

We further extended the application of these design principles to 
R$\cdot$ = cycloheptadienyl ($cy$-hep-1, $cy$-hep-2) and its
methylated forms ($cy$-hep-3 and $cy$-hep-4) as shown  in
 FIG.~\ref{fig: Design cyclohepta}). 
 
It is noteworthy that the $\cdot$QOOH radicals of 
$cy$-hep-1 and $cy$-hep-2 have been experimentally studied 
by Savee \textit{et al.}\cite{savee2015direct}. 
Among the four $cy$-heptadienyl systems, $cy$-hep-1 shows the highest barrier with a 5-membered TS. On the other hand, low-energy 6-membered TSs are observed for $cy$-hep-2, $cy$-hep-3, and $cy$-hep-4 systems. Since $cy$-heptadienyl is a flexible 7-membered ring, fusing with another ring does not impose a significant strain. Therefore, in the systems presented in the FIG.~\ref{fig: Design cyclohepta}, the 6-membered TS is consistently ({\it i.e.} irrespective of whether it is of spiro or fused type) found to have lower energy than the 5-membered TS due to the lack of ring strain. As predicted, the inductive (+I) effect in $cy$-hep-4
lowers the barrier height and stabilizes the $\cdot$QOOH radical, leading to a more exothermic \rxnone{} step.

\begin{figure}[!hbt]
        \centering
        \includegraphics[width=1.0\linewidth]{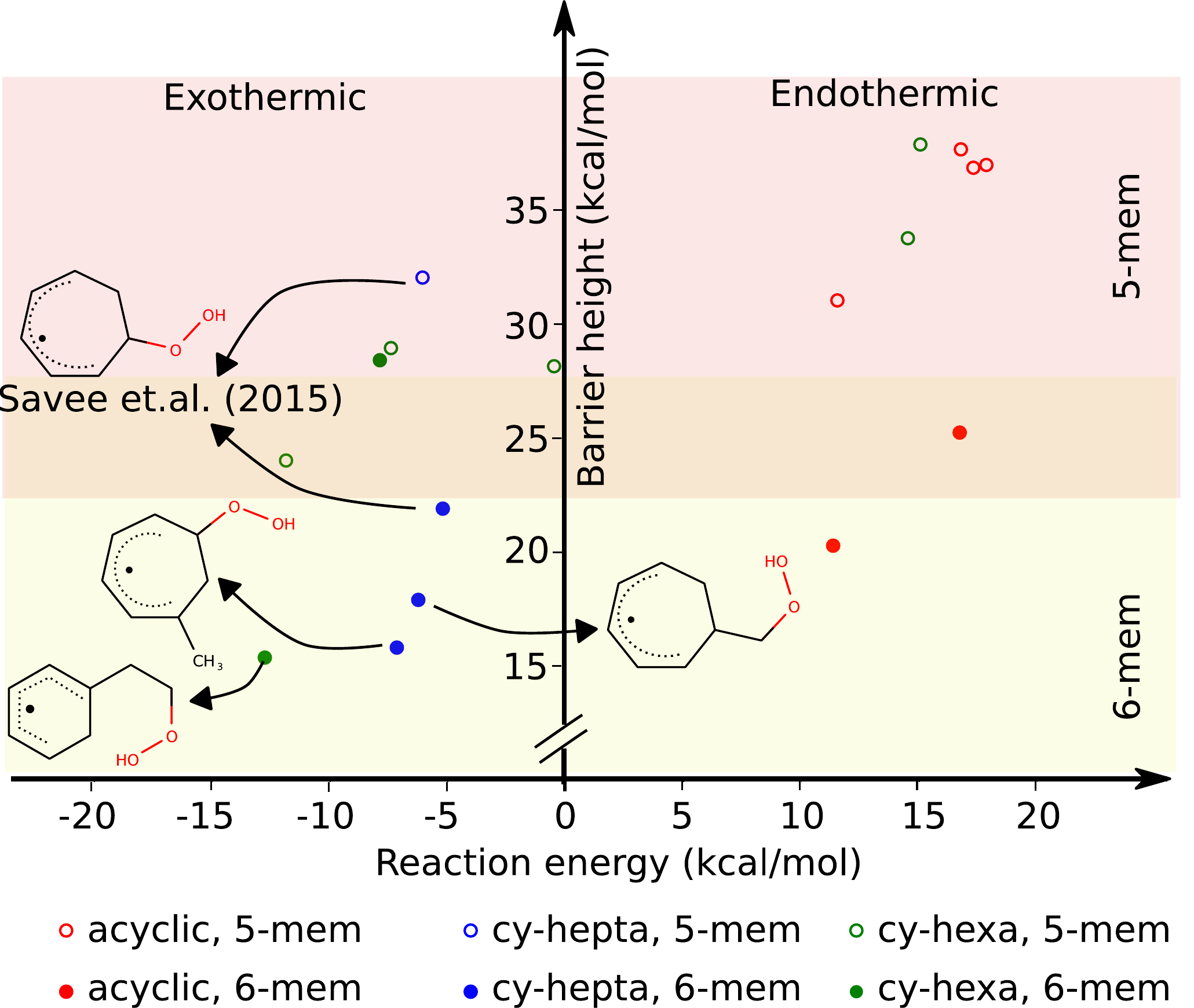}
       \caption{Joint distribution of reaction energies and barrier heights
       along the \rxnone{} path for all acyclic and alicyclic hydrocarbons studied in this work. 
       `$cy$-hepta' represents cycloheptadienyl and substituted 
       cycloheptadienyl systems. 
       `$cy$-hex' represents cyclohexyl, cyclohexenyl, cyclohexadienyl, and substituted cyclohexadienyl systems. Structures of thermodynamically
       and kinetically stable $\cdot$QOOH are shown.}
       \label{summary_all}
\end{figure}

Summarizing the results for the \rxnone{} path for acyclic and alicyclic hydrocarbons across all model and validation systems, we have plotted the reaction energies along with the barrier heights 
in FIG.~\ref{summary_all}. In this Bell--Evans--Polanyi plot\cite{bell1936theory,evans1936further}, the linear relationship between the reaction energies and barrier heights is not evident due to the diversity of the dataset. In this plot, the regions spanned by 5-membered TSs (pink area) and 6-membered TSs (yellow area) are shaded. The central zone (peach area) consists of both types of TS1. 

It is evident from the plot that all acyclic systems (red circles) lie in the endothermic region (right side of FIG.~\ref{summary_all}). 
For alicyclic systems, except for cyclohexyl and cyclohexenyl hydrocarbons, all the other systems exothermically produce $\cdot$QOOH. The four low-lying points on the left side of the plot are the rationally designed systems:
$cy$-hep-2 (blue circle), 
$cy$-hep-3 (blue circle), 
$cy$-hep-4 (blue circle), and $cy$-hex-2 (green circle). 
The lowest barrier height and the most exothermic \rxnone{} corresponds to the $cy$-hex-2 system. Three points have a lower barrier height and a more exothermic reaction than $cy$-hep-2 reported by Savee \textit{et al.} (as highlighted in the plot). Based on the energetics, one may expect these three systems to have long lifetimes facilitating their experimental detection.

\begin{figure*}
        \centering
        \includegraphics[width=1.0\linewidth]{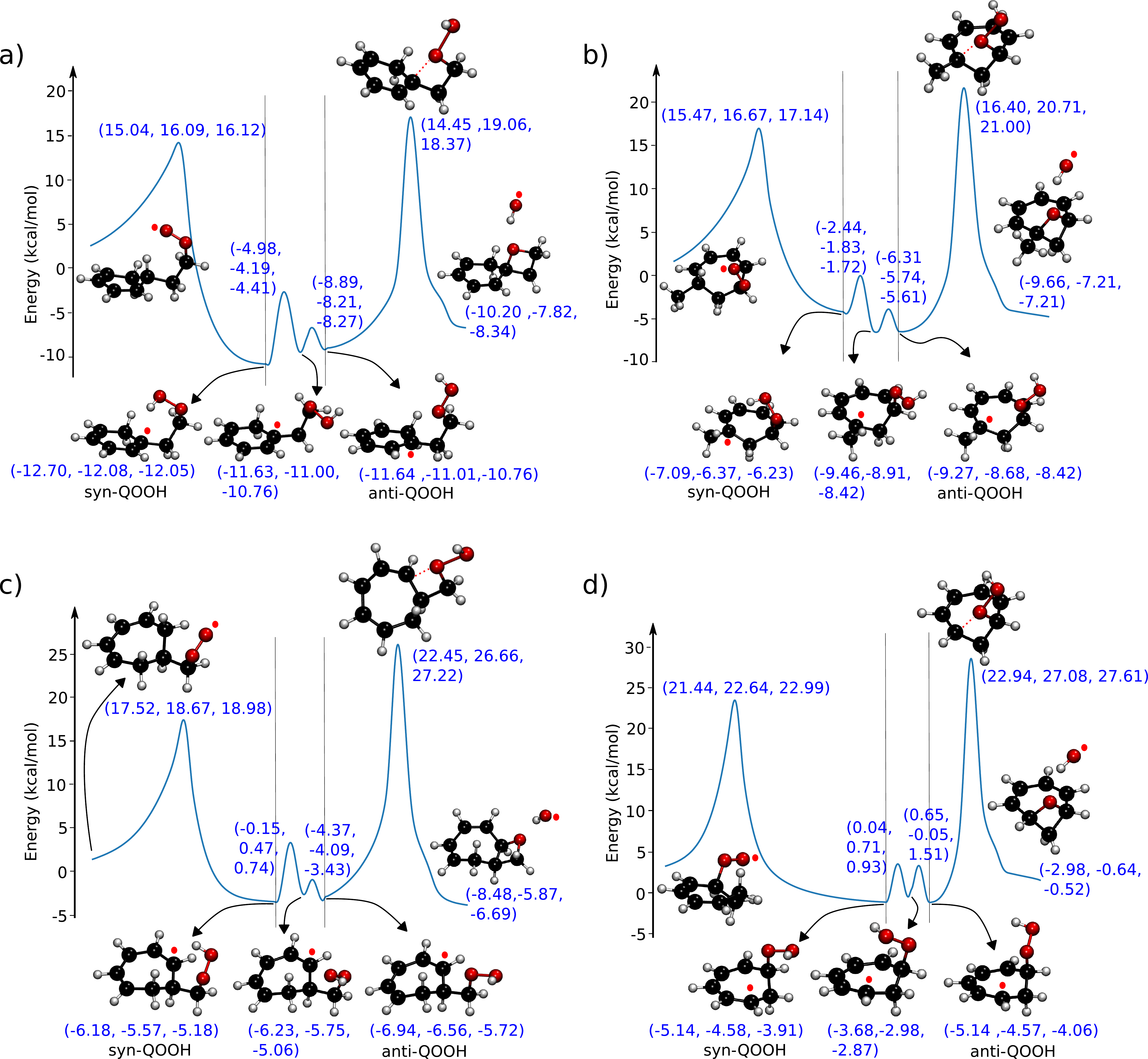}
       \caption{IRC profiles for ROO$\cdot$ $\xrightarrow{\rm TS1}$ $\cdot$QOOH $\xrightarrow{\rm TS4}$ $cy$-ether + $\cdot$OH, where R is   a) $cy$-hex-2, 
        b) $cy$-hep-4, 
        c) $cy$-hep-3, and 
        d) $cy$-hep-2. See FIG.~\ref{fig: Design cyclohepta} for the 
        IUPAC names. 
        The IRC profiles are calculated with B3LYP/6-31G(2{\it df},{\it p}) method and shown with respect to the ROO$\cdot$ radical. The energies highlighted in blue text are calculated using the CCSD(T)-F12a/cc-pVDZ-F12, G4, and RI-$\omega$B97M-V, respectively, and are given with respect to the energy of ROO$\cdot$. 
        The middle part of IRC profiles between the two black vertical lines connects {\it syn}-$\cdot$QOOH to {\it anti}-$\cdot$QOOH and are 
        calculated through constrained optimization for fixed CCOO torsion angles. The structures of ROO$\cdot$, {\it syn}-$\cdot$QOOH, {\it anti}-$\cdot$QOOH, TS4 and $cy$-ether + $\cdot$OH are shown in the profile. 
        }
       \label{IRC}
\end{figure*}

\subsection{Stability of $\cdot$QOOH along the dissociation channel}
Two primary strategies are commonly employed for the experimental detection and kinetic analysis of $\cdot$QOOH radicals.  
(1) Infrared (IR) activation of the $\cdot$QOOH radical, where sufficient energy is provided to surpass the dissociation barrier or reach the tunnelling regime near the barrier\cite{hansen2021watching}. 
The resulting dissociation products,  $cy$-ether + $\cdot$OH, are typically investigated. The energy required for IR-induced dissociation of the $\cdot$QOOH radical yields an IR fingerprint characteristic of the $\cdot$QOOH radical. Systems featuring $\cdot$QOOH $\longrightarrow$ $cy$-ether + $\cdot$OH dissociation barriers with energies comparable to a vibrational excitation are more favourable for studying the dissociation reaction. (2) Trapping $\cdot$QOOH in the potential well between the TS1 and TS4\cite{savee2015direct}. In this case, systems that exhibit exothermic ROO$\cdot$ to $\cdot$QOOH conversion along with a low barrier in the forward direction facilitate the formation of $\cdot$QOOH. Furthermore, a significantly high barrier for $\cdot$QOOH towards unimolecular dissociation is desirable to impede subsequent dissociation and favour direct detection of $\cdot$QOOH. Both strategies highlight the pivotal role of the dissociation barrier in these reactions. Therefore, in addition to studying the \rxnone{} pathway, we have also examined the \rxnfour{} dissociation pathway for the four stable $\cdot$QOOH systems highlighted in FIG.~\ref{summary_all}.  

This dissociation pathway reveals a TS where the O$_1$ atom in $\cdot$QOOH ($\cdot$Q-O$_1$-O$_2$-H) approaches the carbon-radical centre of $\cdot$QOOH, leading to the formation of a ring structure and sufficiently long O$_1$-O$_2$ bond to facilitate the release of $\cdot$OH. The attack at the $\gamma$ radical site yields 
an oxetane-type (4-membered heterocycle) TS4, where a cage formation is seen by bonding with the ring of the RH. The formation of a constrained 4-membered ring in TS4 may be the reason for a very high barrier height of $30-33$ kcal/mol for \rxnfour{} in these four systems. In systems where the attack occurs on $\beta$ radical centre, one can expect an oxirane (3-membered heterocycle) formation in the TS4. On the other hand, if the attack takes place on the $\delta$ radical centre, a tetrahydrofuran-type (5-membered heterocycle) system is expected to form in TS4. For a radical at $\alpha$ carbon, the dissociation product will be a carbonyl compound instead of a cyclic ether. Since the four systems discussed above involve only the 4-membered oxetane-type TS structures, the ring strain is uniform. More elaborate studies are required to quantify the trends in the barrier height of TS4 in terms of the structural
variations. 

Upon closer examination of this dissociation process, it is apparent that the orientation of the OOH fragment in $\cdot$QOOH plays a crucial role in the formation of $cy$-ether. To verify this observation, we have calculated the IRC profiles for ROO$\cdot$ $\xrightarrow{\rm TS1}$ $\cdot$QOOH $\xrightarrow{\rm TS4}$ $cy$-ether + $\cdot$OH, where R is 
a) $cy$-hex-2, 
b) $cy$-hep-4, 
c) $cy$-hep-3, and 
d) $cy$-hep-2 as shown in FIG.~\ref{IRC}. The IRC profiles were calculated with the B3LYP/6-31G(2{\it df},{\it p}) method, and the energy is plotted with respect to ROO$\cdot$. The energies highlighted in blue text are calculated using the CCSD(T)-F12a/cc-pVDZ-F12, G4, and RI-$\omega$B97M-V method and are given with respect to the energy of ROO$\cdot$.

All three methods agree very well ($\sim 1$ kcal/mol) for predicting the 
energy of TS1 and $\cdot$QOOH (with respect to ROO$\cdot$), see FIG.~\ref{IRC}. For $cy$-ether+$\cdot$OH, the prediction of CCSD(T)-F12a/cc-pVDZ-F12 differs from G4 and RI-$\omega$B97M-V 
by 2-3 kcal/mol. The discrepancy is maximum for TS4 (4--5 kcal/mol) due to an over-stabilization of ROO$\cdot$ in G4 and RI-$\omega$B97M-V and an underestimation of TS4's energy as explained in \ref{Benchmarks}. All the absolute energies are available in the SI. The complete IRC comprises three stages:
(1) The first stage involves the formation of {\it syn}-$\cdot$QOOH from ROO$\cdot$. 
(2) The second stage entails the internal rotation of the CCOO bond in {\it syn}-$\cdot$QOOH, leading to the formation of {\it anti}-$\cdot$QOOH via a {\it gauche}-minimum. 
(3) The third stage involves the formation of $cy$-ether + $\cdot$OH from {\it anti}-$\cdot$QOOH. 

The designations of {\it syn}-$\cdot$QOOH and {\it anti}-$\cdot$QOOH are based on the criteria of the CCOO torsion angle being acute for {\it syn}-$\cdot$QOOH and ranging between 170$^{\circ}$ and 180$^{\circ}$ for {\it anti}-$\cdot$QOOH. The trans orientation of the CCOO bond in {\it anti}-$\cdot$QOOH appears to facilitate the breaking of the $\cdot$QO$_1$-O$_2$H bond and the subsequent ring cyclization 
through bonding between the carbon radical site in $\cdot$QOOH and the O$_1$ atom of $\cdot$QO$_1$O$_2$H. This orientation provides a favourable alignment and proximity between the carbon radical centre and the O$_1$ atom, promoting ring structure formation during the dissociation process. Moreover, the \rxnfour{} step is endothermic, 
implying the reverse barrier for this step to be lower than the forward barrier, imparting better stability to $\cdot$QOOH.
In summary, $\cdot$QOOH is found to be more stable than both its precursor ROO$\cdot$ and the dissociation products $cy$-ether + $\cdot$OH for $cy$-hex-2, $cy$-hep-4, $cy$-hep-3 and $cy$-hep-2. The high dissociation barrier  enhances the suitability of these radicals for experimental studies that necessitate the 
trapping of $\cdot$QOOH in a potential well.

\section{Conclusions}
In this study, we have systematically benchmarked popular DFT methods and selected wavefunction methods using reference energies calculated at the very accurate (but computationally costly) W1 method. We found that quantitatively accurate predictions require empirically calibrating DFT predictions using a representative dataset. We
found $\omega$B97M-V, along with empirically determined dressed-atom corrections and the  RI approximation, to offer the most accurate predictions among the tested DFT methods comparable to that of G4. When compared to W1 predictions, we found CCSD(T)-F12a/cc-pVDZ-F12 to have a mean absolute error of 0.23 kcal/mol in agreement with the trends shown in previous studies. The reasonably affordable cost of the method extends the scope of quantitatively precise elucidation to 
small hydrocarbons. For modeling larger hydrocarbons derived from chemical space datasets\cite{ramakrishnan2014quantum,von2020exploring,senthil2021troubleshooting,kayastha2022resolution}, G4 and RI-$\omega$B97M-V are appropriate choices.

Furthermore, in this study, we assess the thermodynamic and kinetic stability of the $\cdot$QOOH radical in comparison to its precursor 
in hydrocarbon combustion (the ROO$\cdot$ radical) and its dissociation to $cy$-ether + $\cdot$OH. 
To ensure a comprehensive analysis, we carefully selected prototype hydrocarbon radicals, R$\cdot$, with 2--6 C atoms exhibiting inductive and resonance effects that are known to stabilize radicals.

Using CCSD(T)-F12a/cc-pVDZ-F12 energies of acyclic hydrocarbons, our analysis revealed $\cdot$QOOH to gain a net stability of $\sim$10 kcal/mol due to the +I effect imparted by a neighbouring methyl group. 
In the case of alicyclic hydrocarbons with two double bonds in conjugation with the radical C site, we found $\cdot$QOOH to be resonance stabilized by $\sim$25 kcal/mol, resulting in an overall exothermic reaction. Furthermore, the abstraction of H in $cy$-hex-2 peroxy radical led to a tertiary C radical site, which was further stabilized by the +I effect of the adjacent methyl groups. 
In this alicyclic system, we observed the 6-membered cyclic TS structure to be lower in energy than a 5-membered structure by 
$\sim$10 kcal/mol---as long as the ring is connected
to cyclohexadiene in a spiro-fashion. We found the activation barrier to rise by $\sim$5 kcal/mol due to steric strain when the cyclic TS structure is fused to an alicyclic cyclohexyl hydrocarbon ring, with two atoms shared in common. However, we found the fused 6-membered TS to be relieved of strain when fused with the cycloheptadienyl ring.
We also  explored the stability of these $\cdot$QOOH radicals with respect to the dissociation path towards cyclic ether+$\cdot$OH, and
have shown these structures to be trapped between barriers
in the range of 15--30 kcal/mol.
Overall, we find $\cdot$QOOH radicals of
$cy$-hex-2,
$cy$-hep-3, and
$cy$-hep-4 
to have favorable stabilities, suitable to enhance their lifetimes, hence enhanced detectability,  
compared to the previously detected resonance-stabilized cycloheptadienyl-based radical.

Understanding the stability and behaviour of $\cdot$QOOH radicals is crucial for comprehending complex atmospheric chemistry processes, including pollutant formation/degradation, and the production of secondary organic aerosols\cite{jokinen2014rapid,glowacki2010unimolecular,barber2021chemistry,wang2017formation}. Our findings provide insights into the chemical physics of free-radical reactive intermediates and offer potential for various applications. This approach can be further expanded to a broader range of alkanes and other compounds, enabling the discovery of novel factors stabilizing free-radical reactive intermediates and facilitating the design of more efficient and controllable combustion processes\cite{gerdroodbary2020scramjets}. TS theory and molecular dynamics simulations can be employed in future research to gain a deeper understanding of radical processes\cite{doentgen2015automated,chenoweth2008reaxff,truhlar1996current}. 

\section{Acknowledgments}
We gratefully acknowledge funding through the Indo-SA joint research project supported by the Department of Science and Technology (India) and the National Research Foundation (South Africa). 
RR acknowledges the support of the Department of Atomic Energy, Government
of India, under Project Identification No.~RTI~4007. 
Calculations have been performed using the Helios computer cluster, 
which is an integral part of the MolDis Big Data facility, TIFR Hyderabad \href{http://moldis.tifrh.res.in}{(http://moldis.tifrh.res.in)}. We also thank the Centre for High-Performance Computing (CHPC) in South Africa for providing access to their computer resources.

\section{Author Declarations}

\subsection{Author Contributions}
{\bf SCK}: Data collection and curation (main); Investigation (main); Visualization (main); Writing (main).
{\bf KPO}: Data collection and curation (supporting); Investigation (supporting).
{\bf SJ}: Data collection and curation (supporting); Writing (supporting).
{\bf SS}: Preliminary investigation (main).
{\bf CM}: Data collection and curation (supporting).
{\bf SC}: Data collection and curation (supporting); Investigation (supporting); Writing (supporting).
{\bf LVM}: Conceptualization (main); Funding acquisition (main); Project administration (main); Supervision (main); Writing (supporting).
{\bf RR}: Conceptualization (main); Funding acquisition (main); Project administration (main); Resources (main); Supervision (main); Writing (main).
\subsection{Conflicts of Interest}
The author has no conflicts of interest to disclose.
\subsection{Data Availability}
The data supporting the findings of this study are available within 
the article and its supplementary materials.

\bibliographystyle{apsrev4-1}
\bibliography{up_doi}

\end{document}


\clearpage

\singlespacing
\subsection*{Table of Contents}
\begin{enumerate}

\item DFT benchmarking (p.~10-11)
\item Spin contamination analysis (p.~12-14)

    \item  Low-temperature hydrocarbon combustion mechanism and the associated reactive
intermediates for R$\cdot$, where R is
    \begin{itemize}
        \item {\bf Figure S1} ethyl (p.~15)
        \item {\bf Figure S2} isopropyl (p.~15) 
        \item {\bf Figure S3} isobutyl (p.~15)
        \item {\bf Figure S4} tertbutyl(p.~16)
        \item {\bf Figure S5} neopentyl (p.~16)
        \item {\bf Figure S6} cyclohexyl (p.~16)
        \item {\bf Figure S7} $\alpha$-cyclohexenyl (p.~17)
        \item {\bf Figure S8} $\beta$-cyclohexenyl (p.~17)
        \item {\bf Figure S9} cyclohexadienyl (p.~17)
        
    \end{itemize}

    \item Intrinsic Reaction Coordinate (IRC) Profiles for the different reactions in low-temperature combustion for R$\cdot$, where R is
    \begin{itemize}
        \item {\bf Figure S10} ethyl (p.~18)
        \item {\bf Figure S11} isopropyl (p.~19)  
        \item {\bf Figure S12} isobutyl (p.~19)  
        \item {\bf Figure S13} tertbutyl (p.~20)  
        \item {\bf Figure S14} neopentyl (p.~20)  
        \item {\bf Figure S15} cyclohexyl (p.~21)  
         \item {\bf Figure S16} $\alpha$-cyclohexenyl (p.~21)
        \item {\bf Figure S17} $\beta$-cyclohexenyl (p.~22)  
        \item {\bf Figure S18} cyclohexadienyl (p.~22)  
    \end{itemize}
    
    \item W1-level\cite{martin1999towards} equilibrium geometry, harmonic frequencies, and thermochemistry energies of different reaction species involved in combustion reaction of R$\cdot$, where R is
    
    \begin{itemize}
    
        \item {\bf Hydrocarbon 1:} ethyl  (pp.~23--31)
        \begin{itemize}
          \item {\bf Species 1:} $\cdot$R (p.~23)
          \item {\bf Species 2:} ROO$\cdot$ (p.~24) 
          \item {\bf Species 3:} $\cdot$QOOH (p.~25)
          \item {\bf Species 4:} {\it cy}-ether (p.~26)
          \item {\bf Species 5:} alkene (p.~27)
          \item {\bf Species 6:} Transition state along \rxnone{} (p.~28)
          \item {\bf Species 7:} Transition state along \rxntwo{} (p.~29)
          \item {\bf Species 8:} Transition state along \rxnthree{} (p.~30)
          \item {\bf Species 9:} Transition state along \rxnfour{} (p.~31)
      \end{itemize} 
      
      \item {\bf Hydrocarbon 2:} isopropyl  (pp.~32--40)
        \begin{itemize}
          \item {\bf Species 1:} $\cdot$R (p.~32)
          \item {\bf Species 2:} ROO$\cdot$ (p.~33) 
          \item {\bf Species 3:} $\cdot$QOOH (p.~34)
          \item {\bf Species 4:} {\it cy}-ether (p.~35)
          \item {\bf Species 5:} alkene (p.~36)
          \item {\bf Species 6:} Transition state along \rxnone{} (p.~37)
          \item {\bf Species 7:} Transition state along \rxntwo{} (p.~38)
          \item {\bf Species 8:} Transition state along \rxnthree{} (p.~39)
          \item {\bf Species 9:} Transition state along \rxnfour{} (p.~40) 
      \end{itemize}

     \item {\bf Hydrocarbon 3:} isobutyl (pp.~41--49)
        \begin{itemize}
          \item {\bf Species 1:} $\cdot$R (p.~41)
          \item {\bf Species 2:} ROO$\cdot$ (p.~42) 
          \item {\bf Species 3:} $\cdot$QOOH (p.~43)
          \item {\bf Species 4:} {\it cy}-ether (p.~44)
          \item {\bf Species 5:} alkene (p.~45)
          \item {\bf Species 6:} Transition state along \rxnone{} (p.~46)
          \item {\bf Species 7:} Transition state along \rxntwo{} (p.~47)
          \item {\bf Species 8:} Transition state along \rxnthree{} (p.~48)
          \item {\bf Species 9:} Transition state along \rxnfour{} (p.~49) 
      \end{itemize}

     \item {\bf Hydrocarbon 4:} tertbutyl (pp.~50--58)
        \begin{itemize}
          \item {\bf Species 1:} $\cdot$R (p.~50)
          \item {\bf Species 2:} ROO$\cdot$ (p.~51) 
          \item {\bf Species 3:} $\cdot$QOOH (p.~52)
          \item {\bf Species 4:} {\it cy}-ether (p.~53)
          \item {\bf Species 5:} alkene (p.~54)
          \item {\bf Species 6:} Transition state along \rxnone{} (p.~55)
          \item {\bf Species 7:} Transition state along \rxntwo{} (p.~56)
          \item {\bf Species 8:} Transition state along \rxnthree{} (p.~57)
          \item {\bf Species 9:} Transition state along \rxnfour{} (p.~58) 
      \end{itemize}

   \item {\bf Hydrocarbon 5:} neopentyl (pp.~59--67)
        \begin{itemize}
          \item {\bf Species 1:} $\cdot$R (p.~59)
          \item {\bf Species 2:} ROO$\cdot$ (p.~60) 
          \item {\bf Species 3:} $\cdot$QOOH (p.~61)
          \item {\bf Species 4:} {\it cy}-ether (p.~61)
          \item {\bf Species 5:} dimetyhlcyclopropane (p.~63)
          \item {\bf Species 6:} Transition state along \rxnone{} (p.~64)
          \item {\bf Species 7:} Transition state along \rxntwo{} (p.~65)
          \item {\bf Species 8:} Transition state along \rxnthree{} (p.~66)
          \item {\bf Species 9:} Transition state along \rxnfour{} (p.~67) 
      \end{itemize}

 \item {\bf Hydrocarbon 6:} cyclohexyl (pp.~68--76)
        \begin{itemize}
          \item {\bf Species 1:} $\cdot$R (p.~68)
          \item {\bf Species 2:} ROO$\cdot$ (p.~69) 
          \item {\bf Species 3:} $\cdot$QOOH (p.~70)
          \item {\bf Species 4:} {\it cy}-ether (p.~71)
          \item {\bf Species 5:} alkene (p.~72)
          \item {\bf Species 6:} Transition state along \rxnone{} (p.~73)
          \item {\bf Species 7:} Transition state along \rxntwo{} (p.~74)
          \item {\bf Species 8:} Transition state along \rxnthree{} (p.~75)
          \item {\bf Species 9:} Transition state along \rxnfour{} (p.~76) 
      \end{itemize}

  \item {\bf Hydrocarbon 7:} $\alpha$-cyclohexenyl (pp.~77--85)
        \begin{itemize}
          \item {\bf Species 1:} $\cdot$R (p.~77)
          \item {\bf Species 2:} ROO$\cdot$ (p.~78) 
          \item {\bf Species 3:} $\cdot$QOOH (p.~79)
          \item {\bf Species 4:} {\it cy}-ether (p.~80)
          \item {\bf Species 5:} alkene (p.~81)
          \item {\bf Species 6:} Transition state along \rxnone{} (p.~82)
          \item {\bf Species 7:} Transition state along \rxntwo{} (p.~83)
          \item {\bf Species 8:} Transition state along \rxnthree{} (p.~84)
          \item {\bf Species 9:} Transition state along \rxnfour{} (p.~85) 
      \end{itemize}

\item {\bf Hydrocarbon 8:} cyclohexadienyl (pp.~86--94)
        \begin{itemize}
          \item {\bf Species 1:} $\cdot$R (p.~86)
          \item {\bf Species 2:} ROO$\cdot$ (p.~87) 
          \item {\bf Species 3:} $\cdot$QOOH (p.~88)
          \item {\bf Species 4:} {\it cy}-ether (p.~89)
          \item {\bf Species 5:} alkene (p.~90)
          \item {\bf Species 6:} Transition state along \rxnone{} (p.~91)
          \item {\bf Species 7:} Transition state along \rxntwo{} (p.~92)
          \item {\bf Species 8:} Transition state along \rxnthree{} (p.~93)
          \item {\bf Species 9:} Transition state along \rxnfour{} (p.~94) 
      \end{itemize}

    \end{itemize} 

   \item G4-level\cite{curtiss2007gaussian} equilibrium geometry, harmonic frequencies, and thermochemistry energies of different reaction species involved in combustion reaction of R$\cdot$, where R is
 \begin{itemize}
    
        \item {\bf Hydrocarbon 1:} ethyl  (pp.~95--103)
        \begin{itemize}
          \item {\bf Species 1:} $\cdot$R (p.~95)
          \item {\bf Species 2:} ROO$\cdot$ (p.~96) 
          \item {\bf Species 3:} $\cdot$QOOH (p.~97)
          \item {\bf Species 4:} {\it cy}-ether (p.~98)
          \item {\bf Species 5:} alkene (p.~99)
          \item {\bf Species 6:} Transition state along \rxnone{} (p.~100)
          \item {\bf Species 7:} Transition state along \rxntwo{} (p.~101)
          \item {\bf Species 8:} Transition state along \rxnthree{} (p.~102)
          \item {\bf Species 9:} Transition state along \rxnfour{} (p.~103)
      \end{itemize} 
      
      \item {\bf Hydrocarbon 2:} isopropyl  (pp.~104--112)
        \begin{itemize}
          \item {\bf Species 1:} $\cdot$R (p.~104)
          \item {\bf Species 2:} ROO$\cdot$ (p.~105) 
          \item {\bf Species 3:} $\cdot$QOOH (p.~106)
          \item {\bf Species 4:} {\it cy}-ether (p.~107)
          \item {\bf Species 5:} alkene (p.~108)
          \item {\bf Species 6:} Transition state along \rxnone{} (p.~109)
          \item {\bf Species 7:} Transition state along \rxntwo{} (p.~110)
          \item {\bf Species 8:} Transition state along \rxnthree{} (p.~111)
          \item {\bf Species 9:} Transition state along \rxnfour{} (p.~112) 
      \end{itemize}

     \item {\bf Hydrocarbon 3:} isobutyl (pp.~113--121)
        \begin{itemize}
          \item {\bf Species 1:} $\cdot$R (p.~113)
          \item {\bf Species 2:} ROO$\cdot$ (p.~114) 
          \item {\bf Species 3:} $\cdot$QOOH (p.~115)
          \item {\bf Species 4:} {\it cy}-ether (p.~116)
          \item {\bf Species 5:} alkene (p.~117)
          \item {\bf Species 6:} Transition state along \rxnone{} (p.~118)
          \item {\bf Species 7:} Transition state along \rxntwo{} (p.~119)
          \item {\bf Species 8:} Transition state along \rxnthree{} (p.~120)
          \item {\bf Species 9:} Transition state along \rxnfour{} (p.~121) 
      \end{itemize}

     \item {\bf Hydrocarbon 4:} tertbutyl (pp.~122--130)
        \begin{itemize}
          \item {\bf Species 1:} $\cdot$R (p.~122)
          \item {\bf Species 2:} ROO$\cdot$ (p.~123) 
          \item {\bf Species 3:} $\cdot$QOOH (p.~124)
          \item {\bf Species 4:} {\it cy}-ether (p.~125)
          \item {\bf Species 5:} alkene (p.~126)
          \item {\bf Species 6:} Transition state along \rxnone{} (p.~127)
          \item {\bf Species 7:} Transition state along \rxntwo{} (p.~128)
          \item {\bf Species 8:} Transition state along \rxnthree{} (p.~129)
          \item {\bf Species 9:} Transition state along \rxnfour{} (p.~130) 
      \end{itemize}

   \item {\bf Hydrocarbon 5:} neopentyl (pp.~131--139)
        \begin{itemize}
          \item {\bf Species 1:} $\cdot$R (p.~131)
          \item {\bf Species 2:} ROO$\cdot$ (p.~132) 
          \item {\bf Species 3:} $\cdot$QOOH (p.~133)
          \item {\bf Species 4:} {\it cy}-ether (p.~134)
          \item {\bf Species 5:} alkene (p.~135)
          \item {\bf Species 6:} Transition state along \rxnone{} (p.~136)
          \item {\bf Species 7:} Transition state along \rxntwo{} (p.~137)
          \item {\bf Species 8:} Transition state along \rxnthree{} (p.~138)
          \item {\bf Species 9:} Transition state along \rxnfour{} (p.~139) 
      \end{itemize}

 \item {\bf Hydrocarbon 6:} cyclohexyl (pp.~140--148)
        \begin{itemize}
          \item {\bf Species 1:} $\cdot$R (p.~140)
          \item {\bf Species 2:} ROO$\cdot$ (p.~141) 
          \item {\bf Species 3:} $\cdot$QOOH (p.~142)
          \item {\bf Species 4:} {\it cy}-ether (p.~143)
          \item {\bf Species 5:} alkene (p.~144)
          \item {\bf Species 6:} Transition state along \rxnone{} (p.~145)
          \item {\bf Species 7:} Transition state along \rxntwo{} (p.~146)
          \item {\bf Species 8:} Transition state along \rxnthree{} (p.~147)
          \item {\bf Species 9:} Transition state along \rxnfour{} (p.~148) 
      \end{itemize}

  \item {\bf Hydrocarbon 7:} $\alpha$-cyclohexenyl (pp.~149--157)
        \begin{itemize}
          \item {\bf Species 1:} $\cdot$R (p.~149)
          \item {\bf Species 2:} ROO$\cdot$ (p.~150) 
          \item {\bf Species 3:} $\cdot$QOOH (p.~151)
          \item {\bf Species 4:} {\it cy}-ether (p.~152)
          \item {\bf Species 5:} alkene (p.~153)
          \item {\bf Species 6:} Transition state along \rxnone{} (p.~154)
          \item {\bf Species 7:} Transition state along \rxntwo{} (p.~155)
          \item {\bf Species 8:} Transition state along \rxnthree{} (p.~156)
          \item {\bf Species 9:} Transition state along \rxnfour{} (p.~157) 
      \end{itemize}

\item {\bf Hydrocarbon 8:} $\beta$-cyclohexenyl (pp.~158--166)
        \begin{itemize}
          \item {\bf Species 1:} $\cdot$R (p.~158)
          \item {\bf Species 2:} ROO$\cdot$ (p.~159) 
          \item {\bf Species 3:} $\cdot$QOOH (p.~160)
          \item {\bf Species 4:} {\it cy}-ether (p.~161)
          \item {\bf Species 5:} alkene (p.~162)
          \item {\bf Species 6:} Transition state along \rxnone{} (p.~163)
          \item {\bf Species 7:} Transition state along \rxntwo{} (p.~164)
          \item {\bf Species 8:} Transition state along \rxnthree{} (p.~165)
          \item {\bf Species 9:} Transition state along \rxnfour{} (p.~166) 
      \end{itemize}

\item {\bf Hydrocarbon 9:} cyclohexadienyl (pp.~167--175)
        \begin{itemize}
          \item {\bf Species 1:} $\cdot$R (p.~167)
          \item {\bf Species 2:} ROO$\cdot$ (p.~168) 
          \item {\bf Species 3:} $\cdot$QOOH (p.~169)
          \item {\bf Species 4:} {\it cy}-ether (p.~170)
          \item {\bf Species 5:} benzene (p.~171)
          \item {\bf Species 6:} Transition state along \rxnone{} (p.~172)
          \item {\bf Species 7:} Transition state along \rxntwo{} (p.~173)
          \item {\bf Species 8:} Transition state along \rxnthree{} (p.~174)
          \item {\bf Species 9:} Transition state along \rxnfour{} (p.~175) 
      \end{itemize}


\item {\bf Hydrocarbon 10:} 2-(cyclohexa-2,4-dien-1-yl)ethyl (pp.~176--183)
        \begin{itemize}
          \item {\bf Species 1:} $\cdot$R (p.~176)
          \item {\bf Species 2:} ROO$\cdot$ (p.~177) 
          \item {\bf Species 3:} $syn$-$\cdot$QOOH (p.~178)
          \item {\bf Species 4:} $anti$-$\cdot$QOOH (p.~179)
          \item {\bf Species 5:} {\it cy}-ether (p.~180)
          \item {\bf Species 6:} Transition state along \rxnone{} (p.~181)
          \item {\bf Species 7:} Transition state along \rxnfour{} (p.~182-183) 
      \end{itemize}

\item {\bf Hydrocarbon 11:} 6-methyl cyclohepta-2,4-dien-1-yl (pp.~184--190)
        \begin{itemize}
           \item {\bf Species 1:} $\cdot$R (p.~184)
          \item {\bf Species 2:} ROO$\cdot$ (p.~185) 
          \item {\bf Species 3:} $syn$-$\cdot$QOOH (p.~186)
          \item {\bf Species 4:} $anti$-$\cdot$QOOH (p.~187)
          \item {\bf Species 5:} {\it cy}-ether (p.~188)
          \item {\bf Species 6:} Transition state along \rxnone{} (p.~189)
          \item {\bf Species 7:} Transition state along \rxnfour{} (p.~190) 
      \end{itemize}
      
\item {\bf Hydrocarbon 12:} (cyclohepta-3,5-dien-1-yl)methyl (pp.~191--198)
        \begin{itemize}
           \item {\bf Species 1:} $\cdot$R (p.~191)
          \item {\bf Species 2:} ROO$\cdot$ (p.~192) 
          \item {\bf Species 3:} $syn$-$\cdot$QOOH (p.~193)
          \item {\bf Species 4:} $anti$-$\cdot$QOOH (p.~194)
          \item {\bf Species 5:} {\it cy}-ether (p.~195)
          \item {\bf Species 6:} Transition state along \rxnone{} (p.~196)
          \item {\bf Species 7:} Transition state along \rxnfour{} (p.~197-198) 
      \end{itemize}

\item {\bf Hydrocarbon 13:} cyclohepta-2,4-dien-1-yl (pp.~199--205)
        \begin{itemize}
           \item {\bf Species 1:} $\cdot$R (p.~199)
          \item {\bf Species 2:} ROO$\cdot$ (p.~200) 
          \item {\bf Species 3:} $syn$-$\cdot$QOOH (p.~201)
          \item {\bf Species 4:} $anti$-$\cdot$QOOH (p.~202)
          \item {\bf Species 5:} {\it cy}-ether (p.~203)
          \item {\bf Species 6:} Transition state along \rxnone{} (p.~204)
          \item {\bf Species 7:} Transition state along \rxnfour{} (p.~205) 
      \end{itemize}
      
\item {\bf Hydrocarbon 14:} cyclohepta-3,5-dien-1-yl (pp.~206--209)
        \begin{itemize}
           \item {\bf Species 1:} $\cdot$R (p.~206)
          \item {\bf Species 2:} ROO$\cdot$ (p.~207) 
          \item {\bf Species 3:} $\cdot$QOOH (p.~208)
          \item {\bf Species 4:} Transition state along \rxnone{} (p.~209)
      \end{itemize}

\item {\bf Hydrocarbon 15:} (cyclohexa-2,4-dien-1-yl)methyl (5-membered path) (pp.~210--213)
        \begin{itemize}
           \item {\bf Species 1:} $\cdot$R (p.~210)
          \item {\bf Species 2:} ROO$\cdot$ (p.~211) 
          \item {\bf Species 3:} $\cdot$QOOH (p.~212)
          \item {\bf Species 4:} Transition state along \rxnone{} (p.~213)
      \end{itemize}
      
\item {\bf Hydrocarbon 16:} (cyclohexa-2,4-dien-1-yl)methyl (6-membered path) (pp.~214--217)
        \begin{itemize}
           \item {\bf Species 1:} $\cdot$R (p.~214)
          \item {\bf Species 2:} ROO$\cdot$ (p.~215) 
          \item {\bf Species 3:} $\cdot$QOOH (p.~216)
          \item {\bf Species 4:} Transition state along \rxnone{} (p.~217)
      \end{itemize}

    \end{itemize}

    \item References (pp.~218)
\end{enumerate}
\onehalfspacing

\clearpage


\section{DFT Benchmarking}
\begin{figure*}
        \centering
        \includegraphics[width=1.0\linewidth]{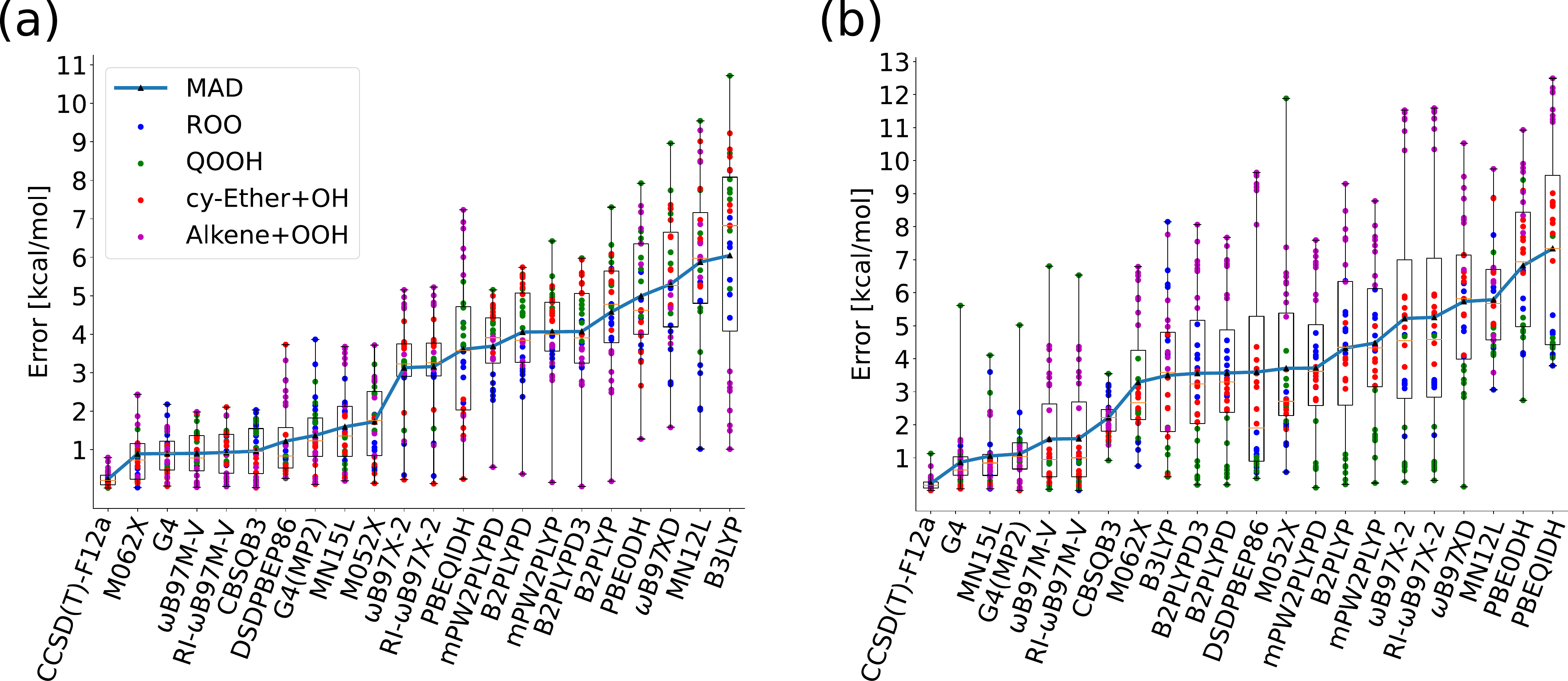}
         \caption{Ranking of the prediction accuracy of DFAs and cWFTs for reaction barriers using box-and-whisker plot. For a given method, the Box-and-whisker plot depicts the deviation in energy with respect to W1 energy of (a) Reaction species and (b) Reaction barriers for all the benchmark systems. For boxplots, the centre line denotes the median, the box displays the interquartile range, and the whiskers indicate the range. The MAD for each method is connected by blue line. Methods are sorted in ascending order of MAD. RI-$\omega$B97M-V and RI-$\omega$B97X-2 represent the methods where RI approximation was activated.}
        \label{fig:DFT_Performance} 
    \end{figure*}  

\begin{table}[ht]
\centering
\caption{$\Delta U^{\rm ROO}$ and $\Delta U^{\rm TS4}$ energies (with respect to R$\cdot$ + O$_2$) for dataset systems calculated with  W1, CCSD(T)-F12a/cc-pVDZ-F12, G4, and RI-$\omega$B97M-V methods. In the table, CC and DFT represent CCSD(T)-F12a/cc-pVDZ-F12 and RI-$\omega$B97M-V, respectively.} 
\small\addtolength{\tabcolsep}{1.2pt}              
\begin{tabular}[t]{l ccc ccc cc}
\hline

 \multicolumn{1}{l}{Species} &
 \multicolumn{1}{l}{$\Delta U_{\rm W1}^{\rm ROO}$} &
\multicolumn{1}{l}{$\Delta U_{\rm CC}^{\rm ROO}$} &
\multicolumn{1}{l}{$\Delta U_{\rm G4}^{\rm ROO}$} &
\multicolumn{1}{l}{$\Delta U_{\rm DFT}^{\rm ROO}$} &
\multicolumn{1}{l}{$\Delta U_{\rm W1}^{\rm TS4}$} &
\multicolumn{1}{l}{$\Delta U_{\rm CC}^{\rm TS4}$} &
\multicolumn{1}{l}{$\Delta U_{\rm G4}^{\rm TS4}$} &
\multicolumn{1}{l}{$\Delta U_{\rm DFT}^{\rm TS4}$} \\
\hline
Ethyl            & -32.82  & -33.03 & -33.45 & -34.23 &  -1.52 & -1.70 & -0.30 & 2.08 \\
Isopropyl        & -34.58  & -34.75 & -35.74 & -35.55 &  -4.94 & -5.06 & -3.97 & -0.46 \\
Isobutyl         & -33.66  & -33.70 & -34.53 & -35.01 & -11.99 & -11.90& -10.44& -9.48 \\
Tertbutyl        & -36.65  & -35.94 & -37.20 & -36.24 &  -6.74 & -7.18 & -6.25 & -2.37 \\
Neopentyl        & -33.73  & -33.74 & -34.67 & -35.13 &   0.71 &  0.47 &  2.20 &  3.96\\
{\it cy}-hexyl       & -33.45  & -35.58 & -36.83 & -36.36 &  -8.55 & -8.71 & -7.54 &  -5.10\\
{\it cy}-hexenyl     & -19.63  & -19.84 & -21.52 & -20.29 &   7.29 &  7.07 &  7.91 &  11.28\\
{\it cy}-hexadienyl  & -11.13  & -11.50 & -13.31 & -11.21 &   3.67 &  2.92 &  4.89 &  6.92\\
                                      
\hline
\end{tabular}
\label{tab:W1}
\end{table}%

\begin{table}[ht]
\centering
\caption{Comparison of performance of CCSD(T)-F12a/cc-pVDZ-F12, G4 and RI-$\omega$B97M-V with respect to W1 for ROO$\cdot$ and TS4. Here, 
$\Delta\Delta U_1^{\rm ROO}$ = $\Delta {U^{\rm W1}_{\rm ROO}}$- $\Delta {U^{\rm CC}_{\rm ROO}}$,
$\Delta\Delta U_2^{\rm ROO}$ = $\Delta {U^{\rm W1}_{\rm ROO}}$- $\Delta {U^{\rm G4}_{\rm ROO}}$,
$\Delta\Delta U_3^{\rm ROO}$ = $\Delta {U^{\rm W1}_{\rm ROO}}$- $\Delta {U^{\rm RI}_{\rm ROO}}$,
$\Delta\Delta U_1^{\rm TS4}$ = $\Delta {U^{\rm W1}_{\rm TS4}}$- $\Delta {U^{\rm CC}_{\rm TS4}}$,
$\Delta\Delta U_2^{\rm TS4}$ = $\Delta {U^{\rm W1}_{\rm TS4}}$- $\Delta {U^{\rm G4}_{\rm TS4}}$, and 
$\Delta\Delta U_3^{\rm TS4}$ = $\Delta {U^{\rm W1}_{\rm TS4}}$- $\Delta {U^{\rm RI}_{\rm TS4}}$.
} 
\small\addtolength{\tabcolsep}{1.2pt}              
\begin{tabular}[t]{l ccc ccc cc cccc }
\hline
\multicolumn{1}{l}{Species} &
\multicolumn{2}{l}{$\Delta\Delta U_1^{\rm ROO}$} &
\multicolumn{2}{l}{$\Delta\Delta U_2^{\rm ROO}$} &
\multicolumn{2}{l}{$\Delta\Delta U_3^{\rm ROO}$} &
\multicolumn{2}{l}{$\Delta\Delta U_1^{\rm TS4}$} &
\multicolumn{2}{l}{$\Delta\Delta U_2^{\rm TS4}$} &
\multicolumn{2}{l}{$\Delta\Delta U_3^{\rm TS4}$} \\
\hline
Ethyl & 0.21 && 0.63 && 1.41 &&0.18 && -1.18 && -3.60 \\
Isopropyl &0.17 &&1.16 && 0.97&&0.32 && -0.77 && -4.28 \\
Isobutyl &0.04 &&0.87 && 1.35&&-0.09 && -1.55 && -2.51 \\
Tertbutyl &0.29 &&1.55 && 0.59&&0.44 && -0.49 && -4.27\\
Neopentyl &0.01 &&0.94 && 1.40&&0.24 && -1.49 && -3.25 \\
{\it cy}-hexyl &0.13 &&1.38 && 0.91&&0.16 && -1.01 && -3.45 \\
{\it cy}-hexenyl&0.21 &&1.89 && 0.66&&0.22 && -0.62 && -3.99\\
{\it cy}-hexadienyl&0.37 &&2.18 && 0.08&&0.75 &&-1.22 && -3.25\\
                                        
\hline
\end{tabular}
\label{tab:W1}
\end{table}%
  
\clearpage
\subsection{Spin contamination}

Based on the results presented in Table.~IV in the main text, it is evident that the MAD is highest for TS4 for all the density functional theory (DFT) functionals considered, except for MN15L. This significant deviation for TS4 is the primary contributor to the considerably high MAD values observed for energy barriers for all the functionals. To understand the cause of this high deviation for TS4, the $<S^{2}>$ values for all reaction species and functionals were examined. The ideal value of $<S^{2}>$ for a singlet radical is 0.75. When $<S^{2}>$ deviates from this value, it is known as spin-contamination or spin error. The higher the deviation, the higher the spin contamination. The results in Figure\ref{fig:Spin contamination} demonstrate that the spin mean absolute error (MAE) for TS4 is higher than that for the other reaction species. The Pearson's correlation coefficient between spin MAE and energy MAE is 0.81 for TS4, indicating that the higher energy MAE of TS4 is due to its higher spin MAE. Furthermore, the $<S^{2}>$ values for TS4 exhibit a more significant deviation from the ideal value compared to the other TS. While TS1, TS2, and TS3 have all their data points between 0.75 and 1.0, TS4 has $<S^{2}>$ values higher than 1.0 for many functionals, which indicates higher spin contamination for TS4. The higher the spin contamination for TS4, the higher the MAD for all the functionals. Hence, this high spin contamination for TS4 may be one of the main reasons for the very high MAD values.\\
When different functionals were compared, MN15L was found to be the best-performing functional for TS, showing very low spin contamination for all the TS. In this study, all the Minnesota functionals (M06-2X, M05-2X, MN15L, MN12L) exhibited very low spin contamination compared to other functionals. Additionally, $\omega$B97M-V also had low spin contamination and low MAD for TS4. However, DSDPBEP86, $\omega$B97X-2, and PBEQIDH exhibited high spin contamination for TS4 and very high MAD values of 9.25 kcal/mol, 11.21, and 11.77 kcal/mol, respectively.

\begin{figure}
        \centering 
        \includegraphics[width=1.0\linewidth]{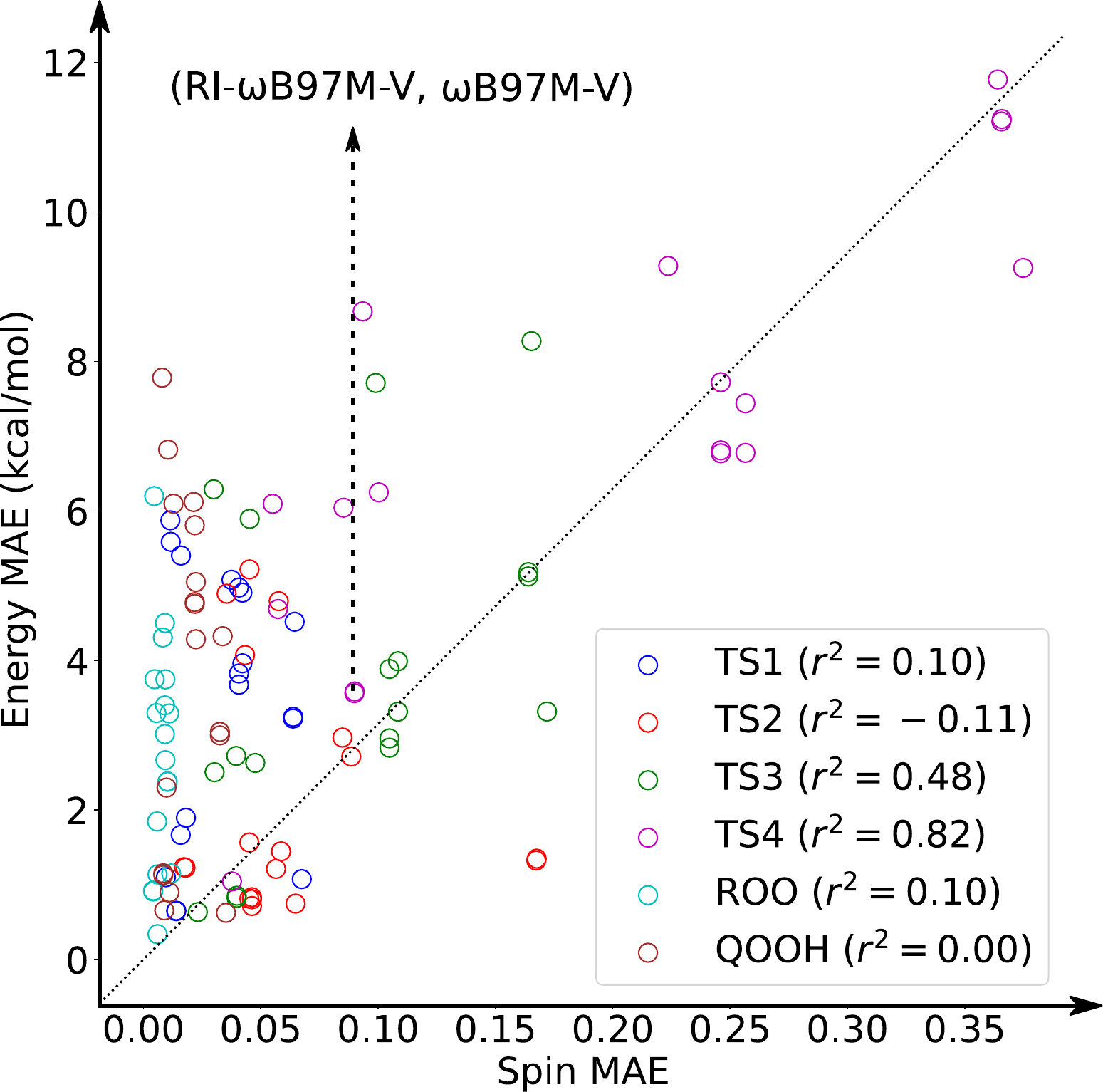}
        \caption{Scatter plot between energy mean absolute error (Energy MAE) and spin mean absolute error (Spin MAE) for reaction species. Each point in the plot represents a DFT functional for the given species. Values in the parenthesis are Pearson's correlation coefficient.}
        \label{fig:Spin contamination}
    \end{figure}

\begin{figure}
        \centering 
        \includegraphics[width=1.0\linewidth]{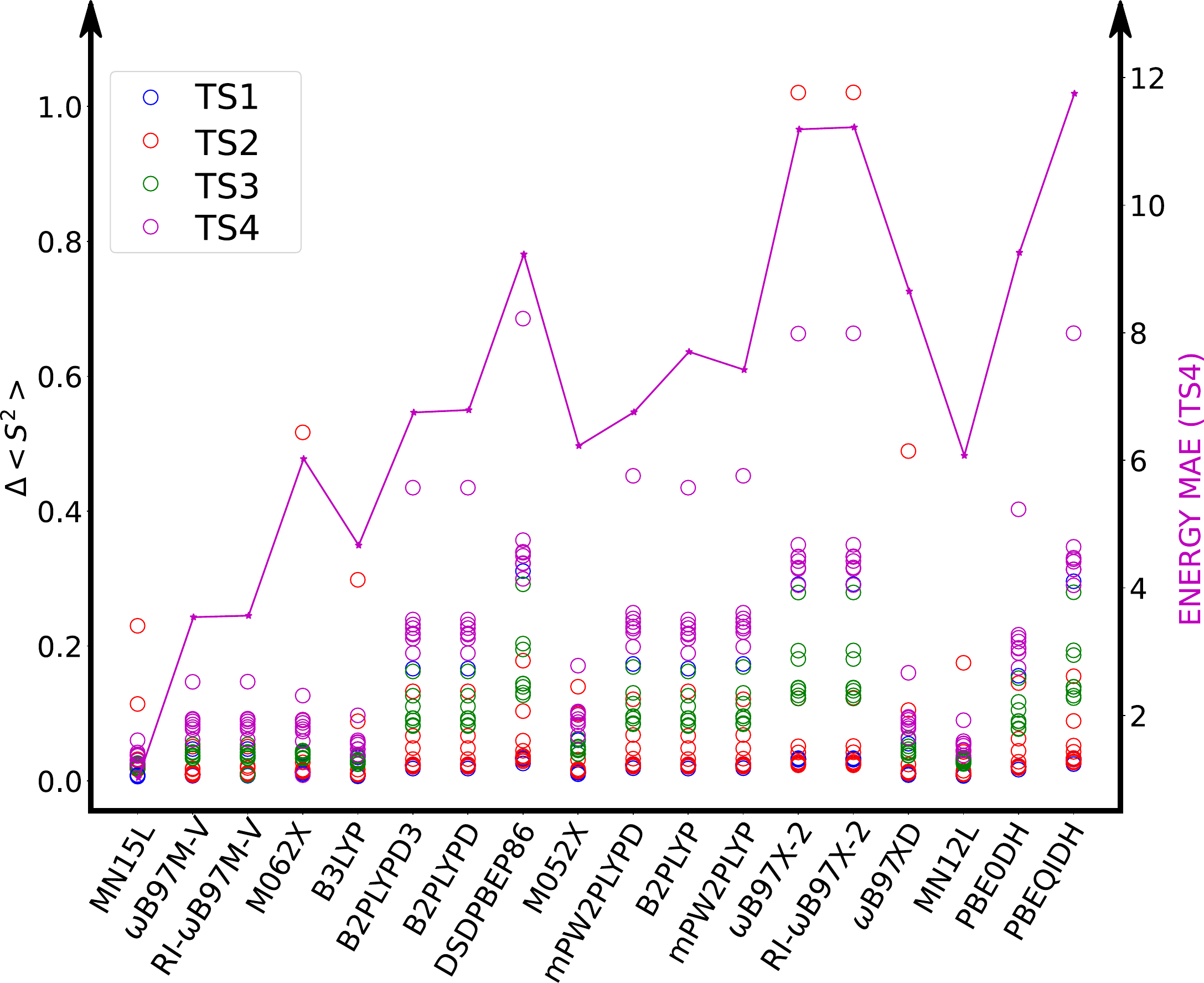}
         \caption{Scatter plot showing the difference between ideal and calculated $<S^{2}>$ values of TS of all systems for all DFT functionals. The left-hand side y-axis represents $\Delta <S^{2}>$ values. The solid pink line and the right-hand side y-axis represent the MAD for TS4. Methods are sorted in ascending order of MAD for TS.}
        \label{fig:Spin contamination}
    \end{figure}

    \clearpage

\clearpage

\section{Low-temperature hydrocarbon combustion mechanism and the associated reactive
intermediates for R$\cdot$}
    \begin{figure}[H]
           \centering
           \includegraphics[width=1.0\linewidth]{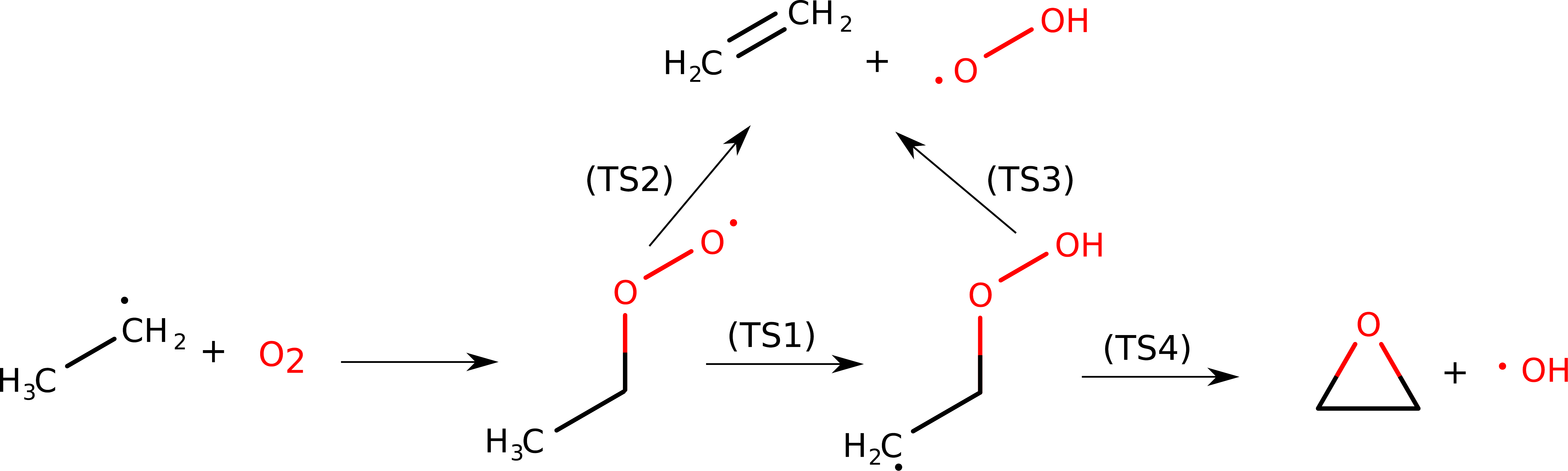}
           \caption{The reactive intermediates for ethyl radical combustion}
           \label{Ethyl Reaction Scheme}
    \end{figure}

    \begin{figure}[H]
           \centering
           \includegraphics[width=1.0\linewidth]{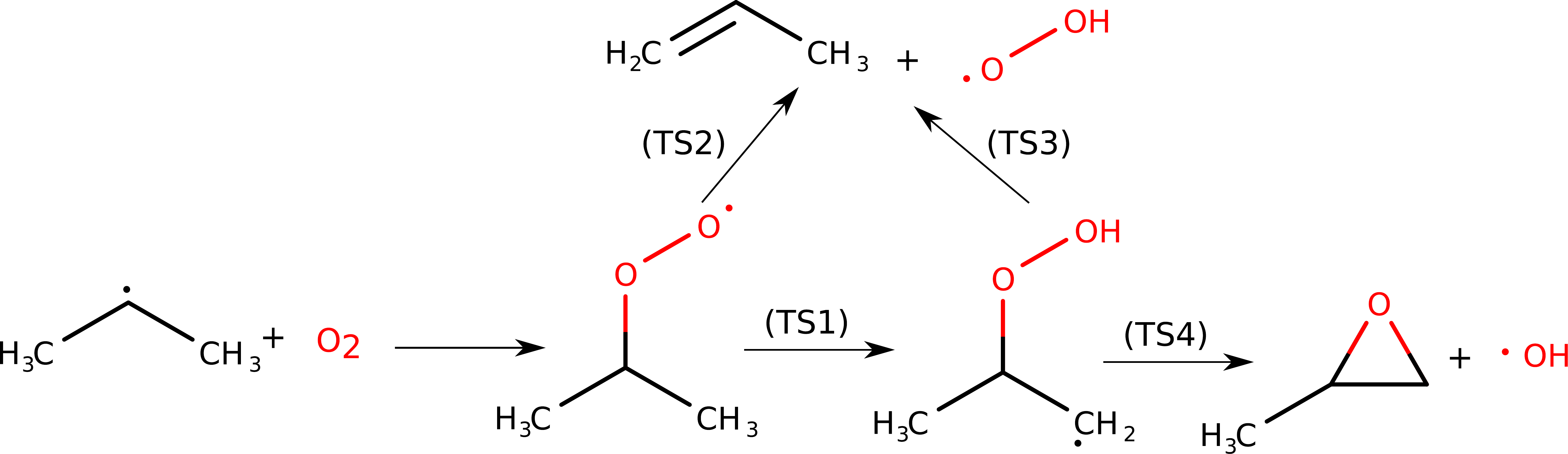}
           \caption{The reactive intermediates for isopropyl radical combustion}
           \label{Isopropyl Reaction Scheme}
    \end{figure}

    \begin{figure}[H]
           \centering
           \includegraphics[width=1.0\linewidth]{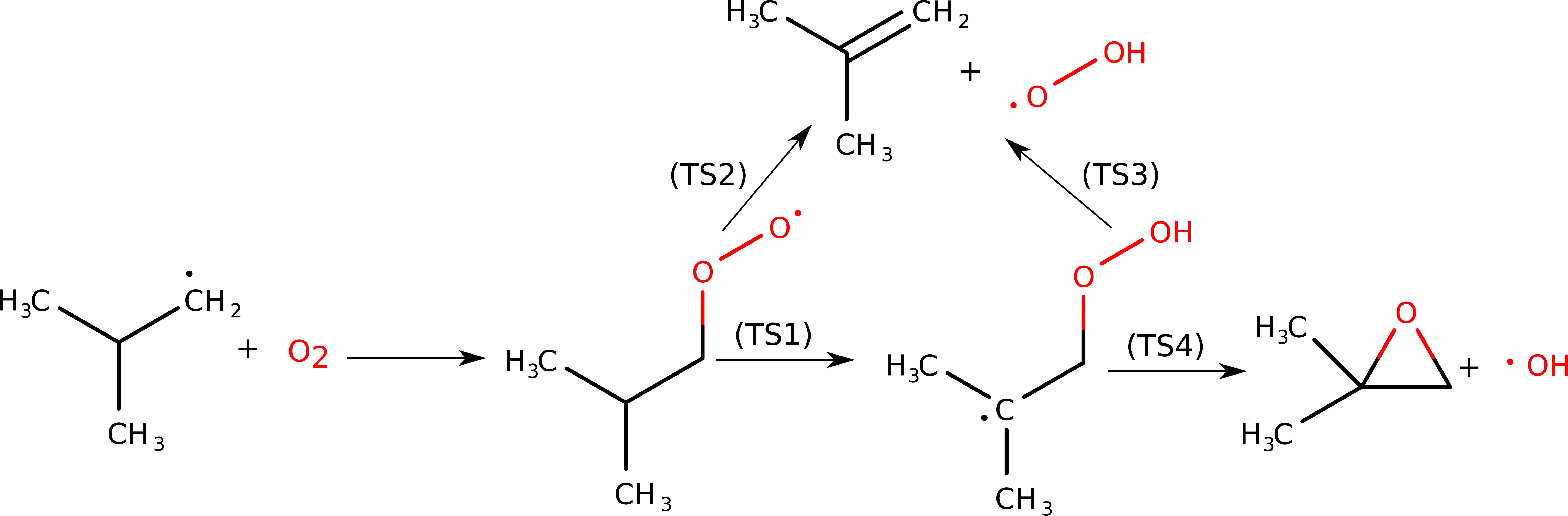}
           \caption{The reactive intermediates for isobutyl radical combustion}
           \label{Isobutyl Reaction Scheme}
    \end{figure}

    \begin{figure}[H]
           \centering
           \includegraphics[width=1.0\linewidth]{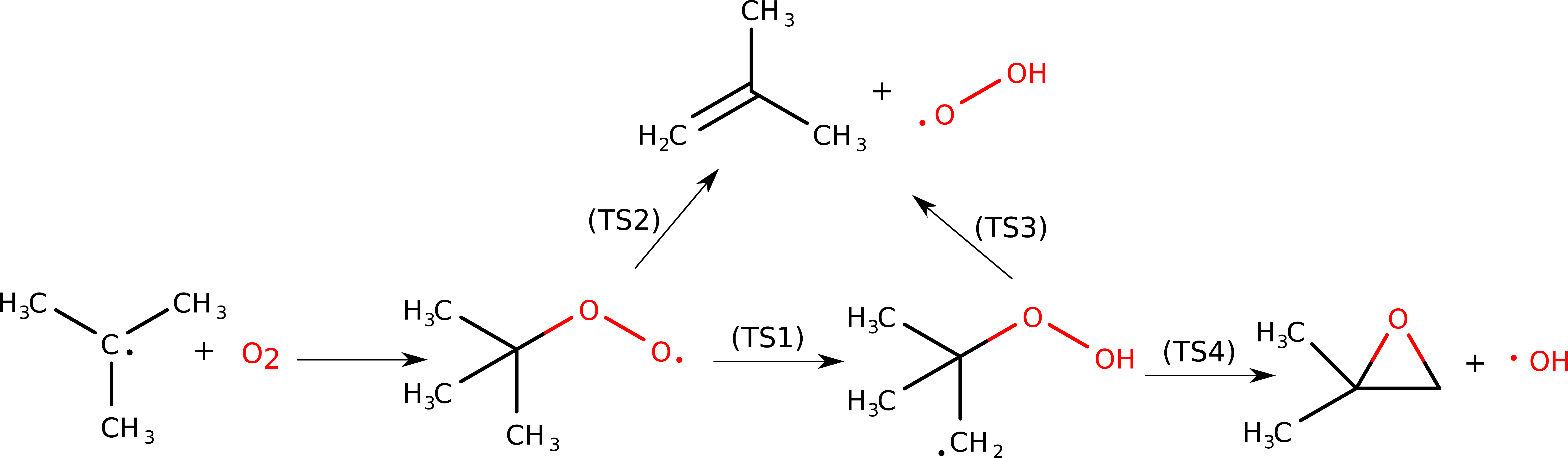}
           \caption{The reactive intermediates for tertbutyl radical combustion}
           \label{Tertbutyl Reaction Scheme}
    \end{figure}

    \begin{figure}[H]
           \centering
           \includegraphics[width=1.0\linewidth]{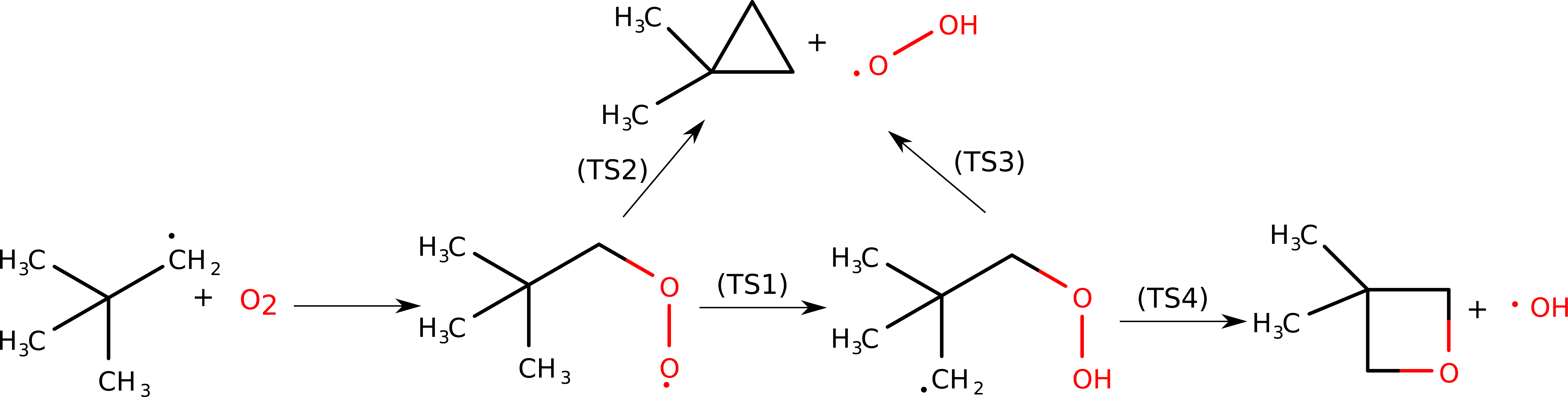}
           \caption{The reactive intermediates for neopentyl radical combustion}
           \label{Neopentyl Reaction Scheme}
    \end{figure}

    \begin{figure}[H]
           \centering
           \includegraphics[width=1.0\linewidth]{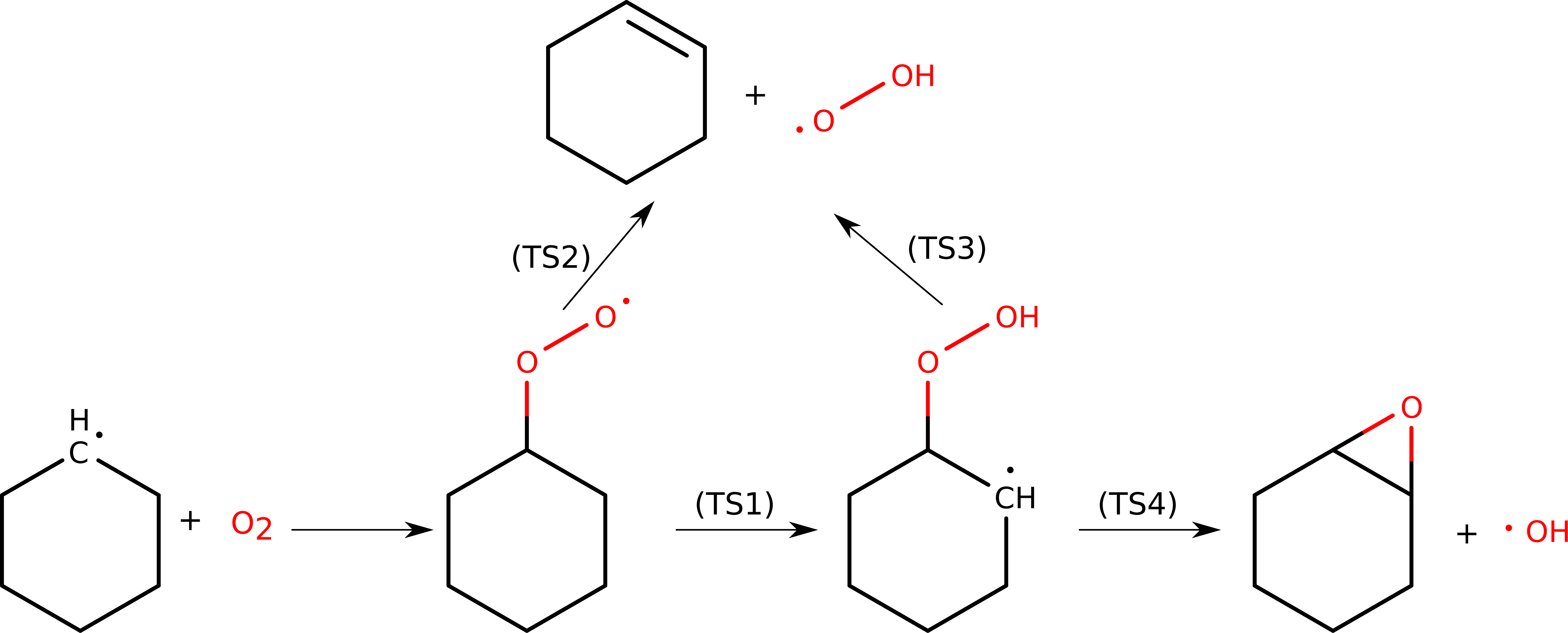}
           \caption{The reactive intermediates for cyclohexyl radical combustion}
           \label{Cyclohexyl Reaction Scheme}
    \end{figure}

    \begin{figure}[H]
           \centering
           \includegraphics[width=1.0\linewidth]{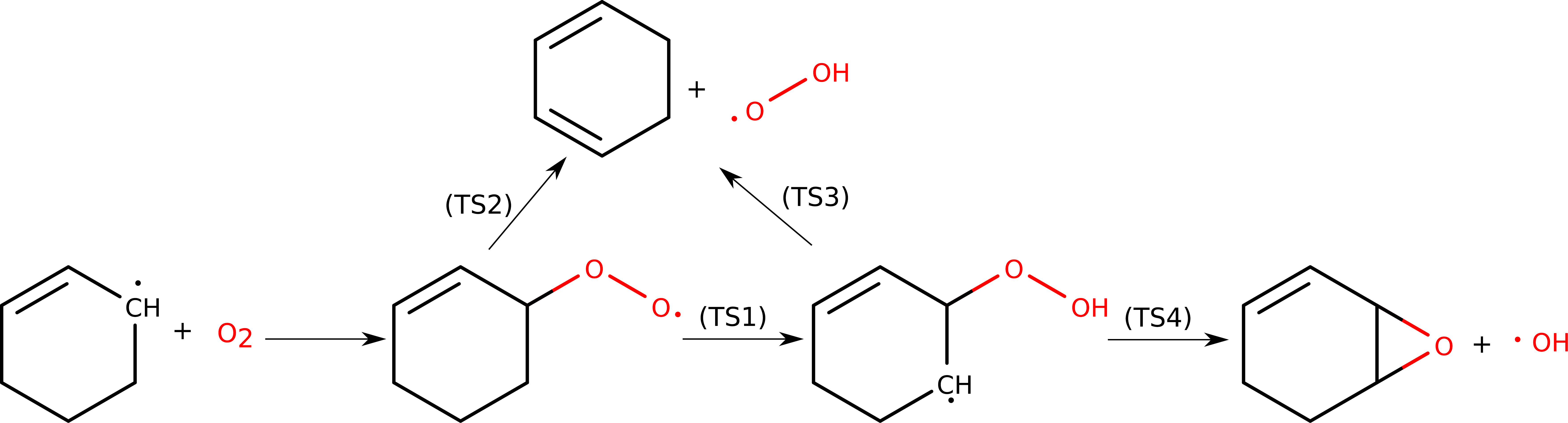}
           \caption{The reactive intermediates for $\alpha$-cyclohexenyl radical combustion}
           \label{Cyclohexenyl Reaction Scheme}
    \end{figure}
    
    \begin{figure}[H]
           \centering
           \includegraphics[width=1.0\linewidth]{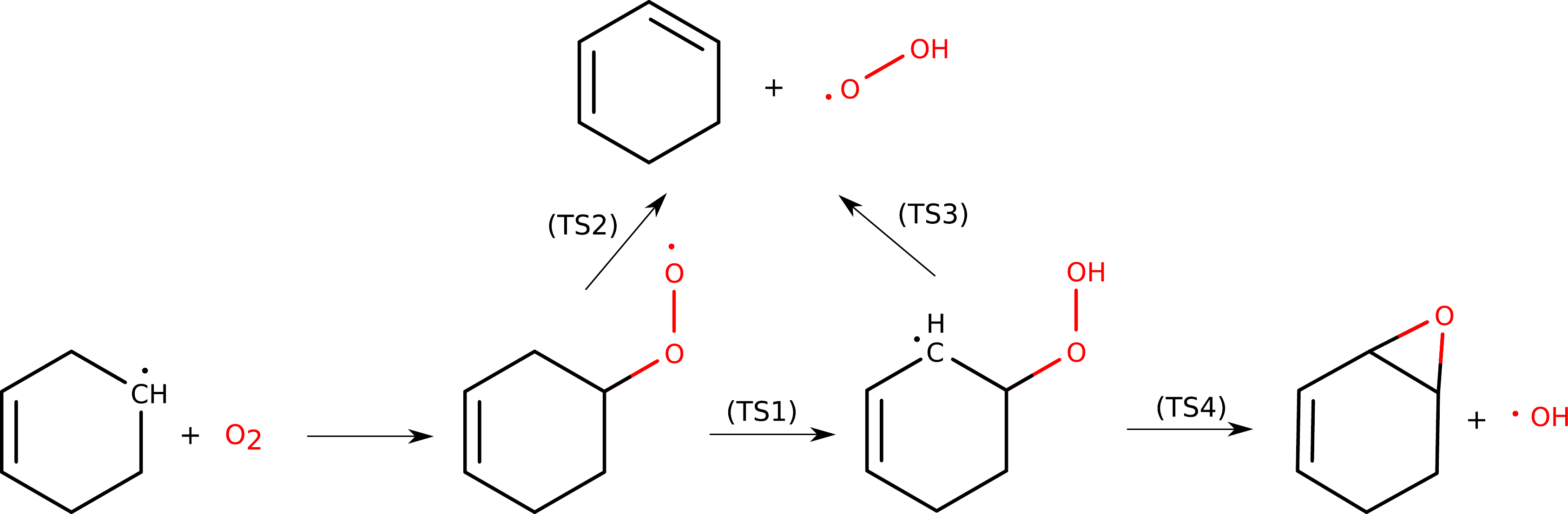}
           \caption{The reactive intermediates for $\beta$-cyclohexenyl radical combustion}
           \label{Cyclohexenyl Reaction Scheme}
    \end{figure}
    
    \begin{figure}[H]
           \centering
           \includegraphics[width=1.0\linewidth]{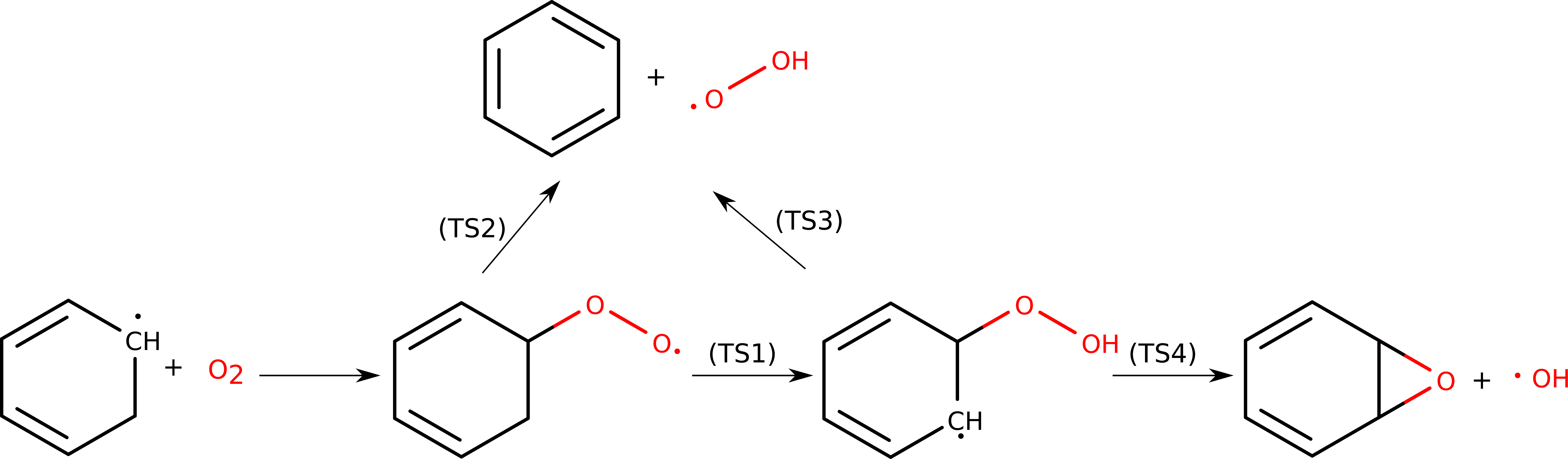}
           \caption{The reactive intermediates for cyclohexadienyl radical combustion}
           \label{Cyclohexadienyl Reaction Scheme}
    \end{figure}
\clearpage
\section{IRC Profiles for the different reactions in low-temperature combustion}
The IRC profiles are represented in the following order for all the hydrocarbons\\
a) \rxnone{} \\
b) \rxntwo{}\\
c) \rxnthree{} \\
d) \rxnfour{} \\
    \begin{figure}[H]
        \centering
        \includegraphics[width=0.7\linewidth]{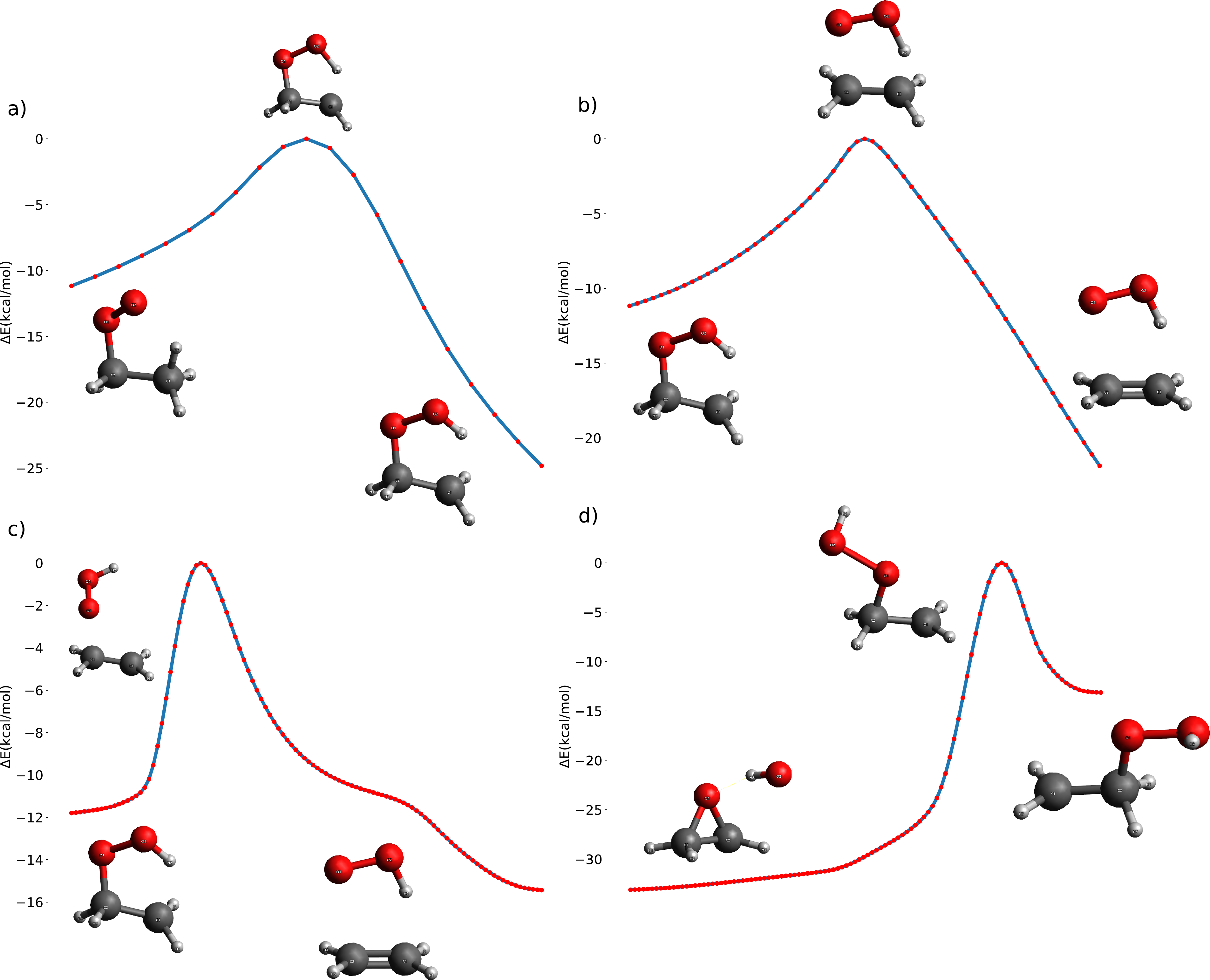}
        \caption{The IRC profile for the combustion of ethyl radical}
        \label{Ethyl IRC}
    \end{figure} 

    \begin{figure}[H]
        \centering
        \includegraphics[width=0.7\linewidth]{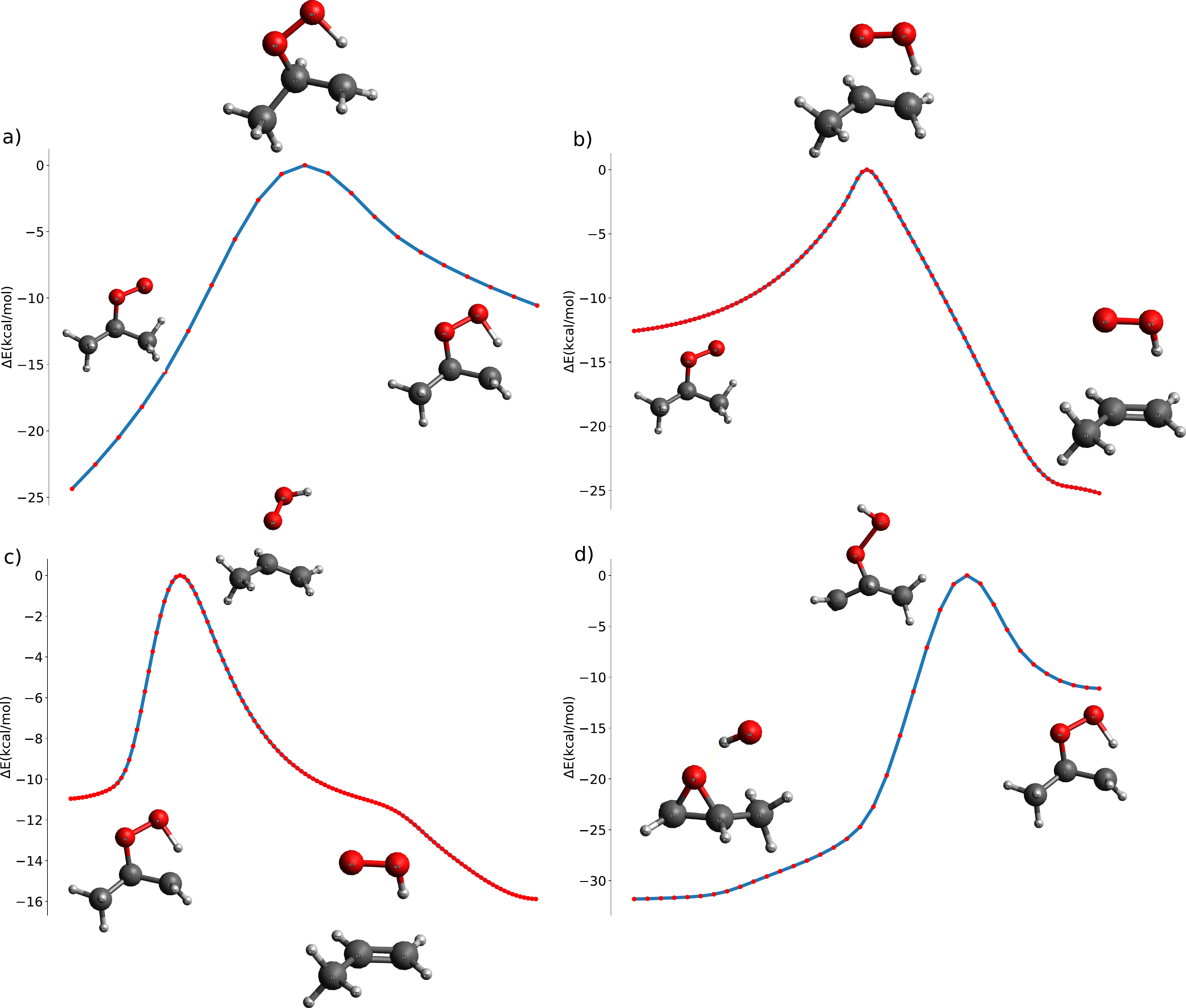}
        \caption{The IRC profile for the combustion of isopropyl radical}
        \label{Isopropyl IRC}
    \end{figure} 

    \begin{figure}[H]
        \centering
        \includegraphics[width=0.7\linewidth]{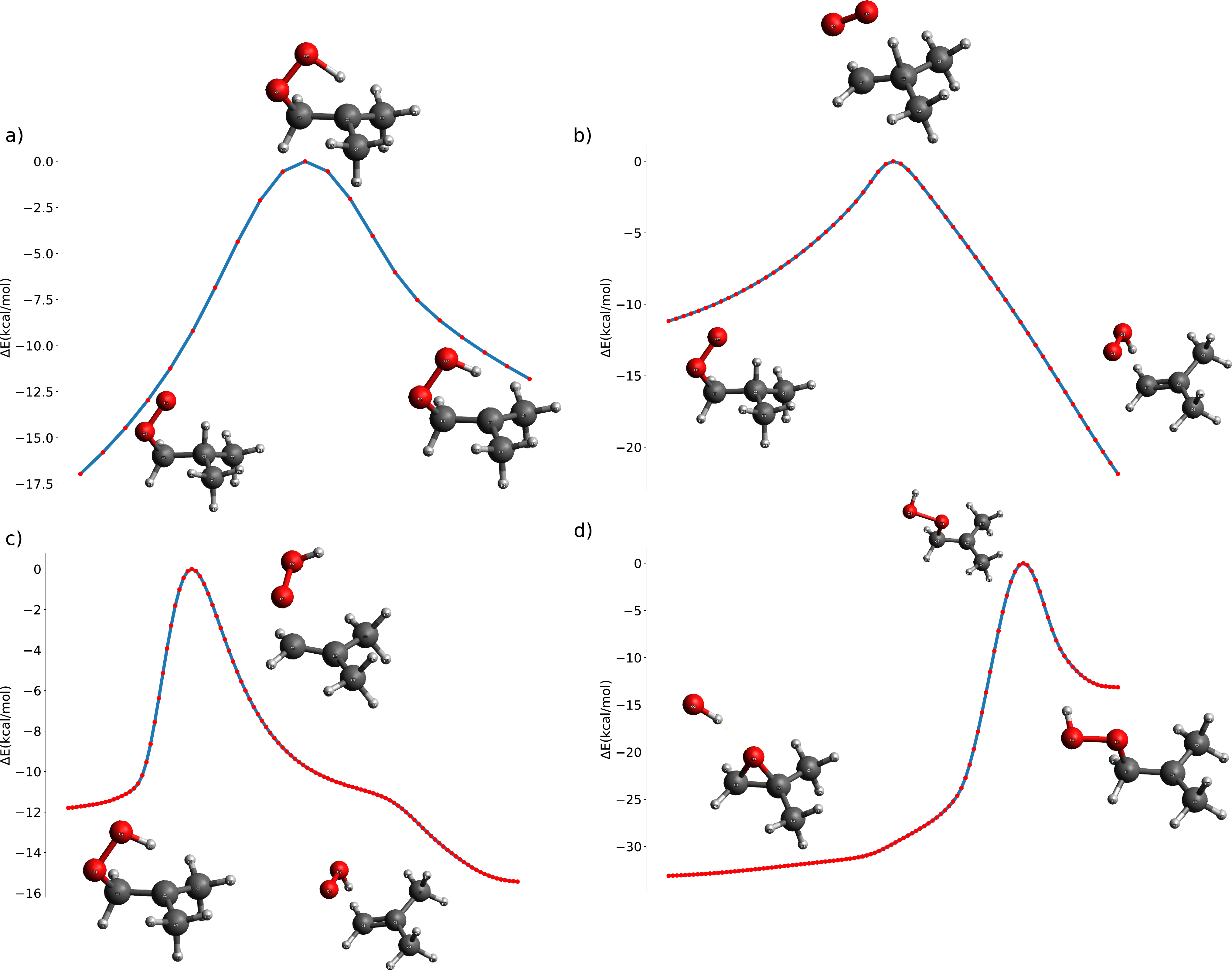}
         \caption{The IRC profile for the combustion of isobutyl radical}
        \label{Isobutyl IRC}
    \end{figure}

    \begin{figure}[H]
        \centering
        \includegraphics[width=0.7\linewidth]{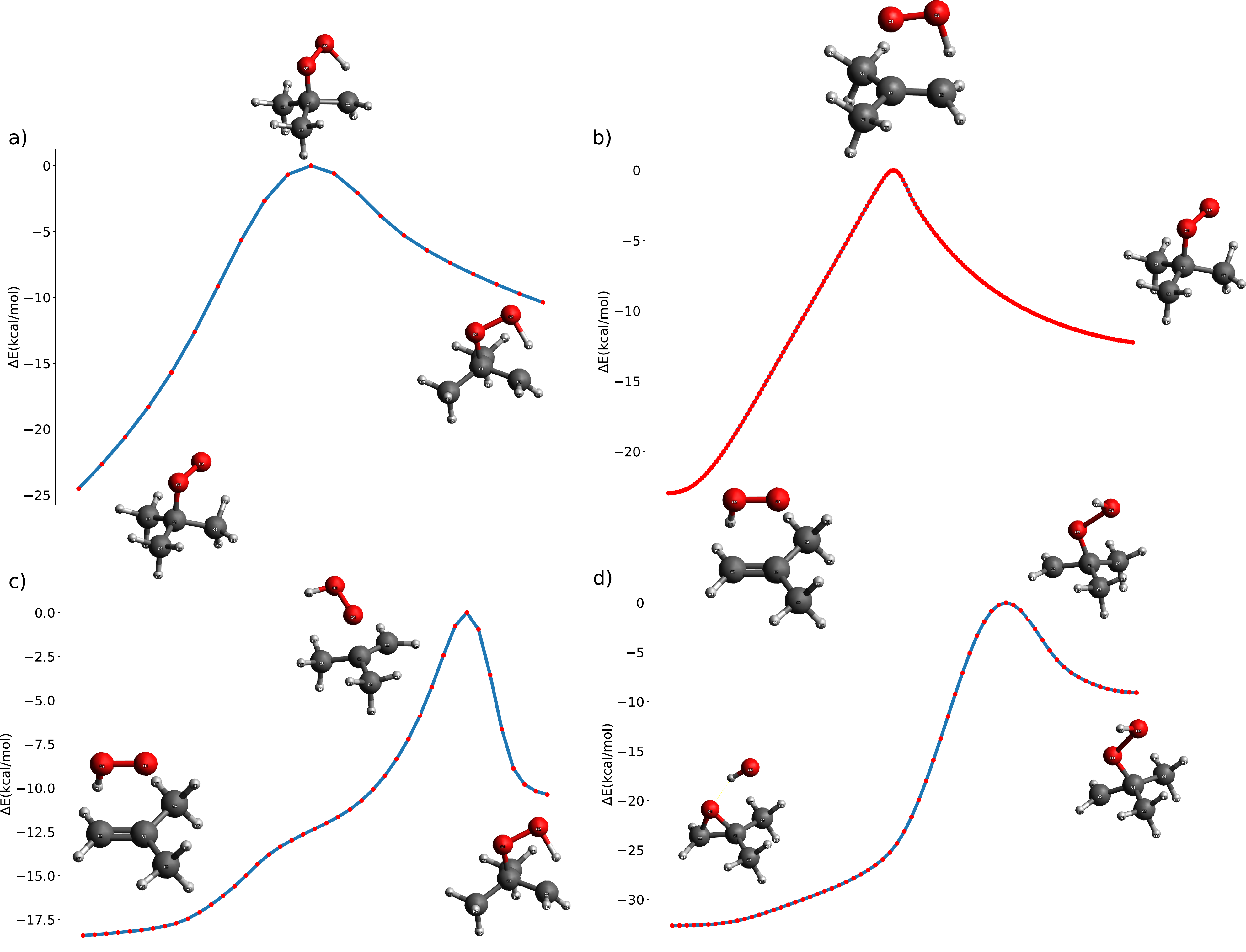}
         \caption{The IRC profile for the combustion of tertbutyl radical}
        \label{Tertbutyl IRC}
    \end{figure}

    \begin{figure}[H]
        \centering
        \includegraphics[width=0.7\linewidth]{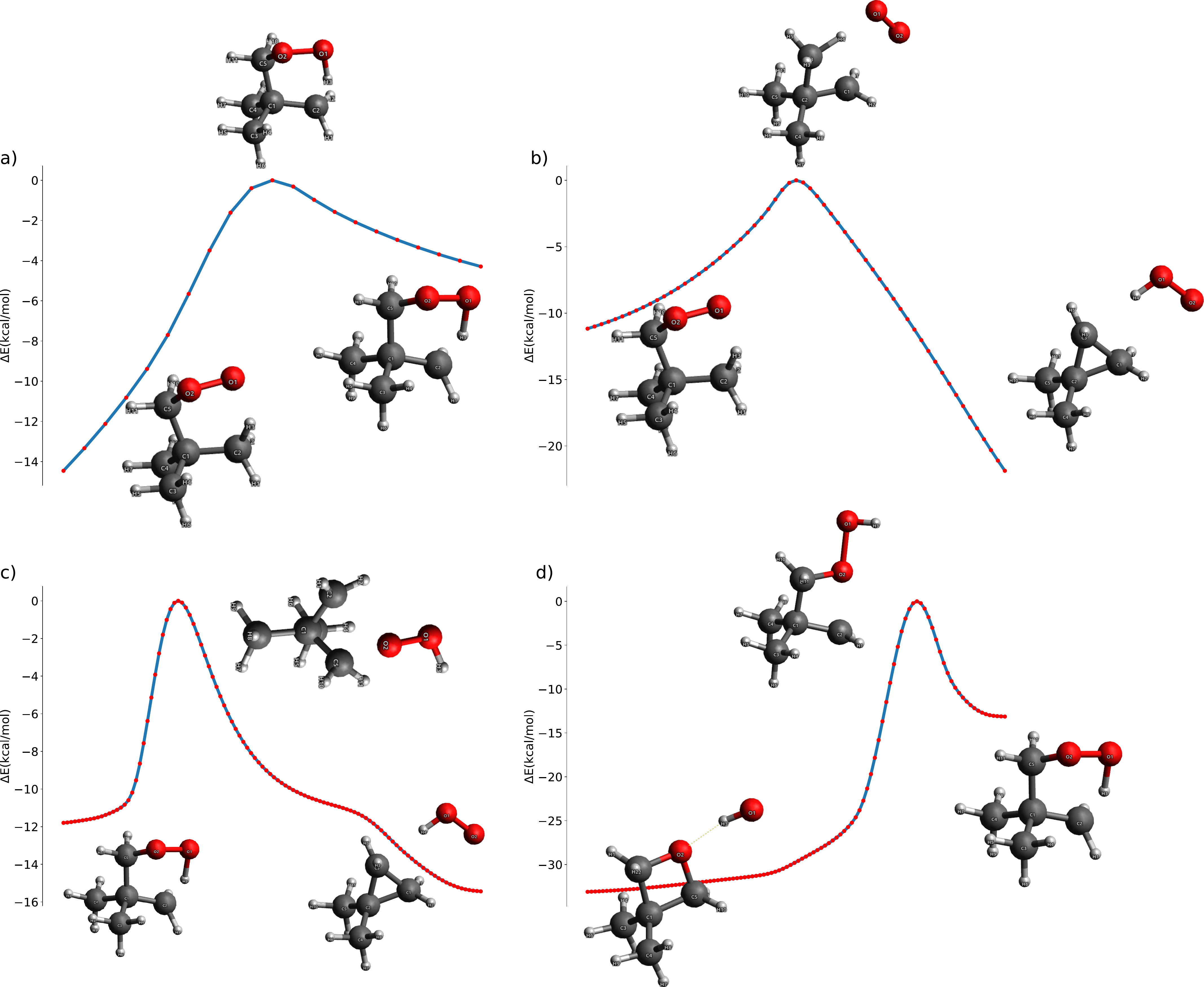}
         \caption{The IRC profile for the combustion of neopentyl radical}
        \label{Neopentyl IRC}
    \end{figure}

    \begin{figure}[H]
        \centering
        \includegraphics[width=0.7\linewidth]{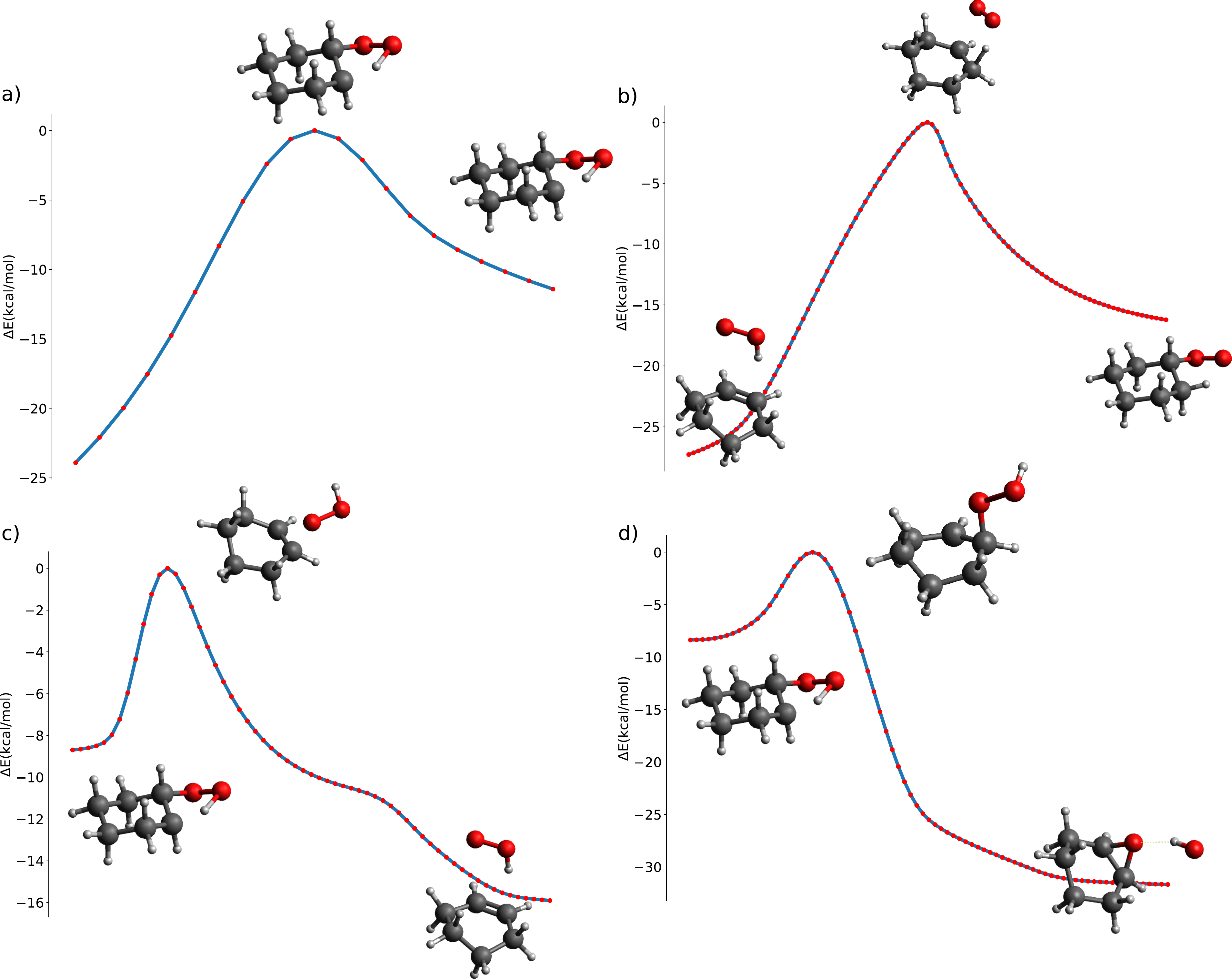}
         \caption{The IRC profile for the combustion of cyclohexyl radical}
        \label{Cyclohexyl IRC}
    \end{figure}

    \begin{figure}[H]
        \centering
        \includegraphics[width=0.7\linewidth]{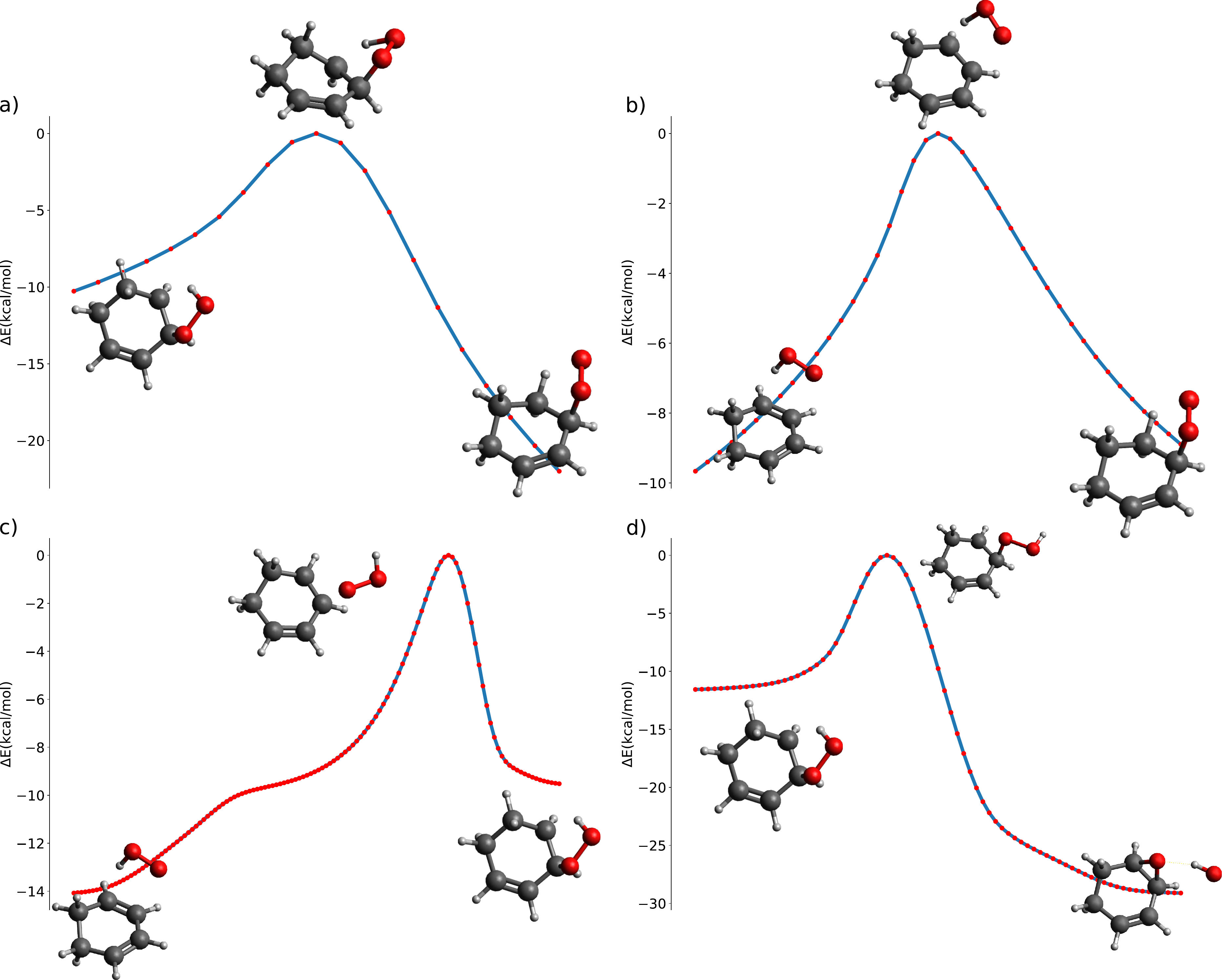}
         \caption{The IRC profile for the combustion of $\alpha$-cyclohexenyl radical}
        \label{Cyclohexenyl IRC}
    \end{figure}

   \begin{figure}[H]
        \centering
        \includegraphics[width=0.7\linewidth]{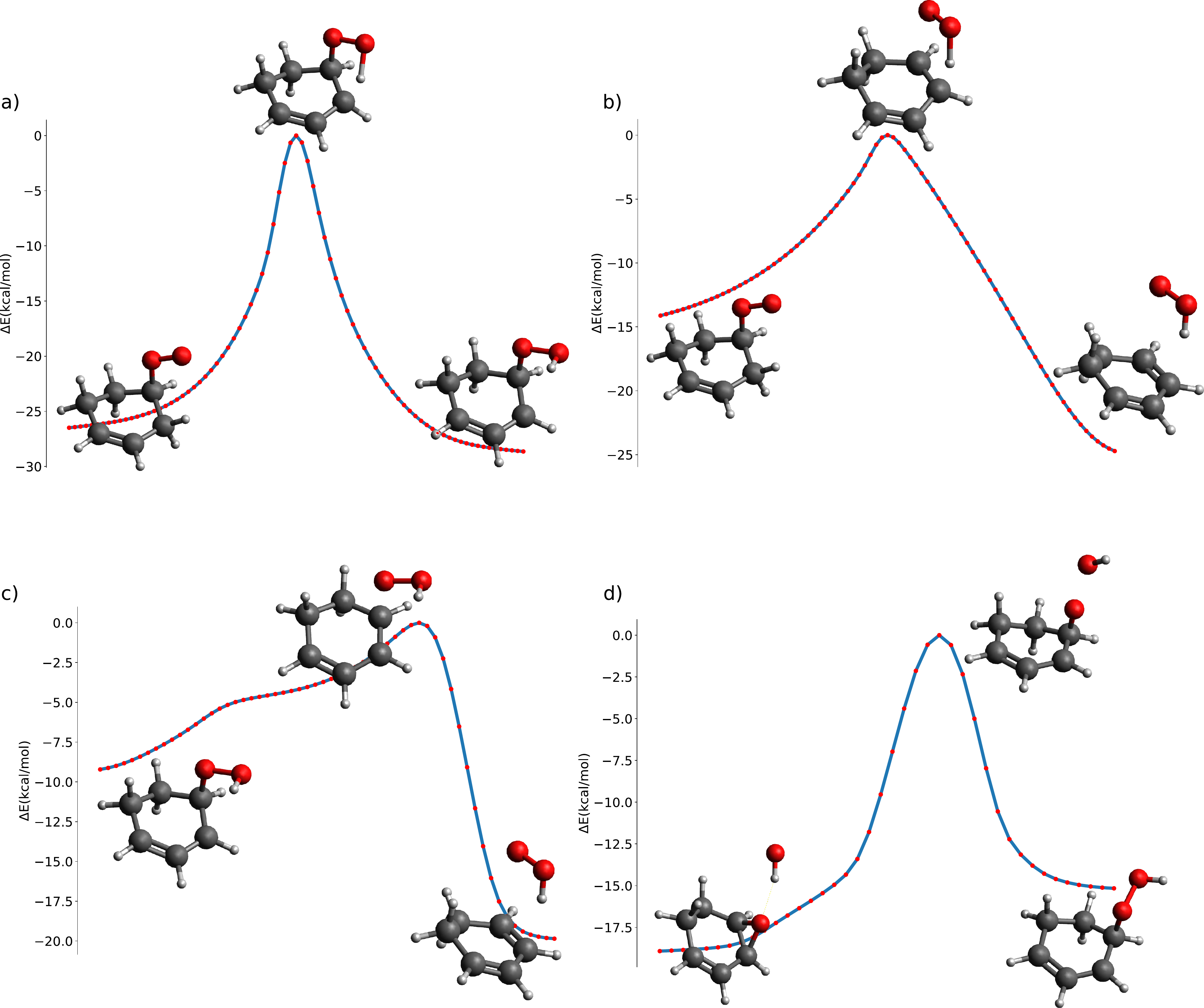}
         \caption{The IRC profile for the combustion of $\beta$-cyclohexenyl radical}
        \label{Cyclohexenyl IRC}
    \end{figure}

    \begin{figure}[H]
        \centering
        \includegraphics[width=0.7\linewidth]{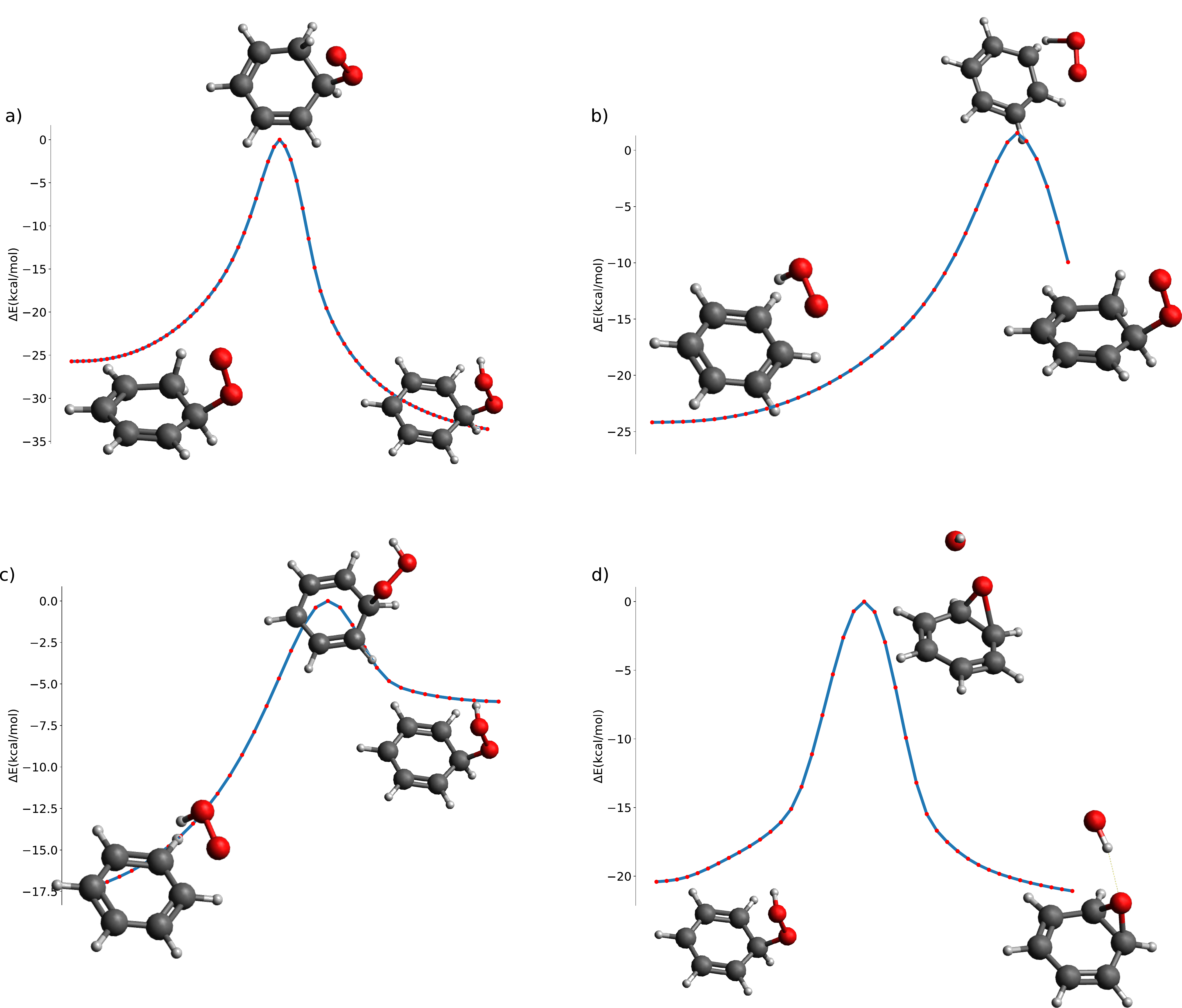}
         \caption{The IRC profile for the combustion of cyclohexadienyl radical}
        \label{Cyclohexadienyl IRC}
    \end{figure}

\section{W1-level equilibrium geometry, harmonic frequencies, and thermochemistry energies of different reaction species involved in combustion reaction of R$\cdot$}

\subsection{Hydrocarbon 1: ethyl}

    \textbf{Species 1: $\cdot$R}
\begin{Verbatim}
[ W1 data ]

== B3LYP/cc-pVTZ+d equilibrium geometry (Angstrom) ==
C     0.691636   -0.000000   -0.002049
C    -0.792081    0.000000   -0.017585
H     1.103759   -0.884476   -0.491892
H     1.091851    0.000004    1.022920
H     1.103759    0.884472   -0.491898
H    -1.348350   -0.924419    0.039336
H    -1.348349    0.924419    0.039336

== B3LYP/cc-pVTZ+d harmonic frequencies (cm^-1) ==
    104.8073    486.2223    813.8449    982.2743   1061.8581   1195.8772   1402.1395
   1467.3672   1481.5538   1482.2557   2943.0673   3034.7866   3077.3432   3142.6027
   3240.6016


== Thermochemistry energies (hartree) ==
  ZPVE =     0.058156  (after scaling by 0.985)
W1  Ee =  -79.168722  (total electronic energy)
W1  U0 =  -79.110566  (Zero-kelvin internal energy)
 
[ CCSDT-F12a/cc-pVDZ-F12 data ]
 
CCSD(T)-F12a/cc-pVDZ-F12  Ee =  -79.029415  (total electronic energy)
\end{Verbatim}
    \clearpage

    \textbf{Species 2: ROO$\cdot$}
\begin{Verbatim}
[ W1 data ]

== B3LYP/cc-pVTZ+d equilibrium geometry (Angstrom) ==
C     1.744457   -0.181342   -0.000005
C     0.414201    0.535217    0.000005
H     1.852261   -0.807843    0.884837
H     2.551771    0.551299   -0.000002
H     1.852256   -0.807830   -0.884857
O    -0.633958   -0.480438    0.000009
H     0.265409    1.148873    0.887848
H     0.265397    1.148879   -0.887833
O    -1.833422    0.060860   -0.000007

== B3LYP/cc-pVTZ+d harmonic frequencies (cm^-1) ==
     74.2528    223.5669    307.1919    502.3475    805.6244    841.9603   1015.2529
   1133.9237   1150.2037   1193.6478   1278.2306   1381.0735   1420.0700   1486.8332
   1496.2760   1508.8877   3043.2166   3055.1942   3096.0954   3109.4166   3120.8248



== Thermochemistry energies (hartree) ==
  ZPVE =     0.070111  (after scaling by 0.985)
W1  Ee =  -229.643821  (total electronic energy)
W1  U0 =  -229.573710  (Zero-kelvin internal energy)
 
[ CCSDT-F12a/cc-pVDZ-F12 data ]
 
CCSD(T)-F12a/cc-pVDZ-F12  Ee =  -229.268433  (total electronic energy)
\end{Verbatim}
    \clearpage

    \textbf{Species 3: $\cdot$QOOH}
\begin{Verbatim}
[ W1 data ]

== B3LYP/cc-pVTZ+d equilibrium geometry (Angstrom) ==
C    -1.852200   -0.166980    0.027775
C    -0.531369    0.493419    0.058741
H    -2.708700    0.334573   -0.394654
H    -1.980039   -1.139938    0.478434
O     0.457468   -0.513383   -0.173682
H    -0.450756    1.278182   -0.699368
H    -0.333842    0.959118    1.037207
O     1.752335    0.130903   -0.036404
H     2.096334   -0.330729    0.739984

== B3LYP/cc-pVTZ+d harmonic frequencies (cm^-1) ==
    110.0437    129.5381    203.3557    300.0234    469.0284    512.6005    819.4553
    933.8400    986.9561   1071.2743   1143.1600   1225.4651   1359.2805   1383.6467
   1454.7737   1496.5921   2921.5501   3022.7943   3156.2382   3264.6239   3758.8937



== Thermochemistry energies (hartree) ==
  ZPVE =     0.066698  (after scaling by 0.985)
W1  Ee =  -229.614038  (total electronic energy)
W1  U0 =  -229.547339  (Zero-kelvin internal energy)
 
[ CCSDT-F12a/cc-pVDZ-F12 data ]
 
CCSD(T)-F12a/cc-pVDZ-F12  Ee =  -229.238190  (total electronic energy)
\end{Verbatim}
    \clearpage

    \textbf{Species 4: {\it cy}-ether}
\begin{Verbatim}
[ W1 data ]

== B3LYP/cc-pVTZ+d equilibrium geometry (Angstrom) ==
C     0.731430   -0.371702    0.000000
C    -0.731430   -0.371702   -0.000000
O    -0.000000    0.853606   -0.000000
H     1.266004   -0.592106   -0.918210
H     1.266004   -0.592106    0.918209
H    -1.266003   -0.592106    0.918210
H    -1.266004   -0.592106   -0.918209

== B3LYP/cc-pVTZ+d harmonic frequencies (cm^-1) ==
    820.9816    846.5212    892.4287   1047.1821   1150.3912   1156.3546   1171.7311
   1177.3731   1301.7028   1507.0326   1541.1250   3078.8490   3085.5370   3157.5797
   3173.3447


== Thermochemistry energies (hartree) ==
  ZPVE =     0.056343  (after scaling by 0.985)
W1  Ee =  -153.855513  (total electronic energy)
W1  U0 =  -153.799170  (Zero-kelvin internal energy)
 
[ CCSDT-F12a/cc-pVDZ-F12 data ]
 
CCSD(T)-F12a/cc-pVDZ-F12  Ee =  -153.598131  (total electronic energy)
\end{Verbatim}
    \clearpage

    \textbf{Species 5: alkene}
\begin{Verbatim}
[ W1 data ]

== B3LYP/cc-pVTZ+d equilibrium geometry (Angstrom) ==
C     0.000000   -0.661966   -0.000000
C     0.000000    0.661966    0.000000
H    -0.920567   -1.231546    0.000000
H     0.920567   -1.231546   -0.000000
H    -0.920567    1.231546    0.000000
H     0.920567    1.231546   -0.000000

== B3LYP/cc-pVTZ+d harmonic frequencies (cm^-1) ==
    835.8298    978.9608    983.5714   1067.0956   1246.4512   1382.3409   1478.8937
   1693.3624   3126.1146   3139.9061   3195.1940   3223.6535


== Thermochemistry energies (hartree) ==
  ZPVE =     0.050156  (after scaling by 0.985)
W1  Ee =  -78.604857  (total electronic energy)
W1  U0 =  -78.554701  (Zero-kelvin internal energy)
 
[ CCSDT-F12a/cc-pVDZ-F12 data ]
 
CCSD(T)-F12a/cc-pVDZ-F12  Ee =  -78.465674  (total electronic energy)
\end{Verbatim}
    \clearpage

    \textbf{Species 6: Transition state along \rxnone{}}
\begin{Verbatim}
[ W1 data ]

== B3LYP/cc-pVTZ+d equilibrium geometry (Angstrom) ==
C    -1.222388   -0.566466   -0.119613
H    -1.941196   -0.966413    0.586382
H     0.022943   -1.137914    0.059018
H    -1.471076   -0.736313   -1.162951
C    -0.540324    0.752654    0.211403
H    -0.961280    1.623724   -0.293709
H    -0.528184    0.915931    1.291989
O     0.791214    0.598096   -0.290043
O     1.140669   -0.700114    0.161109

== B3LYP/cc-pVTZ+d harmonic frequencies (cm^-1) ==
  -2274.2560    274.9541    438.7388    554.4706    687.2618    868.9591    899.9893
    906.2836    962.5398   1047.0938   1110.8987   1169.2613   1244.6992   1346.9688
   1456.7294   1499.5115   1723.6265   3022.5170   3077.4755   3101.5997   3192.2292



== Thermochemistry energies (hartree) ==
  ZPVE =     0.064146  (after scaling by 0.985)
W1  Ee =  -229.578911  (total electronic energy)
W1  U0 =  -229.514765  (Zero-kelvin internal energy)
 
[ CCSDT-F12a/cc-pVDZ-F12 data ]
 
CCSD(T)-F12a/cc-pVDZ-F12  Ee =  -229.203690  (total electronic energy)
\end{Verbatim}
    \clearpage

    \textbf{Species 7: Transition state along \rxntwo{}}
\begin{Verbatim}
[ W1 data ]

== B3LYP/cc-pVTZ+d equilibrium geometry (Angstrom) ==
C    -1.285569   -0.616825    0.000027
H    -1.630771   -1.085342    0.915429
H     0.026458   -0.950626   -0.000075
H    -1.630877   -1.085201   -0.915406
C    -1.044096    0.743790    0.000118
H    -1.023461    1.316997   -0.914815
H    -1.023353    1.316856    0.915137
O     1.145733    0.615227   -0.000019
O     1.261766   -0.649536   -0.000123

== B3LYP/cc-pVTZ+d harmonic frequencies (cm^-1) ==
  -1089.2415    216.0393    354.2849    478.9129    516.9775    643.5291    832.9459
    891.5177   1016.0131   1040.6236   1233.2915   1299.5412   1315.1097   1334.4709
   1474.4510   1571.0968   1599.0782   3102.4777   3160.7059   3177.7927   3246.4129



== Thermochemistry energies (hartree) ==
  ZPVE =     0.063966  (after scaling by 0.985)
W1  Ee =  -229.587475  (total electronic energy)
W1  U0 =  -229.523509  (Zero-kelvin internal energy)
 
[ CCSDT-F12a/cc-pVDZ-F12 data ]
 
CCSD(T)-F12a/cc-pVDZ-F12  Ee =  -229.211904  (total electronic energy)
\end{Verbatim}
    \clearpage

    \textbf{Species 8: Transition state along \rxnthree{}}
\begin{Verbatim}
[ W1 data ]

== B3LYP/cc-pVTZ+d equilibrium geometry (Angstrom) ==
C    -1.625492    0.535263   -0.115726
H    -2.181955    0.478982   -1.040579
H     1.676417    1.144824   -0.288345
H    -1.556027    1.502362    0.362906
C    -0.927126   -0.548362    0.362641
H    -0.522174   -0.530436    1.363983
H    -1.134012   -1.532305   -0.034874
O     0.760953   -0.468441   -0.484652
O     1.618229    0.345337    0.254079

== B3LYP/cc-pVTZ+d harmonic frequencies (cm^-1) ==
   -533.1097     91.2057    227.0998    306.5308    415.9603    470.6679    817.9757
    835.6601    918.9880   1030.2604   1084.6426   1244.5559   1263.4251   1382.3208
   1470.8280   1549.3441   3140.8926   3144.8996   3216.0621   3241.8410   3717.4656



== Thermochemistry energies (hartree) ==
  ZPVE =     0.066356  (after scaling by 0.985)
W1  Ee =  -229.586871  (total electronic energy)
W1  U0 =  -229.520514  (Zero-kelvin internal energy)
 
[ CCSDT-F12a/cc-pVDZ-F12 data ]
 
CCSD(T)-F12a/cc-pVDZ-F12  Ee =  -229.211256  (total electronic energy)
\end{Verbatim}
    \clearpage

   \textbf{Species 9: Transition state along \rxnfour{}}
\begin{Verbatim}
[ W1 data ]

== B3LYP/cc-pVTZ+d equilibrium geometry (Angstrom) ==
C     1.694417   -0.258022    0.043055
H     2.126977   -0.605577   -0.881634
H    -2.125997   -0.710448    0.532811
H     1.991309   -0.727872    0.967095
C     0.582893    0.704239    0.030160
H     0.404487    1.231206    0.968323
H     0.540807    1.384108   -0.822162
O    -0.200115   -0.444105   -0.124266
O    -1.875066    0.038015   -0.026199

== B3LYP/cc-pVTZ+d harmonic frequencies (cm^-1) ==
   -753.5035     90.8140    134.8719    258.8011    400.1051    505.8126    781.5253
    838.0150    937.2796    997.2226   1166.4545   1170.4344   1203.4706   1317.1301
   1471.1226   1535.9148   3031.5855   3084.1669   3167.4547   3281.3366   3777.9256



== Thermochemistry energies (hartree) ==
  ZPVE =     0.065416  (after scaling by 0.985)
W1  Ee =  -229.589061  (total electronic energy)
W1  U0 =  -229.523646  (Zero-kelvin internal energy)
 
[ CCSDT-F12a/cc-pVDZ-F12 data ]
 
CCSD(T)-F12a/cc-pVDZ-F12  Ee =  -229.213812  (total electronic energy)
\end{Verbatim}
    \clearpage

\subsection{Hydrocarbon 2: isopropyl}

    \textbf{Species 1: $\cdot$R}
\begin{Verbatim}
[ W1 data ]

== B3LYP/cc-pVTZ+d equilibrium geometry (Angstrom) ==
C     1.294138   -0.197180    0.002337
C    -0.000000    0.532887   -0.041105
C    -1.294138   -0.197180    0.002337
H     2.139669    0.453710   -0.221196
H     1.309964   -1.030634   -0.708112
H     1.483889   -0.641834    0.991608
H     0.000000    1.606355    0.093979
H    -1.309964   -1.030634   -0.708112
H    -2.139669    0.453710   -0.221196
H    -1.483889   -0.641835    0.991608

== B3LYP/cc-pVTZ+d harmonic frequencies (cm^-1) ==
    108.0142    110.0636    351.0519    390.5172    884.7674    941.2064    948.1429
   1029.2895   1146.0139   1181.1098   1371.8999   1409.4355   1414.2704   1468.3527
   1477.4209   1478.6875   1491.0577   2933.1961   2937.2405   3007.7790   3009.0800
   3082.4644   3083.5626   3167.9761


== Thermochemistry energies (hartree) ==
  ZPVE =     0.086220  (after scaling by 0.985)
W1  Ee =  -118.495776  (total electronic energy)
W1  U0 =  -118.409556  (Zero-kelvin internal energy)
 
[ CCSDT-F12a/cc-pVDZ-F12 data ]
 
CCSD(T)-F12a/cc-pVDZ-F12  Ee =  -118.286801  (total electronic energy)
\end{Verbatim}
    \clearpage

    \textbf{Species 2: ROO$\cdot$}
\begin{Verbatim}
[ W1 data ]

== B3LYP/cc-pVTZ+d equilibrium geometry (Angstrom) ==
C     1.557589   -0.849100    0.019749
C     0.367826    0.018772   -0.337445
C     0.486014    1.463512    0.108219
H     1.401385   -1.878604   -0.300108
H     1.732552   -0.841900    1.096080
H     2.451715   -0.467755   -0.473428
H     0.689245    1.522585    1.178287
H    -0.435814    2.000834   -0.107015
H     1.302902    1.950956   -0.425191
O    -0.787241   -0.604232    0.337199
O    -1.931838   -0.151722   -0.125850
H     0.142070   -0.037586   -1.402548

== B3LYP/cc-pVTZ+d harmonic frequencies (cm^-1) ==
     99.0371    199.7081    237.6007    302.5951    343.0845    449.4491    521.4669
    791.9765    886.3322    939.9622    950.6077   1123.5137   1153.2153   1195.9904
   1204.7216   1344.1207   1368.1951   1403.3738   1421.4108   1482.8878   1488.9614
   1494.3705   1511.5021   3036.2506   3040.9106   3062.2310   3099.6125   3109.6361
   3112.3951   3120.3661


== Thermochemistry energies (hartree) ==
  ZPVE =     0.097604  (after scaling by 0.985)
W1  Ee =  -268.972923  (total electronic energy)
W1  U0 =  -268.875319  (Zero-kelvin internal energy)
 
[ CCSDT-F12a/cc-pVDZ-F12 data ]
 
CCSD(T)-F12a/cc-pVDZ-F12  Ee =  -268.527989  (total electronic energy)
\end{Verbatim}
    \clearpage

    \textbf{Species 3: $\cdot$QOOH}
\begin{Verbatim}
[ W1 data ]

== B3LYP/cc-pVTZ+d equilibrium geometry (Angstrom) ==
C     1.690585   -0.689674    0.023809
C     0.415876    0.074284   -0.336531
C     0.439022    1.491088    0.110637
H     1.608619   -1.728445   -0.295243
H     1.860184   -0.664931    1.100346
H     2.547635   -0.234432   -0.470990
H     0.529234    1.713122    1.166125
H     0.388342    2.311646   -0.589110
O    -0.628624   -0.676243    0.322328
O    -1.905372   -0.206703   -0.174531
H     0.238717    0.015634   -1.412699
H    -2.173658    0.396793    0.531707

== B3LYP/cc-pVTZ+d harmonic frequencies (cm^-1) ==
     83.4052    136.1081    217.1285    232.7340    279.5208    336.2052    456.4842
    482.5275    583.7069    829.8867    906.5006    926.8397    947.8389   1035.7493
   1154.3802   1175.6436   1319.9764   1366.9110   1378.9031   1398.3131   1458.5600
   1484.4061   1496.7041   3037.0159   3038.9094   3108.8988   3112.6193   3135.6699
   3246.0430   3747.7208


== Thermochemistry energies (hartree) ==
  ZPVE =     0.094507  (after scaling by 0.985)
W1  Ee =  -268.942482  (total electronic energy)
W1  U0 =  -268.847975  (Zero-kelvin internal energy)
 
[ CCSDT-F12a/cc-pVDZ-F12 data ]
 
CCSD(T)-F12a/cc-pVDZ-F12  Ee =  -268.497205  (total electronic energy)
\end{Verbatim}
    \clearpage

    \textbf{Species 4: {\it cy}-ether}
\begin{Verbatim}
[ W1 data ]

== B3LYP/cc-pVTZ+d equilibrium geometry (Angstrom) ==
C    -1.506883    0.098067   -0.148522
C    -0.152156   -0.036390    0.486086
C     1.039267    0.615939   -0.060866
H    -2.078321   -0.824652   -0.035345
H    -1.413569    0.313775   -1.212601
H    -2.073128    0.903919    0.322991
O     0.827897   -0.788134   -0.239781
H    -0.154185   -0.249774    1.552381
H     1.863894    0.881375    0.592649
H     0.950765    1.214734   -0.962017

== B3LYP/cc-pVTZ+d harmonic frequencies (cm^-1) ==
    206.0392    366.6399    410.5318    771.8821    844.8421    907.5872    974.0105
   1042.0992   1130.0974   1156.6732   1165.3168   1188.0691   1294.1605   1407.0987
   1439.9307   1483.5220   1497.8665   1531.5306   3028.4132   3078.6941   3085.5630
   3087.1776   3109.8463   3161.6612


== Thermochemistry energies (hartree) ==
  ZPVE =     0.083856  (after scaling by 0.985)
W1  Ee =  -193.185543  (total electronic energy)
W1  U0 =  -193.101687  (Zero-kelvin internal energy)
 
[ CCSDT-F12a/cc-pVDZ-F12 data ]
 
CCSD(T)-F12a/cc-pVDZ-F12  Ee =  -192.858655  (total electronic energy)
\end{Verbatim}
    \clearpage

    \textbf{Species 5: alkene}
\begin{Verbatim}
[ W1 data ]

== B3LYP/cc-pVTZ+d equilibrium geometry (Angstrom) ==
C     1.230760   -0.161964    0.000017
C    -0.134285    0.452265   -0.000173
C    -1.277679   -0.219887   -0.000007
H     1.802769    0.153368    0.876702
H     1.803971    0.153838   -0.875707
H     1.179896   -1.250881   -0.000295
H    -0.166820    1.538156    0.000026
H    -1.299008   -1.303499   -0.000161
H    -2.233587    0.286535    0.000411

== B3LYP/cc-pVTZ+d harmonic frequencies (cm^-1) ==
    205.3893    426.3125    592.5848    924.6092    947.7973    950.8368   1030.2464
   1074.7583   1193.2901   1332.4547   1411.1422   1453.1586   1480.3100   1495.1727
   1713.2537   3014.3956   3055.4048   3091.0867   3122.4458   3129.0737   3209.1532



== Thermochemistry energies (hartree) ==
  ZPVE =     0.078210  (after scaling by 0.985)
W1  Ee =  -117.931883  (total electronic energy)
W1  U0 =  -117.853673  (Zero-kelvin internal energy)
 
[ CCSDT-F12a/cc-pVDZ-F12 data ]
 
CCSD(T)-F12a/cc-pVDZ-F12  Ee =  -117.723135  (total electronic energy)
\end{Verbatim}
    \clearpage

    \textbf{Species 6: Transition state along \rxnone{}}
\begin{Verbatim}
[ W1 data ]

== B3LYP/cc-pVTZ+d equilibrium geometry (Angstrom) ==
C     0.107902    1.380235   -0.117394
H     0.044709    2.231466    0.551560
H    -0.078164    1.613755   -1.162745
H     1.361734    0.815538   -0.082115
C    -0.417122    0.033843    0.373739
H    -0.339473   -0.000158    1.465112
C    -1.811936   -0.342302   -0.084470
H    -2.540596    0.358274    0.324370
H    -1.878365   -0.321418   -1.171945
H    -2.070937   -1.342749    0.263113
O     1.770664   -0.308768    0.036747
O     0.507839   -0.914402   -0.189072

== B3LYP/cc-pVTZ+d harmonic frequencies (cm^-1) ==
  -2229.5380    193.6070    206.6409    316.4591    401.9669    482.9566    570.3859
    653.8259    840.7772    873.2173    922.5457    956.2967    972.8956   1092.7208
   1110.1643   1139.2020   1180.3636   1334.1392   1377.9033   1409.8930   1457.0231
   1484.4302   1496.3130   1718.6786   3009.6166   3037.1843   3090.8801   3103.0547
   3110.6763   3184.3479


== Thermochemistry energies (hartree) ==
  ZPVE =     0.091394  (after scaling by 0.985)
W1  Ee =  -268.909011  (total electronic energy)
W1  U0 =  -268.817617  (Zero-kelvin internal energy)
 
[ CCSDT-F12a/cc-pVDZ-F12 data ]
 
CCSD(T)-F12a/cc-pVDZ-F12  Ee =  -268.464270  (total electronic energy)
\end{Verbatim}
    \clearpage

    \textbf{Species 7: Transition state along \rxntwo{}}
\begin{Verbatim}
[ W1 data ]

== B3LYP/cc-pVTZ+d equilibrium geometry (Angstrom) ==
C    -0.318352    1.405754   -0.045578
H    -0.077361    2.213988    0.636823
H    -0.660059    1.718665   -1.027834
H     0.910852    0.914651   -0.306183
C    -0.820927    0.225485    0.478778
H    -0.755081    0.060857    1.546583
C    -1.666571   -0.735539   -0.283845
H    -1.534288   -0.620544   -1.359456
H    -1.450086   -1.767406   -0.009176
H    -2.722395   -0.552349   -0.056129
O     1.794402   -0.002177   -0.295321
O     1.096038   -0.915580    0.255227

== B3LYP/cc-pVTZ+d harmonic frequencies (cm^-1) ==
  -1078.3680    149.4909    175.6899    207.7079    365.7842    420.3225    508.3866
    624.2410    636.7650    894.9108    956.1438    962.3985   1035.8341   1062.6771
   1200.0591   1287.2995   1308.5527   1317.7436   1406.7636   1442.8220   1476.7399
   1488.8662   1571.5485   1599.6502   3008.8532   3072.6314   3093.7832   3108.9174
   3161.6502   3180.5679


== Thermochemistry energies (hartree) ==
  ZPVE =     0.091390  (after scaling by 0.985)
W1  Ee =  -268.917052  (total electronic energy)
W1  U0 =  -268.825662  (Zero-kelvin internal energy)
 
[ CCSDT-F12a/cc-pVDZ-F12 data ]
 
CCSD(T)-F12a/cc-pVDZ-F12  Ee =  -268.471990  (total electronic energy)
\end{Verbatim}
    \clearpage

    \textbf{Species 8: Transition state along \rxnthree{}}
\begin{Verbatim}
[ W1 data ]

== B3LYP/cc-pVTZ+d equilibrium geometry (Angstrom) ==
C    -0.717212    1.454957   -0.119912
H    -0.127007    2.256686    0.300522
H    -1.260894    1.658666   -1.032845
H     2.434326    0.474479   -0.368593
C    -0.671273    0.181161    0.408776
H    -0.244311    0.065758    1.396922
C    -1.649450   -0.877960   -0.004827
H    -1.939309   -0.761269   -1.049111
H    -1.223661   -1.870435    0.131882
H    -2.552731   -0.809223    0.605864
O     2.023440   -0.157469    0.238085
O     0.869209   -0.537982   -0.449193

== B3LYP/cc-pVTZ+d harmonic frequencies (cm^-1) ==
   -502.0347     77.2151    144.3460    199.9043    296.8304    319.6959    417.2000
    440.8612    520.7855    810.8888    895.4051    942.9991    956.5096   1024.0479
   1082.1822   1191.7425   1268.2948   1380.4795   1401.4365   1434.1805   1479.7537
   1495.4993   1535.6085   3030.7300   3087.3208   3114.3451   3138.0173   3159.0095
   3231.4749   3721.2202


== Thermochemistry energies (hartree) ==
  ZPVE =     0.093795  (after scaling by 0.985)
W1  Ee =  -268.915747  (total electronic energy)
W1  U0 =  -268.821952  (Zero-kelvin internal energy)
 
[ CCSDT-F12a/cc-pVDZ-F12 data ]
 
CCSD(T)-F12a/cc-pVDZ-F12  Ee =  -268.470670  (total electronic energy)
\end{Verbatim}
    \clearpage

   \textbf{Species 9: Transition state along \rxnfour{}}
\begin{Verbatim}
[ W1 data ]

== B3LYP/cc-pVTZ+d equilibrium geometry (Angstrom) ==
C     1.328750   -1.111615   -0.010989
H     1.337250   -2.018033    0.573212
H     1.817421   -1.104394   -0.973549
C     0.509882    0.039305    0.410748
H     0.292989    0.046625    1.481818
C     0.866820    1.414864   -0.103616
H     1.101080    1.377743   -1.167827
H     0.023219    2.088105    0.041361
H     1.727759    1.813636    0.435997
O    -0.499496   -0.620103   -0.314111
O    -2.007577    0.169848    0.018478
H    -2.475845   -0.656969    0.197195

== B3LYP/cc-pVTZ+d harmonic frequencies (cm^-1) ==
   -740.6327    105.6354    150.0838    203.3205    251.6169    350.5830    415.1727
    428.6591    538.0679    769.1875    845.9072    907.3394    927.5228    986.1707
   1091.8859   1166.2383   1198.0876   1288.0605   1396.4421   1422.0378   1466.9220
   1487.4752   1498.2347   3029.8699   3033.9197   3099.3422   3116.4493   3160.2450
   3275.2054   3782.5489


== Thermochemistry energies (hartree) ==
  ZPVE =     0.092884  (after scaling by 0.985)
W1  Ee =  -268.920648  (total electronic energy)
W1  U0 =  -268.827765  (Zero-kelvin internal energy)
 
[ CCSDT-F12a/cc-pVDZ-F12 data ]
 
CCSD(T)-F12a/cc-pVDZ-F12  Ee =  -268.475958  (total electronic energy)
\end{Verbatim}
    \clearpage

\subsection{Hydrocarbon 3: isobutyl}

    \textbf{Species 1: $\cdot$R}
\begin{Verbatim}
[ W1 data ]

== B3LYP/cc-pVTZ+d equilibrium geometry (Angstrom) ==
C    -1.268755   -0.679206    0.101502
C    -0.000003    0.059715   -0.346127
C    -0.000020    1.480077    0.101187
C     1.268776   -0.679160    0.101437
H    -2.166973   -0.166159   -0.245760
H    -1.316887   -0.738399    1.190674
H    -1.289134   -1.696251   -0.293322
H    -0.925833    2.030515    0.198930
H     0.925807    2.030322    0.199678
H    -0.000067    0.052067   -1.449569
H     1.317008   -0.738263    1.190615
H     2.166963   -0.166147   -0.245950
H     1.289122   -1.696240   -0.293291

== B3LYP/cc-pVTZ+d harmonic frequencies (cm^-1) ==
    122.6150    228.7966    253.5757    360.0523    373.7450    411.5581    542.9511
    810.2077    905.9566    947.5623    967.1558    982.2091   1091.6563   1180.4174
   1205.7344   1318.8453   1328.2073   1399.9761   1415.1436   1467.2682   1489.8184
   1493.3116   1503.5523   1511.3797   2893.7842   3019.5467   3023.4866   3080.7408
   3084.6451   3085.0696   3087.5311   3131.4525   3230.3489


== Thermochemistry energies (hartree) ==
  ZPVE =     0.114328  (after scaling by 0.985)
W1  Ee =  -157.815778  (total electronic energy)
W1  U0 =  -157.701450  (Zero-kelvin internal energy)
 
[ CCSDT-F12a/cc-pVDZ-F12 data ]
 
CCSD(T)-F12a/cc-pVDZ-F12  Ee =  -157.537629  (total electronic energy)
\end{Verbatim}
    \clearpage

    \textbf{Species 2: ROO$\cdot$}
\begin{Verbatim}
[ W1 data ]

== B3LYP/cc-pVTZ+d equilibrium geometry (Angstrom) ==
C     1.984858   -0.899222   -0.165417
C     0.742102   -0.013183   -0.287945
C    -0.357775   -0.558890    0.614675
C     1.056898    1.451677    0.021474
H     1.768017   -1.936283   -0.426410
H     2.382673   -0.884290    0.852243
H     2.773082   -0.545688   -0.830684
H     0.365338   -0.084380   -1.311772
H     1.451260    1.560560    1.034976
H     0.169099    2.077616   -0.062996
H     1.807262    1.837690   -0.669400
O    -1.609731    0.169672    0.452971
O    -2.213630   -0.153516   -0.673092
H    -0.116441   -0.429904    1.671111
H    -0.569894   -1.606855    0.407174

== B3LYP/cc-pVTZ+d harmonic frequencies (cm^-1) ==
     76.3854    109.0916    219.3306    234.7808    250.5676    332.9370    415.6778
    418.0864    564.9671    817.4196    873.4656    937.6780    946.8789    956.3792
    975.5436   1139.3979   1152.2855   1197.0320   1201.2073   1290.4553   1327.6167
   1376.5753   1384.3715   1407.0402   1427.6585   1476.2450   1491.5985   1496.6487
   1507.2019   1512.7898   3020.2410   3021.6721   3030.2480   3050.1113   3078.3494
   3084.5650   3093.0823   3103.3409   3108.1994


== Thermochemistry energies (hartree) ==
  ZPVE =     0.125904  (after scaling by 0.985)
W1  Ee =  -308.291652  (total electronic energy)
W1  U0 =  -308.165749  (Zero-kelvin internal energy)
 
[ CCSDT-F12a/cc-pVDZ-F12 data ]
 
CCSD(T)-F12a/cc-pVDZ-F12  Ee =  -307.777334  (total electronic energy)
\end{Verbatim}
    \clearpage

    \textbf{Species 3: $\cdot$QOOH}
\begin{Verbatim}
[ W1 data ]

== B3LYP/cc-pVTZ+d equilibrium geometry (Angstrom) ==
C     1.747753   -1.148068   -0.170438
C     0.828672   -0.004333    0.084194
C    -0.384257   -0.177993    0.931730
C     1.258994    1.375988   -0.282916
H     1.242046   -2.107979   -0.061594
H     2.603225   -1.152270    0.522119
H     2.178246   -1.100358   -1.175375
H     2.047683    1.744907    0.390221
H     0.431620    2.083978   -0.230015
H     1.680466    1.414577   -1.291300
O    -1.580275    0.388399    0.371828
O    -1.970741   -0.435894   -0.756588
H    -0.307879    0.380120    1.875954
H    -0.563479   -1.228642    1.169149
H    -1.610767    0.072063   -1.496499

== B3LYP/cc-pVTZ+d harmonic frequencies (cm^-1) ==
     33.4358     70.5341    104.9496    129.9819    204.1928    235.8002    310.0464
    367.7000    400.7589    551.2879    774.8480    858.6426    939.9628    942.3565
    972.3378   1000.4202   1005.5271   1051.9266   1239.0216   1293.7248   1307.7913
   1360.1532   1376.1121   1400.4718   1416.3914   1460.1408   1469.8108   1475.6848
   1485.6146   1496.4919   2940.1077   2944.7109   2963.5689   3017.8420   3027.1571
   3053.2430   3084.9746   3089.4938   3746.8587


== Thermochemistry energies (hartree) ==
  ZPVE =     0.122531  (after scaling by 0.985)
W1  Ee =  -308.270527  (total electronic energy)
W1  U0 =  -308.147996  (Zero-kelvin internal energy)
 
[ CCSDT-F12a/cc-pVDZ-F12 data ]
 
CCSD(T)-F12a/cc-pVDZ-F12  Ee =  -307.755469  (total electronic energy)
\end{Verbatim}
    \clearpage

    \textbf{Species 4: {\it cy}-ether}
\begin{Verbatim}
[ W1 data ]

== B3LYP/cc-pVTZ+d equilibrium geometry (Angstrom) ==
C    -0.930698   -1.279992   -0.049036
C    -0.131144    0.000000   -0.054913
C     1.237579   -0.000000   -0.582716
C    -0.930698    1.279992   -0.049036
H    -0.272060   -2.146381   -0.014050
H    -1.555426   -1.351676   -0.941587
H    -1.587027   -1.314167    0.822839
H    -1.555426    1.351676   -0.941587
H    -0.272060    2.146381   -0.014050
H    -1.587027    1.314167    0.822839
O     1.004512   -0.000000    0.830447
H     1.661345    0.916464   -0.981890
H     1.661345   -0.916464   -0.981890

== B3LYP/cc-pVTZ+d harmonic frequencies (cm^-1) ==
    182.2996    218.0260    356.0528    360.6338    410.6438    412.8239    700.0132
    807.9259    908.0013    923.8222    959.9394   1020.4768   1074.6552   1131.2834
   1150.3799   1172.9890   1284.2192   1391.2651   1413.0982   1423.4775   1476.3986
   1488.0208   1495.2216   1507.0339   1534.6608   3024.8752   3030.1894   3074.6200
   3079.8161   3084.9296   3109.1838   3111.0373   3155.6317


== Thermochemistry energies (hartree) ==
  ZPVE =     0.111018  (after scaling by 0.985)
W1  Ee =  -232.515736  (total electronic energy)
W1  U0 =  -232.404717  (Zero-kelvin internal energy)
 
[ CCSDT-F12a/cc-pVDZ-F12 data ]
 
CCSD(T)-F12a/cc-pVDZ-F12  Ee =  -232.119425  (total electronic energy)
\end{Verbatim}
    \clearpage

    \textbf{Species 5: alkene}
\begin{Verbatim}
[ W1 data ]

== B3LYP/cc-pVTZ+d equilibrium geometry (Angstrom) ==
C    -1.272636   -0.676750    0.000001
C    -0.000007    0.124048    0.000004
C    -0.000110    1.454215   -0.000002
C     1.272735   -0.676582    0.000001
H    -2.153991   -0.037135    0.000036
H    -1.322215   -1.329609    0.876239
H    -1.322239   -1.329555   -0.876275
H     1.322437   -1.329441    0.876233
H     2.154004   -0.036844    0.000024
H     1.322420   -1.329365   -0.876288
H    -0.922249    2.021111   -0.000001
H     0.921944    2.021249    0.000008

== B3LYP/cc-pVTZ+d harmonic frequencies (cm^-1) ==
    172.8162    210.8870    381.2455    440.1560    442.7760    707.2396    814.6234
    925.1425    961.6709    988.4662   1023.0717   1086.4536   1108.9179   1297.4171
   1410.9842   1416.3024   1445.7769   1472.4631   1485.0281   1489.9416   1503.6531
   1717.8689   3007.7226   3013.3153   3048.1450   3050.8984   3101.8017   3103.4482
   3130.2230   3207.6813


== Thermochemistry energies (hartree) ==
  ZPVE =     0.105840  (after scaling by 0.985)
W1  Ee =  -157.260142  (total electronic energy)
W1  U0 =  -157.154302  (Zero-kelvin internal energy)
 
[ CCSDT-F12a/cc-pVDZ-F12 data ]
 
CCSD(T)-F12a/cc-pVDZ-F12  Ee =  -156.981896  (total electronic energy)
\end{Verbatim}
    \clearpage

    \textbf{Species 6: Transition state along \rxnone{}}
\begin{Verbatim}
[ W1 data ]

== B3LYP/cc-pVTZ+d equilibrium geometry (Angstrom) ==
C     1.729982   -1.026856   -0.142334
C     0.632718   -0.006702   -0.035631
C    -0.496792   -0.299777    0.966070
C     1.050402    1.437910   -0.120253
H     1.333278   -2.039470   -0.223676
H     2.376129   -0.998500    0.744486
H     2.369501   -0.832714   -1.004641
H    -0.304917   -0.199741   -0.981413
H     1.667877    1.706944    0.746098
H     0.184029    2.098691   -0.128733
H     1.646542    1.629578   -1.013186
O    -1.673057    0.198285    0.336461
O    -1.550679   -0.303552   -0.986789
H    -0.402736    0.226914    1.918965
H    -0.577677   -1.377008    1.137599

== B3LYP/cc-pVTZ+d harmonic frequencies (cm^-1) ==
  -2074.1999    131.0424    187.6082    220.8176    259.6493    282.2091    363.7549
    401.9509    549.7213    660.9296    815.6935    899.6138    933.3573    948.1618
    968.9846    978.9501   1014.4557   1127.7462   1138.4629   1227.3442   1252.0236
   1282.2910   1351.4149   1400.4985   1414.2542   1471.4576   1477.7350   1491.3574
   1491.6597   1499.8779   1720.4371   2982.0842   2987.2194   3010.8712   3056.8370
   3065.2979   3066.2637   3095.5233   3106.0826


== Thermochemistry energies (hartree) ==
  ZPVE =     0.119680  (after scaling by 0.985)
W1  Ee =  -308.237025  (total electronic energy)
W1  U0 =  -308.117344  (Zero-kelvin internal energy)
 
[ CCSDT-F12a/cc-pVDZ-F12 data ]
 
CCSD(T)-F12a/cc-pVDZ-F12  Ee =  -307.722677  (total electronic energy)
\end{Verbatim}
    \clearpage

    \textbf{Species 7: Transition state along \rxntwo{}}
\begin{Verbatim}
[ W1 data ]

== B3LYP/cc-pVTZ+d equilibrium geometry (Angstrom) ==
C    -0.103419    0.000576    1.260337
H    -0.371480   -0.915374    1.766499
H    -0.371297    0.917222    1.765367
C     0.704754    0.000003    0.129033
H    -0.335511    0.000470   -0.666622
C     1.428132   -1.286073   -0.258379
H     1.616142   -1.325179   -1.332347
H     2.396191   -1.352011    0.244308
H     0.847036   -2.167747    0.012042
C     1.429550    1.285173   -0.258734
H     1.617564    1.323813   -1.332720
H     0.849467    2.167571    0.011506
H     2.397693    1.350129    0.243910
O    -2.052038   -0.000121    0.225095
O    -1.622950    0.000500   -0.968280

== B3LYP/cc-pVTZ+d harmonic frequencies (cm^-1) ==
   -920.2987    122.2922    179.9021    192.5073    208.2988    257.9623    366.5070
    372.9122    416.7113    546.4520    683.1347    756.3789    843.0233    943.4251
    972.8798    989.9748   1017.4941   1074.9199   1132.9713   1209.6971   1288.8704
   1327.1116   1408.8980   1414.5012   1424.5769   1483.1828   1495.0807   1498.6456
   1508.6753   1568.5048   1588.1585   3019.0304   3022.3352   3073.2892   3074.7033
   3096.5168   3097.7930   3153.4326   3239.1244


== Thermochemistry energies (hartree) ==
  ZPVE =     0.119088  (after scaling by 0.985)
W1  Ee =  -308.234418  (total electronic energy)
W1  U0 =  -308.115330  (Zero-kelvin internal energy)
 
[ CCSDT-F12a/cc-pVDZ-F12 data ]
 
CCSD(T)-F12a/cc-pVDZ-F12  Ee =  -307.720203  (total electronic energy)
\end{Verbatim}
    \clearpage

    \textbf{Species 8: Transition state along \rxnthree{}}
\begin{Verbatim}
[ W1 data ]

== B3LYP/cc-pVTZ+d equilibrium geometry (Angstrom) ==
C     0.070187   -0.537266    0.995685
H     0.611884    0.083767    1.694306
H     0.054349   -1.594035    1.225013
C    -0.951568    0.006922    0.245689
H     1.951756    0.770387   -1.193525
C    -1.089518    1.487428    0.083045
H    -0.922252    1.785568   -0.959046
H    -2.103726    1.817719    0.329673
H    -0.386937    2.035112    0.708192
C    -1.869988   -0.854459   -0.562024
H    -1.723492   -0.687467   -1.634888
H    -1.715286   -1.913818   -0.364020
H    -2.917461   -0.613472   -0.355460
O     1.550311   -0.751100   -0.185898
O     2.224249    0.463661   -0.317178

== B3LYP/cc-pVTZ+d harmonic frequencies (cm^-1) ==
   -499.1799     27.6767     97.4718    120.9997    126.7134    173.2424    293.4801
    375.2122    418.0195    440.4780    456.4588    800.3667    875.0434    936.6090
    964.4024    982.8232   1031.2558   1057.0041   1070.4450   1078.0214   1310.7476
   1377.1609   1397.1779   1407.4586   1415.2549   1466.4683   1477.5072   1483.2692
   1498.4608   1555.7415   2987.5213   2998.0116   3025.3999   3032.5624   3104.3266
   3109.1281   3139.4021   3223.2736   3714.8010


== Thermochemistry energies (hartree) ==
  ZPVE =     0.121287  (after scaling by 0.985)
W1  Ee =  -308.245487  (total electronic energy)
W1  U0 =  -308.124200  (Zero-kelvin internal energy)
 
[ CCSDT-F12a/cc-pVDZ-F12 data ]
 
CCSD(T)-F12a/cc-pVDZ-F12  Ee =  -307.730694  (total electronic energy)
\end{Verbatim}
    \clearpage

   \textbf{Species 9: Transition state along \rxnfour{}}
\begin{Verbatim}
[ W1 data ]

== B3LYP/cc-pVTZ+d equilibrium geometry (Angstrom) ==
C     0.365522   -0.041112    0.874498
H     0.608783    0.842364    1.470600
H     0.566038   -0.954671    1.440864
C    -0.933526    0.002640    0.175068
H     2.843522    0.598261   -0.854025
C    -1.559262    1.308674   -0.158355
H    -1.916489    1.334855   -1.190242
H    -2.436990    1.476166    0.480001
H    -0.870538    2.136541   -0.003271
C    -1.611276   -1.259903   -0.218232
H    -1.992082   -1.214122   -1.240760
H    -0.946062   -2.116033   -0.128815
H    -2.478840   -1.436138    0.431844
O     0.948000   -0.035001   -0.394531
O     2.683739   -0.056126   -0.160977

== B3LYP/cc-pVTZ+d harmonic frequencies (cm^-1) ==
   -651.7600     78.8946     96.5472    111.5576    124.7809    191.1735    203.4085
    282.9919    388.0618    402.8686    434.3460    766.4297    923.0601    935.8733
    980.8702    992.8149   1017.6637   1054.6418   1098.7156   1176.3438   1287.3300
   1315.7163   1365.3611   1403.7770   1411.1883   1459.2924   1475.4304   1477.9044
   1499.0488   1535.7380   2976.7452   2981.8922   3004.2764   3047.0586   3050.8414
   3051.2920   3116.2985   3118.9940   3785.1107


== Thermochemistry energies (hartree) ==
  ZPVE =     0.120333  (after scaling by 0.985)
W1  Ee =  -308.251537  (total electronic energy)
W1  U0 =  -308.131204  (Zero-kelvin internal energy)
 
[ CCSDT-F12a/cc-pVDZ-F12 data ]
 
CCSD(T)-F12a/cc-pVDZ-F12  Ee =  -307.737022  (total electronic energy)
\end{Verbatim}
    \clearpage
    
\subsection{Hydrocarbon 4: tertbutyl}

    \textbf{Species 1: $\cdot$R}
\begin{Verbatim}
[ W1 data ]

== B3LYP/cc-pVTZ+d equilibrium geometry (Angstrom) ==
C     0.587488   -1.361150    0.014532
C    -0.000027   -0.000011   -0.148270
C    -1.472550    0.171814    0.014623
C     0.885071    1.189347    0.014314
H     1.581371   -1.430125   -0.433166
H    -0.043758   -2.131690   -0.433583
H     0.705018   -1.634109    1.076509
H    -1.824088    1.104079   -0.432927
H    -1.767707    0.205859    1.076623
H    -2.029355   -0.654072   -0.433583
H     1.867952    1.027801   -0.433852
H     1.062775    1.427714    1.076254
H     0.447901    2.084541   -0.433467

== B3LYP/cc-pVTZ+d harmonic frequencies (cm^-1) ==
    132.9968    133.2369    134.8186    262.1915    379.8252    379.9802    757.7653
    937.9257    938.0411    972.6887   1008.0016   1008.0918   1097.1729   1294.2784
   1294.3488   1397.7567   1397.8094   1425.3273   1468.5504   1471.2244   1471.3241
   1488.9202   1491.8940   1491.9784   2914.1698   2914.1873   2922.4951   3028.5368
   3028.6184   3031.6386   3070.4209   3074.9860   3075.0498


== Thermochemistry energies (hartree) ==
  ZPVE =     0.114211  (after scaling by 0.985)
W1  Ee =  -157.823966  (total electronic energy)
W1  U0 =  -157.709756  (Zero-kelvin internal energy)
 
[ CCSDT-F12a/cc-pVDZ-F12 data ]
 
CCSD(T)-F12a/cc-pVDZ-F12  Ee =  -157.545445  (total electronic energy)
\end{Verbatim}
    \clearpage

    \textbf{Species 2: ROO$\cdot$}
\begin{Verbatim}
[ W1 data ]

== B3LYP/cc-pVTZ+d equilibrium geometry (Angstrom) ==
C    -0.667957    0.670834    1.265796
C     0.113820    0.350403   -0.000000
C     1.503478    0.968774    0.000000
C    -0.667957    0.670834   -1.265796
H    -0.887045    1.738171    1.304642
H    -0.093255    0.405232    2.153117
H    -1.607746    0.122322    1.280426
H     1.423454    2.055754    0.000000
H     2.062996    0.666024   -0.884767
H     2.062996    0.666024    0.884767
O     0.410249   -1.118607    0.000000
H    -0.093255    0.405232   -2.153117
H    -0.887045    1.738171   -1.304642
H    -1.607746    0.122322   -1.280426
O    -0.667957   -1.866931    0.000000

== B3LYP/cc-pVTZ+d harmonic frequencies (cm^-1) ==
    123.2015    180.1497    232.4116    243.0958    269.2671    331.3690    360.0768
    402.8449    437.4422    541.8098    728.8411    799.9479    929.0682    931.3197
    971.7406   1041.0430   1053.5585   1169.7132   1219.4434   1262.0575   1292.8167
   1399.5582   1402.8135   1427.7105   1471.6543   1489.4173   1490.9813   1497.9442
   1498.9160   1522.0499   3038.7158   3039.1648   3045.7611   3102.2559   3104.1263
   3110.2291   3111.6928   3120.7871   3124.6407


== Thermochemistry energies (hartree) ==
  ZPVE =     0.124586  (after scaling by 0.985)
W1  Ee =  -308.301806  (total electronic energy)
W1  U0 =  -308.177220  (Zero-kelvin internal energy)
 
[ CCSDT-F12a/cc-pVDZ-F12 data ]
 
CCSD(T)-F12a/cc-pVDZ-F12  Ee =  -307.787529  (total electronic energy)
\end{Verbatim}
    \clearpage

    \textbf{Species 3: $\cdot$QOOH}
\begin{Verbatim}
[ W1 data ]

== B3LYP/cc-pVTZ+d equilibrium geometry (Angstrom) ==
C     0.299981   -0.417790    1.502845
C     0.398461    0.017356    0.081051
C     1.599862   -0.630154   -0.616530
C     0.416065    1.535827   -0.063131
H     0.094122    0.282969    2.298536
H     0.376328   -1.469121    1.746016
H     2.525261   -0.270029   -0.168641
H     1.599808   -0.375432   -1.676334
H     1.564555   -1.714624   -0.518454
O    -0.711641   -0.545907   -0.691519
H     0.478580    1.814196   -1.114656
H     1.276854    1.953596    0.460649
H    -0.489456    1.969532    0.356241
O    -1.968222   -0.013991   -0.203604
H    -2.273367   -0.743339    0.352216

== B3LYP/cc-pVTZ+d harmonic frequencies (cm^-1) ==
    109.2194    141.3243    206.9200    240.7980    252.7215    258.1678    328.1161
    353.2250    402.1434    452.2221    498.9845    601.4172    762.0479    825.1127
    913.5871    932.7689    956.2792    982.0867   1036.1685   1152.8553   1261.4030
   1274.9492   1364.9961   1394.6598   1411.1307   1460.2112   1477.8673   1488.4401
   1494.0838   1512.1049   3037.6502   3042.2023   3102.6808   3109.4591   3113.6103
   3122.9679   3136.1293   3243.2977   3753.6479


== Thermochemistry energies (hartree) ==
  ZPVE =     0.121642  (after scaling by 0.985)
W1  Ee =  -308.270613  (total electronic energy)
W1  U0 =  -308.148971  (Zero-kelvin internal energy)
 
[ CCSDT-F12a/cc-pVDZ-F12 data ]
 
CCSD(T)-F12a/cc-pVDZ-F12  Ee =  -307.756034  (total electronic energy)
\end{Verbatim}
    \clearpage

    \textbf{Species 4: {\it cy}-ether}
\begin{Verbatim}
[ W1 data ]

== B3LYP/cc-pVTZ+d equilibrium geometry (Angstrom) ==
C    -0.930698   -1.279992   -0.049036
C    -0.131144    0.000000   -0.054913
C     1.237579   -0.000000   -0.582716
C    -0.930698    1.279992   -0.049036
H    -0.272060   -2.146381   -0.014050
H    -1.555426   -1.351676   -0.941587
H    -1.587027   -1.314167    0.822839
H    -1.555426    1.351676   -0.941587
H    -0.272060    2.146381   -0.014050
H    -1.587027    1.314167    0.822839
O     1.004512   -0.000000    0.830447
H     1.661345    0.916464   -0.981890
H     1.661345   -0.916464   -0.981890

== B3LYP/cc-pVTZ+d harmonic frequencies (cm^-1) ==
    182.2996    218.0260    356.0528    360.6338    410.6438    412.8239    700.0132
    807.9259    908.0013    923.8222    959.9394   1020.4768   1074.6552   1131.2834
   1150.3799   1172.9890   1284.2192   1391.2651   1413.0982   1423.4775   1476.3986
   1488.0208   1495.2216   1507.0339   1534.6608   3024.8752   3030.1894   3074.6200
   3079.8161   3084.9296   3109.1838   3111.0373   3155.6317


== Thermochemistry energies (hartree) ==
  ZPVE =     0.111018  (after scaling by 0.985)
W1  Ee =  -232.515736  (total electronic energy)
W1  U0 =  -232.404717  (Zero-kelvin internal energy)
 
[ CCSDT-F12a/cc-pVDZ-F12 data ]
 
CCSD(T)-F12a/cc-pVDZ-F12  Ee =  -232.119425  (total electronic energy)
\end{Verbatim}
    \clearpage

    \textbf{Species 5: alkene}
\begin{Verbatim}
[ W1 data ]

== B3LYP/cc-pVTZ+d equilibrium geometry (Angstrom) ==
C    -1.272636   -0.676750    0.000001
C    -0.000007    0.124048    0.000004
C    -0.000110    1.454215   -0.000002
C     1.272735   -0.676582    0.000001
H    -2.153991   -0.037135    0.000036
H    -1.322215   -1.329609    0.876239
H    -1.322239   -1.329555   -0.876275
H     1.322437   -1.329441    0.876233
H     2.154004   -0.036844    0.000024
H     1.322420   -1.329365   -0.876288
H    -0.922249    2.021111   -0.000001
H     0.921944    2.021249    0.000008

== B3LYP/cc-pVTZ+d harmonic frequencies (cm^-1) ==
    172.8162    210.8870    381.2455    440.1560    442.7760    707.2396    814.6234
    925.1425    961.6709    988.4662   1023.0717   1086.4536   1108.9179   1297.4171
   1410.9842   1416.3024   1445.7769   1472.4631   1485.0281   1489.9416   1503.6531
   1717.8689   3007.7226   3013.3153   3048.1450   3050.8984   3101.8017   3103.4482
   3130.2230   3207.6813


== Thermochemistry energies (hartree) ==
  ZPVE =     0.105840  (after scaling by 0.985)
W1  Ee =  -157.260142  (total electronic energy)
W1  U0 =  -157.154302  (Zero-kelvin internal energy)
 
[ CCSDT-F12a/cc-pVDZ-F12 data ]
 
CCSD(T)-F12a/cc-pVDZ-F12  Ee =  -156.981896  (total electronic energy)
\end{Verbatim}
    \clearpage

    \textbf{Species 6: Transition state along \rxnone{}}
\begin{Verbatim}
[ W1 data ]

== B3LYP/cc-pVTZ+d equilibrium geometry (Angstrom) ==
C     0.374528    0.007943    0.041163
C    -0.239334   -0.489943    1.355864
H    -1.472485   -0.247647    0.809877
H    -0.216368   -1.563877    1.518701
H    -0.084712    0.101532    2.251682
C     0.584397    1.520086    0.059182
H     1.318278    1.799473    0.816397
H     0.945472    1.857845   -0.911837
H    -0.352010    2.030992    0.276632
C     1.629428   -0.748555   -0.369766
H     1.437974   -1.819892   -0.408153
H     1.966037   -0.419234   -1.353630
H     2.432224   -0.560296    0.343716
O    -0.642771   -0.355494   -0.933256
O    -1.865795   -0.009016   -0.299499

== B3LYP/cc-pVTZ+d harmonic frequencies (cm^-1) ==
  -2232.9375    188.9270    195.0793    238.4295    299.3179    339.7262    392.5465
    406.8249    507.8713    569.9865    639.6141    782.1498    833.1641    905.6383
    924.3858    946.5922    972.7926   1017.4376   1042.7846   1108.3458   1173.1397
   1248.6215   1270.3820   1399.7768   1414.8222   1455.1113   1479.1162   1488.0157
   1496.2509   1509.6754   1710.7843   3034.9281   3040.1094   3094.7877   3098.0676
   3104.1973   3113.5280   3118.9170   3184.8277


== Thermochemistry energies (hartree) ==
  ZPVE =     0.118364  (after scaling by 0.985)
W1  Ee =  -308.237692  (total electronic energy)
W1  U0 =  -308.119329  (Zero-kelvin internal energy)
 
[ CCSDT-F12a/cc-pVDZ-F12 data ]
 
CCSD(T)-F12a/cc-pVDZ-F12  Ee =  -307.723601  (total electronic energy)
\end{Verbatim}
    \clearpage

    \textbf{Species 7: Transition state along \rxntwo{}}
\begin{Verbatim}
[ W1 data ]

== B3LYP/cc-pVTZ+d equilibrium geometry (Angstrom) ==
C     0.721059   -0.000018    0.227831
C     0.072724   -0.000113    1.458531
H    -1.179341   -0.000070    0.970425
H     0.082107   -0.915377    2.042335
H     0.082110    0.915059    2.042482
C     1.209909    1.274063   -0.387021
H     2.276798    1.395786   -0.167934
H     1.100321    1.262699   -1.471143
H     0.686663    2.141889    0.010713
C     1.209909   -1.274006   -0.387214
H     0.686638   -2.141888    0.010365
H     1.100352   -1.262464   -1.471336
H     2.276790   -1.395780   -0.168114
O    -1.246798    0.000084   -0.950567
O    -2.052458   -0.000011    0.041747

== B3LYP/cc-pVTZ+d harmonic frequencies (cm^-1) ==
  -1066.4599    129.5392    165.2107    194.0853    201.8582    204.6439    366.7876
    432.4375    433.2686    534.7269    633.6676    652.0179    816.9233    939.5041
    989.8160   1022.0894   1024.2148   1072.2748   1100.5524   1286.0098   1305.2299
   1326.8529   1409.7241   1410.8391   1425.9692   1468.9390   1480.3136   1484.5351
   1499.1572   1569.9244   1598.6130   3003.2904   3008.1821   3073.4676   3076.5548
   3091.6281   3114.7785   3117.1111   3168.6948


== Thermochemistry energies (hartree) ==
  ZPVE =     0.118558  (after scaling by 0.985)
W1  Ee =  -308.247514  (total electronic energy)
W1  U0 =  -308.128956  (Zero-kelvin internal energy)
 
[ CCSDT-F12a/cc-pVDZ-F12 data ]
 
CCSD(T)-F12a/cc-pVDZ-F12  Ee =  -307.733017  (total electronic energy)
\end{Verbatim}
    \clearpage

    \textbf{Species 8: Transition state along \rxnthree{}}
\begin{Verbatim}
[ W1 data ]

== B3LYP/cc-pVTZ+d equilibrium geometry (Angstrom) ==
C     0.659424    0.111180    0.154318
C     0.629958   -0.166234    1.513604
H    -2.220744    0.547831   -0.683078
H     1.001960   -1.106607    1.894172
H     0.111527    0.483571    2.204284
C     0.372785    1.513754   -0.313028
H     1.269106    2.130950   -0.215509
H     0.080859    1.524197   -1.363802
H    -0.413588    1.976165    0.282474
C     1.553931   -0.709594   -0.737671
H     1.620258   -1.738368   -0.387526
H     1.179707   -0.715055   -1.760570
H     2.561384   -0.286405   -0.747566
O    -0.945723   -0.789945   -0.372351
O    -2.115159   -0.123919    0.006574

== B3LYP/cc-pVTZ+d harmonic frequencies (cm^-1) ==
   -487.7669     88.9942    151.7565    187.8326    203.4678    238.7849    313.5023
    383.0518    393.9403    436.9883    445.0092    514.1904    794.1423    798.9923
    937.3273    945.8456    975.7839   1020.0629   1059.3166   1077.9920   1303.4883
   1375.2400   1388.7153   1405.2746   1419.3570   1477.5224   1487.4114   1488.9793
   1508.0398   1528.5590   3025.6914   3030.9291   3078.6432   3090.0985   3106.6231
   3117.0108   3145.0795   3238.2692   3701.7344


== Thermochemistry energies (hartree) ==
  ZPVE =     0.120915  (after scaling by 0.985)
W1  Ee =  -308.243556  (total electronic energy)
W1  U0 =  -308.122642  (Zero-kelvin internal energy)
 
[ CCSDT-F12a/cc-pVDZ-F12 data ]
 
CCSD(T)-F12a/cc-pVDZ-F12  Ee =  -307.729235  (total electronic energy)
\end{Verbatim}
    \clearpage

   \textbf{Species 9: Transition state along \rxnfour{}}
\begin{Verbatim}
[ W1 data ]

== B3LYP/cc-pVTZ+d equilibrium geometry (Angstrom) ==
C     0.487789    0.043724    0.026149
C     1.265037   -1.196235   -0.199899
H    -2.492390   -0.559124    0.456318
H     1.565088   -1.476775   -1.197744
H     1.427753   -1.889516    0.610898
C     0.479072    0.552137    1.459441
H     1.414881    1.060642    1.694989
H    -0.343744    1.252754    1.590493
H     0.343867   -0.273327    2.158369
C     0.681171    1.144900   -1.003846
H     0.674859    0.729836   -2.011003
H    -0.133241    1.863715   -0.925230
H     1.625262    1.666943   -0.841161
O    -0.612196   -0.797319   -0.274198
O    -2.082897    0.092031   -0.129177

== B3LYP/cc-pVTZ+d harmonic frequencies (cm^-1) ==
   -723.3235    106.4119    180.1855    192.2029    210.6768    242.7670    335.0221
    348.3646    401.7137    411.4388    480.1318    530.8473    754.2206    764.4269
    868.1073    909.8624    957.4954    983.0920   1029.0132   1036.6748   1158.7278
   1265.3814   1326.8174   1400.6343   1411.9795   1465.1336   1474.2970   1490.3417
   1498.6409   1510.3099   3034.8933   3040.6053   3100.0703   3105.1350   3116.1359
   3120.0357   3158.1296   3273.0329   3784.3896


== Thermochemistry energies (hartree) ==
  ZPVE =     0.120003  (after scaling by 0.985)
W1  Ee =  -308.251146  (total electronic energy)
W1  U0 =  -308.131144  (Zero-kelvin internal energy)
 
[ CCSDT-F12a/cc-pVDZ-F12 data ]
 
CCSD(T)-F12a/cc-pVDZ-F12  Ee =  -307.737109  (total electronic energy)
\end{Verbatim}
    \clearpage

\subsection{Hydrocarbon 5: neopentyl}

    \textbf{Species 1: $\cdot$R}
\begin{Verbatim}
[ W1 data ]

== B3LYP/cc-pVTZ+d equilibrium geometry (Angstrom) ==
C    -0.000005    1.320684   -0.753285
C     0.000000   -0.001686    0.054897
C     1.258699   -0.808909   -0.308833
C    -1.258695   -0.808916   -0.308832
C    -0.000000    0.314303    1.517295
H     0.882453    1.919408   -0.523874
H    -0.882466    1.919403   -0.523873
H    -0.000004    1.112215   -1.826312
H     0.926270    0.515943    2.038541
H    -0.926271    0.515943    2.038541
H    -2.164819   -0.254131   -0.058074
H    -1.284921   -1.756526    0.231099
H    -1.284964   -1.027242   -1.378262
H     2.164820   -0.254117   -0.058076
H     1.284969   -1.027234   -1.378263
H     1.284933   -1.756518    0.231098

== B3LYP/cc-pVTZ+d harmonic frequencies (cm^-1) ==
    106.5750    223.5441    271.5708    276.4896    309.0223    330.4343    389.5571
    415.5578    417.3262    554.9511    736.9411    906.1956    911.9028    947.2156
    952.2161    962.0325   1031.4285   1076.6879   1200.1347   1264.3617   1278.0764
   1391.7524   1399.1188   1419.7938   1465.5041   1480.7889   1487.2601   1489.1176
   1503.1149   1504.4361   1519.1660   3012.6404   3017.4487   3023.6726   3074.5707
   3078.5879   3083.3933   3084.2489   3088.7211   3091.1544   3127.4082   3227.6421



== Thermochemistry energies (hartree) ==
  ZPVE =     0.141668  (after scaling by 0.985)
W1  Ee =  -197.141099  (total electronic energy)
W1  U0 =  -196.999431  (Zero-kelvin internal energy)
 
[ CCSDT-F12a/cc-pVDZ-F12 data ]
 
CCSD(T)-F12a/cc-pVDZ-F12  Ee =  -196.793722  (total electronic energy)
\end{Verbatim}
    \clearpage

    \textbf{Species 2: ROO$\cdot$}
\begin{Verbatim}
[ W1 data ]

== B3LYP/cc-pVTZ+d equilibrium geometry (Angstrom) ==
C    -1.063554   -1.018643   -1.089256
C    -0.747680    0.031173   -0.013335
C    -1.927658    0.145185    0.965624
C    -0.481716    1.394391   -0.666931
C     0.465868   -0.414646    0.814125
H    -1.274865   -1.992396   -0.642476
H    -0.230913   -1.139395   -1.781870
H    -1.940053   -0.718943   -1.665593
H     0.732530    0.316984    1.574524
H     0.282019   -1.386665    1.274502
H     0.361704    1.351582   -1.355179
H    -0.262106    2.157344    0.081515
H    -1.359093    1.716973   -1.229980
H    -2.140251   -0.810148    1.449619
H    -2.828532    0.455943    0.434842
H    -1.729905    0.883449    1.745090
O     1.650641   -0.622680   -0.005948
O     2.464097    0.415244    0.012654

== B3LYP/cc-pVTZ+d harmonic frequencies (cm^-1) ==
     67.2322     96.3604    214.3459    229.7828    260.9030    265.8219    307.0666
    335.9743    389.6075    406.1536    480.1114    553.0345    739.8855    883.4435
    912.3390    932.5342    944.4727    963.9107    968.8674   1048.3496   1085.0899
   1159.5049   1232.5810   1248.1471   1275.8564   1320.9303   1368.2178   1404.2359
   1408.3626   1437.2366   1470.6616   1483.4225   1489.3524   1491.5669   1508.8469
   1509.1239   1521.1851   3020.5321   3024.9991   3031.3061   3048.8968   3080.5358
   3084.0447   3086.2493   3091.1858   3100.3205   3104.8386   3114.9399


== Thermochemistry energies (hartree) ==
  ZPVE =     0.153045  (after scaling by 0.985)
W1  Ee =  -347.616877  (total electronic energy)
W1  U0 =  -347.463832  (Zero-kelvin internal energy)
 
[ CCSDT-F12a/cc-pVDZ-F12 data ]
 
CCSD(T)-F12a/cc-pVDZ-F12  Ee =  -347.033300  (total electronic energy)
\end{Verbatim}
    \clearpage

    \textbf{Species 3: $\cdot$QOOH}
\begin{Verbatim}
[ W1 data ]

== B3LYP/cc-pVTZ+d equilibrium geometry (Angstrom) ==
C    -1.090901    0.819175    1.235392
C    -0.757997   -0.072705    0.030108
C    -1.979130   -0.139922   -0.922839
C    -0.409571   -1.456614    0.477090
C     0.398603    0.558631   -0.783508
H    -1.402166    1.813394    0.910178
H    -0.224900    0.936204    1.886051
H    -1.904533    0.387887    1.820486
H     0.624035   -0.036070   -1.671569
H     0.107767    1.564181   -1.099818
H     1.928120   -0.930355    0.721946
H    -0.066123   -2.188790   -0.242015
H    -0.701939   -1.812020    1.455955
H    -2.242816    0.858228   -1.280755
H    -2.845019   -0.553635   -0.406297
H    -1.771207   -0.769336   -1.789165
O     1.593985    0.772145   -0.046966
O     2.347610   -0.462281   -0.018341

== B3LYP/cc-pVTZ+d harmonic frequencies (cm^-1) ==
     82.2195    145.9221    168.2912    223.1865    238.9078    258.9470    282.1709
    306.1951    341.5138    396.1744    401.9358    453.8458    553.4559    621.5174
    757.7773    867.5341    893.2415    930.4618    947.8231    956.2221   1010.9234
   1024.9251   1056.1846   1164.6371   1242.9597   1253.6327   1328.9846   1371.1282
   1389.0006   1397.1891   1417.2041   1454.5062   1469.8460   1487.3982   1492.2050
   1503.8472   1512.0371   3016.6355   3021.3596   3029.7800   3068.5296   3081.6000
   3087.7546   3098.3088   3103.4782   3126.9234   3227.6601   3662.9664


== Thermochemistry energies (hartree) ==
  ZPVE =     0.150188  (after scaling by 0.985)
W1  Ee =  -347.587649  (total electronic energy)
W1  U0 =  -347.437461  (Zero-kelvin internal energy)
 
[ CCSDT-F12a/cc-pVDZ-F12 data ]
 
CCSD(T)-F12a/cc-pVDZ-F12  Ee =  -347.003692  (total electronic energy)
\end{Verbatim}
    \clearpage

    \textbf{Species 4: {\it cy}-ether}
        \VerbatimInput{SI_data/Neopentyl/Mol_04_W1.txt}
    \clearpage

    \textbf{Species 5: dimethylcyclopropane}
        \VerbatimInput{SI_data/Neopentyl/Mol_05_W1.txt}
    \clearpage

    \textbf{Species 6: Transition state along \rxnone{}}
        \VerbatimInput{SI_data/Neopentyl/TS_01_W1.txt}
    \clearpage

    \textbf{Species 7: Transition state along \rxntwo{}}
        \VerbatimInput{SI_data/Neopentyl/TS_02_W1.txt}
    \clearpage

    \textbf{Species 8: Transition state along \rxnthree{}}
        \VerbatimInput{SI_data/Neopentyl/TS_03_W1.txt}
    \clearpage

   \textbf{Species 9: Transition state along \rxnfour{}}
        \VerbatimInput{SI_data/Neopentyl/TS_04_W1.txt}
    \clearpage

\subsection{Hydrocarbon 6: cyclohexyl}

    \textbf{Species 1: $\cdot$R}
        \VerbatimInput{SI_data/Cyclohexyl/Mol_01_W1.txt}
    \clearpage

    \textbf{Species 2: ROO$\cdot$}
        \VerbatimInput{SI_data/Cyclohexyl/Mol_02_W1.txt}
    \clearpage

    \textbf{Species 3: $\cdot$QOOH}
        \VerbatimInput{SI_data/Cyclohexyl/Mol_03_W1.txt}
    \clearpage

    \textbf{Species 4: {\it cy}-ether}
        \VerbatimInput{SI_data/Cyclohexyl/Mol_04_W1.txt}
    \clearpage

    \textbf{Species 5: alkene}
        \VerbatimInput{SI_data/Cyclohexyl/Mol_05_W1.txt}
    \clearpage

    \textbf{Species 6: Transition state along \rxnone{}}
        \VerbatimInput{SI_data/Cyclohexyl/TS_01_W1.txt}
    \clearpage

    \textbf{Species 7: Transition state along \rxntwo{}}
        \VerbatimInput{SI_data/Cyclohexyl/TS_02_W1.txt}
    \clearpage

    \textbf{Species 8: Transition state along \rxnthree{}}
        \VerbatimInput{SI_data/Cyclohexyl/TS_03_W1.txt}
    \clearpage

   \textbf{Species 9: Transition state along \rxnfour{}}
        \VerbatimInput{SI_data/Cyclohexyl/TS_04_W1.txt}
    \clearpage  

\subsection{Hydrocarbon 7: cyclohexenyl}

    \textbf{Species 1: $\cdot$R}
        \VerbatimInput{SI_data/Cyclohexenyl/Mol_01_W1.txt}
    \clearpage

    \textbf{Species 2: ROO$\cdot$}
        \VerbatimInput{SI_data/Cyclohexenyl/Mol_02_W1.txt}
    \clearpage

    \textbf{Species 3: $\cdot$QOOH}
        \VerbatimInput{SI_data/Cyclohexenyl/Mol_03_W1.txt}
    \clearpage

    \textbf{Species 4: {\it cy}-ether}
        \VerbatimInput{SI_data/Cyclohexenyl/Mol_04_W1.txt}
    \clearpage

    \textbf{Species 5: alkene}
        \VerbatimInput{SI_data/Cyclohexenyl/Mol_05_W1.txt}
    \clearpage

    \textbf{Species 6: Transition state along \rxnone{}}
        \VerbatimInput{SI_data/Cyclohexenyl/TS_01_W1.txt}
    \clearpage

    \textbf{Species 8: Transition state along \rxntwo{}}
        \VerbatimInput{SI_data/Cyclohexenyl/TS_02_W1.txt}
    \clearpage

    \textbf{Species 8: Transition state along \rxnthree{}}
        \VerbatimInput{SI_data/Cyclohexenyl/TS_03_W1.txt}
    \clearpage

   \textbf{Species 9: Transition state along \rxnfour{}}
        \VerbatimInput{SI_data/Cyclohexenyl/TS_04_W1.txt}
    \clearpage

\subsection{Hydrocarbon 8: cyclohexadienyl}

    \textbf{Species 1: $\cdot$R}
        \VerbatimInput{SI_data/Cyclohexadienyl/Mol_01_W1.txt}
    \clearpage

    \textbf{Species 2: ROO$\cdot$}
        \VerbatimInput{SI_data/Cyclohexadienyl/Mol_02_W1.txt}
    \clearpage

    \textbf{Species 3: $\cdot$QOOH}
        \VerbatimInput{SI_data/Cyclohexadienyl/Mol_03_W1.txt}
    \clearpage

    \textbf{Species 4: {\it cy}-ether}
        \VerbatimInput{SI_data/Cyclohexadienyl/Mol_04_W1.txt}
    \clearpage

    \textbf{Species 5: alkene}
        \VerbatimInput{SI_data/Cyclohexadienyl/Mol_05_W1.txt}
    \clearpage

    \textbf{Species 6: Transition state along \rxnone{}}
        \VerbatimInput{SI_data/Cyclohexadienyl/TS_01_W1.txt}
    \clearpage

    \textbf{Species 7: Transition state along \rxntwo{}}
        \VerbatimInput{SI_data/Cyclohexadienyl/TS_02_W1.txt}
    \clearpage

    \textbf{Species 8: Transition state along \rxnthree{}}
        \VerbatimInput{SI_data/Cyclohexadienyl/TS_03_W1.txt}
    \clearpage

   \textbf{Species 9: Transition state along \rxnfour{}}
        \VerbatimInput{SI_data/Cyclohexadienyl/TS_04_W1.txt}
    \clearpage  

\section{G4-level equilibrium geometry, harmonic frequencies, and thermochemistry energies of different reaction species involved in combustion reaction of R$\cdot$}

\subsection{Hydrocarbon 1: ethyl}

    \textbf{Species 1: $\cdot$R}
        \VerbatimInput{SI_data/Ethyl/Mol_01_G4.txt}
    \clearpage

    \textbf{Species 2: ROO$\cdot$}
        \VerbatimInput{SI_data/Ethyl/Mol_02_G4.txt}
    \clearpage

    \textbf{Species 3: $\cdot$QOOH}
        \VerbatimInput{SI_data/Ethyl/Mol_03_G4.txt}
    \clearpage

    \textbf{Species 4: {\it cy}-ether}
        \VerbatimInput{SI_data/Ethyl/Mol_04_G4.txt}
    \clearpage

    \textbf{Species 5: alkene}
        \VerbatimInput{SI_data/Ethyl/Mol_05_G4.txt}
    \clearpage

    \textbf{Species 6: Transition state along \rxnone{}}
        \VerbatimInput{SI_data/Ethyl/TS_01_G4.txt}
    \clearpage

    \textbf{Species 7: Transition state along \rxntwo{}}
        \VerbatimInput{SI_data/Ethyl/TS_02_G4.txt}
    \clearpage

    \textbf{Species 8: Transition state along \rxnthree{}}
        \VerbatimInput{SI_data/Ethyl/TS_03_G4.txt}
    \clearpage

   \textbf{Species 9: Transition state along \rxnfour{}}
        \VerbatimInput{SI_data/Ethyl/TS_04_G4.txt}
    \clearpage

\subsection{Hydrocarbon 2: isopropyl}

    \textbf{Species 1: $\cdot$R}
        \VerbatimInput{SI_data/Isopropyl/Mol_01_G4.txt}
    \clearpage

    \textbf{Species 2: ROO$\cdot$}
        \VerbatimInput{SI_data/Isopropyl/Mol_02_G4.txt}
    \clearpage

    \textbf{Species 3: $\cdot$QOOH}
        \VerbatimInput{SI_data/Isopropyl/Mol_03_G4.txt}
    \clearpage

    \textbf{Species 4: {\it cy}-ether}
        \VerbatimInput{SI_data/Isopropyl/Mol_04_G4.txt}
    \clearpage

    \textbf{Species 5: alkene}
        \VerbatimInput{SI_data/Isopropyl/Mol_05_G4.txt}
    \clearpage

    \textbf{Species 6: Transition state along \rxnone{}}
        \VerbatimInput{SI_data/Isopropyl/TS_01_G4.txt}
    \clearpage

    \textbf{Species 7: Transition state along \rxntwo{}}
        \VerbatimInput{SI_data/Isopropyl/TS_02_G4.txt}
    \clearpage

    \textbf{Species 8: Transition state along \rxnthree{}}
        \VerbatimInput{SI_data/Isopropyl/TS_03_G4.txt}
    \clearpage

   \textbf{Species 9: Transition state along \rxnfour{}}
        \VerbatimInput{SI_data/Isopropyl/TS_04_G4.txt}
    \clearpage

\subsection{Hydrocarbon 3: isobutyl}

    \textbf{Species 1: $\cdot$R}
        \VerbatimInput{SI_data/Isobutyl/Mol_01_G4.txt}
    \clearpage

    \textbf{Species 2: ROO$\cdot$}
        \VerbatimInput{SI_data/Isobutyl/Mol_02_G4.txt}
    \clearpage

    \textbf{Species 3: $\cdot$QOOH}
        \VerbatimInput{SI_data/Isobutyl/Mol_03_G4.txt}
    \clearpage

    \textbf{Species 4: {\it cy}-ether}
        \VerbatimInput{SI_data/Isobutyl/Mol_04_G4.txt}
    \clearpage

    \textbf{Species 5: alkene}
        \VerbatimInput{SI_data/Isobutyl/Mol_05_G4.txt}
    \clearpage

    \textbf{Species 6: Transition state along \rxnone{}}
        \VerbatimInput{SI_data/Isobutyl/TS_01_G4.txt}
    \clearpage

    \textbf{Species 7: Transition state along \rxntwo{}}
        \VerbatimInput{SI_data/Isobutyl/TS_02_G4.txt}
    \clearpage

    \textbf{Species 8: Transition state along \rxnthree{}}
        \VerbatimInput{SI_data/Isobutyl/TS_03_G4.txt}
    \clearpage

   \textbf{Species 9: Transition state along \rxnfour{}}
        \VerbatimInput{SI_data/Isobutyl/TS_04_G4.txt}
    \clearpage
    
\subsection{Hydrocarbon 4: tertbutyl}

    \textbf{Species 1: $\cdot$R}
        \VerbatimInput{SI_data/Tertbutyl/Mol_01_G4.txt}
    \clearpage

    \textbf{Species 2: ROO$\cdot$}
        \VerbatimInput{SI_data/Tertbutyl/Mol_02_G4.txt}
    \clearpage

    \textbf{Species 3: $\cdot$QOOH}
        \VerbatimInput{SI_data/Tertbutyl/Mol_03_G4.txt}
    \clearpage

    \textbf{Species 4: {\it cy}-ether}
        \VerbatimInput{SI_data/Tertbutyl/Mol_04_G4.txt}
    \clearpage

    \textbf{Species 5: alkene}
        \VerbatimInput{SI_data/Tertbutyl/Mol_05_G4.txt}
    \clearpage

    \textbf{Species 6: Transition state along \rxnone{}}
        \VerbatimInput{SI_data/Tertbutyl/TS_01_G4.txt}
    \clearpage

    \textbf{Species 7: Transition state along \rxntwo{}}
        \VerbatimInput{SI_data/Tertbutyl/TS_02_G4.txt}
    \clearpage

    \textbf{Species 8: Transition state along \rxnthree{}}
        \VerbatimInput{SI_data/Tertbutyl/TS_03_G4.txt}
    \clearpage

   \textbf{Species 9: Transition state along \rxnfour{}}
        \VerbatimInput{SI_data/Tertbutyl/TS_04_G4.txt}
    \clearpage

\subsection{Hydrocarbon 5: neopentyl}

    \textbf{Species 1: $\cdot$R}
        \VerbatimInput{SI_data/Neopentyl/Mol_01_G4.txt}
    \clearpage

    \textbf{Species 2: ROO$\cdot$}
        \VerbatimInput{SI_data/Neopentyl/Mol_02_G4.txt}
    \clearpage

    \textbf{Species 3: $\cdot$QOOH}
        \VerbatimInput{SI_data/Neopentyl/Mol_03_G4.txt}
    \clearpage

    \textbf{Species 4: {\it cy}-ether}
        \VerbatimInput{SI_data/Neopentyl/Mol_04_G4.txt}
    \clearpage

    \textbf{Species 5: alkene}
        \VerbatimInput{SI_data/Neopentyl/Mol_05_G4.txt}
    \clearpage

    \textbf{Species 6: Transition state along \rxnone{}}
        \VerbatimInput{SI_data/Neopentyl/TS_01_G4.txt}
    \clearpage

    \textbf{Species 7: Transition state along \rxntwo{}}
        \VerbatimInput{SI_data/Neopentyl/TS_02_G4.txt}
    \clearpage

    \textbf{Species 8: Transition state along \rxnthree{}}
        \VerbatimInput{SI_data/Neopentyl/TS_03_G4.txt}
    \clearpage

   \textbf{Species 9: Transition state along \rxnfour{}}
        \VerbatimInput{SI_data/Neopentyl/TS_04_G4.txt}
    \clearpage

\subsection{Hydrocarbon 6:cyclohexyl}

    \textbf{Species 1: $\cdot$R}
        \VerbatimInput{SI_data/Cyclohexyl/Mol_01_G4.txt}
    \clearpage

    \textbf{Species 2: ROO$\cdot$}
        \VerbatimInput{SI_data/Cyclohexyl/Mol_02_G4.txt}
    \clearpage

    \textbf{Species 3: $\cdot$QOOH}
        \VerbatimInput{SI_data/Cyclohexyl/Mol_03_G4.txt}
    \clearpage

    \textbf{Species 4: {\it cy}-ether}
        \VerbatimInput{SI_data/Cyclohexyl/Mol_04_G4.txt}
    \clearpage

    \textbf{Species 5: alkene}
        \VerbatimInput{SI_data/Cyclohexyl/Mol_05_G4.txt}
    \clearpage

    \textbf{Species 6: Transition state along \rxnone{}}
        \VerbatimInput{SI_data/Cyclohexyl/TS_01_G4.txt}
    \clearpage

    \textbf{Species 7: Transition state along \rxntwo{}}
        \VerbatimInput{SI_data/Cyclohexyl/TS_02_G4.txt}
    \clearpage

    \textbf{Species 8: Transition state along \rxnthree{}}
        \VerbatimInput{SI_data/Cyclohexyl/TS_03_G4.txt}
    \clearpage

   \textbf{Species 9: Transition state along \rxnfour{}}
        \VerbatimInput{SI_data/Cyclohexyl/TS_04_G4.txt}
    \clearpage  

\subsection{Hydrocarbon 7: $\alpha$-cyclohexenyl}

    \textbf{Species 1: $\cdot$R}
        \VerbatimInput{SI_data/Cyclohexenyl/Mol_01_G4.txt}
    \clearpage

    \textbf{Species 2: ROO$\cdot$}
        \VerbatimInput{SI_data/Cyclohexenyl/Mol_02_G4.txt}
    \clearpage

    \textbf{Species 3: $\cdot$QOOH}
        \VerbatimInput{SI_data/Cyclohexenyl/Mol_03_G4.txt}
    \clearpage

    \textbf{Species 4: {\it cy}-ether}
        \VerbatimInput{SI_data/Cyclohexenyl/Mol_04_G4.txt}
    \clearpage

    \textbf{Species 5: alkene}
        \VerbatimInput{SI_data/Cyclohexenyl/Mol_05_G4.txt}
    \clearpage

    \textbf{Species 6: Transition state along \rxnone{}}
        \VerbatimInput{SI_data/Cyclohexenyl/TS_01_G4.txt}
    \clearpage

    \textbf{Species 7: Transition state along \rxntwo{}}
        \VerbatimInput{SI_data/Cyclohexenyl/TS_02_G4.txt}
    \clearpage

    \textbf{Species 8: Transition state along \rxnthree{}}
        \VerbatimInput{SI_data/Cyclohexenyl/TS_03_G4.txt}
    \clearpage

   \textbf{Species 9: Transition state along \rxnfour{}}
        \VerbatimInput{SI_data/Cyclohexenyl/TS_04_G4.txt}
    \clearpage

\subsection{Hydrocarbon 8: $\beta$-cyclohexenyl}

    \textbf{Species 1: $\cdot$R}
        \VerbatimInput{SI_data/Cyclohexenyl_beta/Mol_01_G4.txt}
    \clearpage

    \textbf{Species 2: ROO$\cdot$}
        \VerbatimInput{SI_data/Cyclohexenyl_beta/Mol_02_G4.txt}
    \clearpage

    \textbf{Species 3: $\cdot$QOOH}
        \VerbatimInput{SI_data/Cyclohexenyl_beta/Mol_03_G4.txt}
    \clearpage

    \textbf{Species 4: {\it cy}-ether}
        \VerbatimInput{SI_data/Cyclohexenyl_beta/Mol_04_G4.txt}
    \clearpage

    \textbf{Species 5: alkene}
        \VerbatimInput{SI_data/Cyclohexenyl_beta/Mol_05_G4.txt}
    \clearpage

    \textbf{Species 6: Transition state along \rxnone{}}
        \VerbatimInput{SI_data/Cyclohexenyl_beta/TS_01_G4.txt}
    \clearpage

    \textbf{Species 7: Transition state along \rxntwo{}}
        \VerbatimInput{SI_data/Cyclohexenyl_beta/TS_02_G4.txt}
    \clearpage

    \textbf{Species 8: Transition state along \rxnthree{}}
        \VerbatimInput{SI_data/Cyclohexenyl_beta/TS_03_G4.txt}
    \clearpage

   \textbf{Species 9: Transition state along \rxnfour{}}
        \VerbatimInput{SI_data/Cyclohexenyl_beta/TS_04_G4.txt}
    \clearpage

\subsection{Hydrocarbon 9: cyclohexadienyl}

    \textbf{Species 1: $\cdot$R}
        \VerbatimInput{SI_data/Cyclohexadienyl/Mol_01_G4.txt}
    \clearpage

    \textbf{Species 2: ROO$\cdot$}
        \VerbatimInput{SI_data/Cyclohexadienyl/Mol_02_G4.txt}
    \clearpage

    \textbf{Species 3: $\cdot$QOOH}
        \VerbatimInput{SI_data/Cyclohexadienyl/Mol_03_G4.txt}
    \clearpage

    \textbf{Species 4: {\it cy}-ether}
        \VerbatimInput{SI_data/Cyclohexadienyl/Mol_04_G4.txt}
    \clearpage

    \textbf{Species 5: benzene}
        \VerbatimInput{SI_data/Cyclohexadienyl/Mol_05_G4.txt}
    \clearpage

    \textbf{Species 6: Transition state along \rxnone{}}
        \VerbatimInput{SI_data/Cyclohexadienyl/TS_01_G4.txt}
    \clearpage

    \textbf{Species 7: Transition state along \rxntwo{}}
        \VerbatimInput{SI_data/Cyclohexadienyl/TS_02_G4.txt}
    \clearpage

    \textbf{Species 8: Transition state along \rxnthree{}}
        \VerbatimInput{SI_data/Cyclohexadienyl/TS_03_G4.txt}
    \clearpage

   \textbf{Species 9: Transition state along \rxnfour{}}
        \VerbatimInput{SI_data/Cyclohexadienyl/TS_04_G4.txt}
    \clearpage

\subsection{Hydrocarbon 10: 2-(cyclohexa-2,4-dien-1-yl)ethyl}

    \textbf{Species 1: $\cdot$R}
        \VerbatimInput{SI_data/2cyhexa24dien1ylethyl/R_G4.txt}
    \clearpage

    \textbf{Species 2: ROO$\cdot$}
        \VerbatimInput{SI_data/2cyhexa24dien1ylethyl/ROO_G4.txt}
    \clearpage

    \textbf{Species 3: $syn$-$\cdot$QOOH}
        \VerbatimInput{SI_data/2cyhexa24dien1ylethyl/QOOH_syn_G4.txt}
    \clearpage

    \textbf{Species 4: $anti$-$\cdot$QOOH }
        \VerbatimInput{SI_data/2cyhexa24dien1ylethyl/QOOH_anti_G4.txt}
    \clearpage

    \textbf{Species 5: {\it cy}-ether}
        \VerbatimInput{SI_data/2cyhexa24dien1ylethyl/cy_ether_G4.txt}
    \clearpage

    \textbf{Species 6: Transition state along \rxnone{}}
        \VerbatimInput{SI_data/2cyhexa24dien1ylethyl/TS_01_G4.txt}
    \clearpage

   \textbf{Species 7: Transition state along \rxnfour{}}
        \VerbatimInput{SI_data/2cyhexa24dien1ylethyl/TS_04_G4.txt}
    \clearpage  

\subsection{Hydrocarbon 11: 6-methyl cyclohepta-2,4-dien-1-yl }

    \textbf{Species 1: $\cdot$R}
        \VerbatimInput{SI_data/6methylcyhepta24dien1yl/R_G4.txt}
    \clearpage

    \textbf{Species 2: ROO$\cdot$}
        \VerbatimInput{SI_data/6methylcyhepta24dien1yl/ROO_G4.txt}
    \clearpage

    \textbf{Species 3: $syn$-$\cdot$QOOH}
        \VerbatimInput{SI_data/6methylcyhepta24dien1yl/QOOH_syn_G4.txt}
    \clearpage

    \textbf{Species 4: $anti$-$\cdot$QOOH }
        \VerbatimInput{SI_data/6methylcyhepta24dien1yl/QOOH_anti_G4.txt}
    \clearpage

    \textbf{Species 5: {\it cy}-ether}
        \VerbatimInput{SI_data/6methylcyhepta24dien1yl/cy_ether_G4.txt}
    \clearpage

    \textbf{Species 6: Transition state along \rxnone{}}
        \VerbatimInput{SI_data/6methylcyhepta24dien1yl/TS_01_G4.txt}
    \clearpage

   \textbf{Species 7: Transition state along \rxnfour{}}
        \VerbatimInput{SI_data/6methylcyhepta24dien1yl/TS_04_G4.txt}
    \clearpage  

\subsection{Hydrocarbon 12: (cyclohepta-3,5-dien-1-yl)methyl  }

    \textbf{Species 1: $\cdot$R}
        \VerbatimInput{SI_data/cyhepta35dien1ylmethyl/R_G4.txt}
    \clearpage

    \textbf{Species 2: ROO$\cdot$}
        \VerbatimInput{SI_data/cyhepta35dien1ylmethyl/ROO_G4.txt}
    \clearpage

    \textbf{Species 3: $syn$-$\cdot$QOOH}
        \VerbatimInput{SI_data/cyhepta35dien1ylmethyl/QOOH_syn_G4.txt}
    \clearpage

    \textbf{Species 4: $anti$-$\cdot$QOOH }
        \VerbatimInput{SI_data/cyhepta35dien1ylmethyl/QOOH_anti_G4.txt}
    \clearpage

    \textbf{Species 5: {\it cy}-ether}
        \VerbatimInput{SI_data/cyhepta35dien1ylmethyl/cy_ether_G4.txt}
    \clearpage

    \textbf{Species 6: Transition state along \rxnone{}}
        \VerbatimInput{SI_data/cyhepta35dien1ylmethyl/TS_01_G4.txt}
    \clearpage

   \textbf{Species 7: Transition state along \rxnfour{}}
        \VerbatimInput{SI_data/cyhepta35dien1ylmethyl/TS_04_G4.txt}
    \clearpage 

\subsection{Hydrocarbon 13: cyclohepta-2,4-dien-1-yl }

    \textbf{Species 1: $\cdot$R}
         \VerbatimInput{SI_data/cyhepta24dien1yl/R_G4.txt}
    \clearpage

    \textbf{Species 2: ROO$\cdot$}
        \VerbatimInput{SI_data/cyhepta24dien1yl/ROO_G4.txt}
    \clearpage

    \textbf{Species 3: $syn$-$\cdot$QOOH}
        \VerbatimInput{SI_data/cyhepta24dien1yl/QOOH_syn_G4.txt}
    \clearpage

    \textbf{Species 4: $anti$-$\cdot$QOOH }
        \VerbatimInput{SI_data/cyhepta24dien1yl/QOOH_anti_G4.txt}
    \clearpage

    \textbf{Species 5: {\it cy}-ether}
        \VerbatimInput{SI_data/cyhepta24dien1yl/cy_ether_G4.txt}
    \clearpage

    \textbf{Species 6: Transition state along \rxnone{}}
        \VerbatimInput{SI_data/cyhepta24dien1yl/TS_01_G4.txt}
    \clearpage

   \textbf{Species 7: Transition state along \rxnfour{}}
        \VerbatimInput{SI_data/cyhepta24dien1yl/TS_04_G4.txt}
    \clearpage

\subsection{Hydrocarbon 14: cyclohepta-3,5-dien-1-yl }

    \textbf{Species 1: $\cdot$R}
         \VerbatimInput{SI_data/cyhepta35dien1yl/R_G4.txt}
    \clearpage

    \textbf{Species 2: ROO$\cdot$}
        \VerbatimInput{SI_data/cyhepta35dien1yl/ROO_G4.txt}
    \clearpage

    \textbf{Species 3: $\cdot$QOOH}
        \VerbatimInput{SI_data/cyhepta35dien1yl/QOOH_G4.txt}
    \clearpage

    \textbf{Species 4: Transition state along \rxnone{}}
        \VerbatimInput{SI_data/cyhepta35dien1yl/TS_01_G4.txt}
    \clearpage

\subsection{Hydrocarbon 15: (cyclohexa-2,4-dien-1-yl)methyl (5-membered path)}

    \textbf{Species 1: $\cdot$R}
         \VerbatimInput{SI_data/cyhexa24dien1ylmethyl5mem/R_G4.txt}
    \clearpage

    \textbf{Species 2: ROO$\cdot$}
        \VerbatimInput{SI_data/cyhexa24dien1ylmethyl5mem/ROO_G4.txt}
    \clearpage

    \textbf{Species 3: $\cdot$QOOH}
        \VerbatimInput{SI_data/cyhexa24dien1ylmethyl5mem/QOOH_G4.txt}
    \clearpage

    \textbf{Species 4: Transition state along \rxnone{}}
        \VerbatimInput{SI_data/cyhexa24dien1ylmethyl5mem/TS_01_G4.txt}
    \clearpage

\subsection{Hydrocarbon 16: (cyclohexa-2,4-dien-1-yl)methyl (6-membered path)}

    \textbf{Species 1: $\cdot$R}
         \VerbatimInput{SI_data/cyhexa24dien1ylmethyl6mem/R_G4.txt}
    \clearpage

    \textbf{Species 2: ROO$\cdot$}
        \VerbatimInput{SI_data/cyhexa24dien1ylmethyl6mem/ROO_G4.txt}
    \clearpage

    \textbf{Species 3: $\cdot$QOOH}
        \VerbatimInput{SI_data/cyhexa24dien1ylmethyl6mem/QOOH_G4.txt}
    \clearpage

    \textbf{Species 4: Transition state along \rxnone{}}
        \VerbatimInput{SI_data/cyhexa24dien1ylmethyl6mem/TS_01_G4.txt}
    \clearpage
\bibliographystyle{apsrev4-1}
\bibliography{up_doi}